\documentclass[preprint,11pt,fleqn]{elsarticle}
\usepackage{threeparttable}
\usepackage{fancyvrb}
\usepackage{algpseudocode}
\usepackage{lmodern}
\usepackage{amsmath}
\usepackage{slantsc}
\usepackage{bold-extra}
\usepackage[T1]{fontenc}
\usepackage{float}
\usepackage{soul}
\usepackage{color}
\usepackage{bm}
\usepackage{xcolor}
\usepackage{microtype}
\usepackage{amsmath}
\usepackage{bbold}

\usepackage{booktabs}
\usepackage{multirow}
\usepackage{booktabs}
\usepackage{tabularx}

\def\redmem#1#2#3{  \left\langle #1 \left\Vert  
                  #2 \right\Vert #3 \right\rangle   }
\def\ketm#1{  \left\vert  #1   \right\rangle   }
\def\bram#1{  \left\langle  #1   \right\vert   }
 
\NewDocumentCommand{\dn}{e{_^}}{%
  _{\IfValueT{#1}{#1}\raisebox{0.26cm}{\smash[b]{}}}
  ^{\IfValueT{#2}{#2}\vphantom{\smash[t]{|}}}
}
\usepackage{amssymb}
\usepackage{amsthm}
\usepackage{amsmath,amssymb}
\usepackage{bm}
\usepackage[normalem]{ulem} 
\usepackage{hyperref}
\usepackage{multirow}
\usepackage{graphicx}

\usepackage{natbib}

\newcounter{bla}

\usepackage[dvipsnames]{xcolor}
\usepackage{subfig}
\usepackage{listings}
\makeatletter
\AtBeginDocument{%
  \let\c@figure\c@lstlisting
  
  \let\ftype@lstlisting\ftype@figure 
}
\makeatother

\usepackage{tikz}
\usetikzlibrary{shapes}

\usepackage{graphicx}
\makeatletter
\newcommand{\shorteq}{%
  \settowidth{\@tempdima}{-}
  \resizebox{\@tempdima}{\height}{=}%
}

\journal{Computer Physics Communications}

\begin{document}

\begin{frontmatter}



\title{Non-orthogonal extension of {\sc Graspg} -- dynamic electron correlation for large and compact active spaces}


\author[a,d]{Sijie Wu}
\author[a]{Ran Si \texorpdfstring{\corref{author}}{}}
\author[a]{Chongyang Chen}
\author[b]{Gediminas Gaigalas\texorpdfstring{\corref{author}}{}}
\author[c]{Michel~Godefroid}
\author[d]{Per J\"onsson \texorpdfstring{\corref{author}}{}}

\cortext[author] {Corresponding author.\\\textit{E-mail address:} rsi@fudan.edu.cn, gediminas.gaigalas@tfai.vu.lt, per.jonsson@mau.se}
\address[a]{
Shanghai EBIT Lab, Key Laboratory of Nuclear Physics and Ion-Beam Application, Institute of Modern Physics, Department of Nuclear Science and Technology, Fudan University, Shanghai 200433, China}
\address[b]{
Institute of Theoretical Physics and Astronomy, Vilnius University, Saulėtekio av. 3, LT-10222 Vilnius, Lithuania}
\address[c]{Spectroscopy, Quantum Chemistry and Atmospheric Remote Sensing, CP160/09, Universit\'e libre de Bruxelles, B-1050 Brussels, Belgium}
\address[d]{
Department of Materials Science and Applied Mathematics, Malmö University, SE-20506 Malmö, Sweden}

\begin{abstract}
\noindent
Accurate relativistic multiconfiguration calculations of correlation-sensitive atomic properties are often limited by the rapid growth of configuration state function expansions when a single common orthonormal orbital basis is used. In this work, a partitioned correlation function interaction (PCFI)
method is developed for relativistic atomic structure calculations. The correlation space is separated into physically motivated components, which are optimized independently with correlation-specific orbital sets. The interactions between configuration spaces constructed from mutually non-orthogonal orbital sets are evaluated using biorthonormal transformations, allowing different correlation effects to be combined in a compact final interaction calculation. Full details of the method are provided, with particular emphasis on its close connection to the recently introduced concept of configuration state function generators (CSFGs), which have previously been shown to significantly reduce the time required to construct the Hamiltonian matrix in conventional RCI calculations. 
Applications to the neutral Li, Be, and Al atoms are presented for energy levels, mass shifts and hyperfine structure constants. 
Compared with conventional relativistic configuration interaction (RCI) calculations that rely on a single orbital basis, PCFI produces more compact and predictable convergence patterns for both total and transition energies. It also offers greater stability for correlation‑sensitive properties such as specific mass shifts and hyperfine constants. By using property‑oriented partitions, PCFI captures core‑polarization effects more effectively, thereby reducing the oscillatory behavior often observed in standard RCI approaches. Overall, the results demonstrate that PCFI provides a promising and computationally efficient framework for accurate relativistic multiconfiguration calculations of correlation-dependent atomic
properties.
\end{abstract}


\end{frontmatter}

\section{Introduction}
There is an ever-increasing demand for accurate atomic data driven by advances in experimental and observational techniques in, e.g., plasma physics,  astrophysics, nuclear physics and fundamental physics.
To meet this demand, a number of computational methods have been developed. Among these methods are multiconfiguration Dirac-Hartree-Fock (MCDHF) and relativistic configuration interaction (RCI) \cite{IAN,Review,GRASPtheory}.
In the MCDHF and RCI methods, as implemented in the General-purpose Relativistic Atomic Structure Package ({\sc Grasp}) \cite{GRASP2018}, the wave function is expanded in $jj$-coupled configuration state functions (CSFs).
Key to the methods is the computation of the Hamiltonian matrix between the CSFs followed by the solution of the corresponding eigenvalue problem to give the expansion coefficients of the CSFs.  
In {\sc Grasp}, the computation of the matrix elements relies on spin-angular integrations using Racah algebra techniques, where the latter assume that the CSFs are built from a single orthonormal orbital set \cite{G1g,GG2,GRASPGG}. 
Accurate multiconfiguration calculations generally require a very large number of configuration state functions (CSFs) based on extensive orbital sets, making the computation of Hamiltonian matrix highly time-consuming.
To remedy this, configuration state function generators (CSFGs), which minimize the time spent on spin-angular integration, were introduced and previously shown to reduce the execution time for RCI calculations by factors of up to two hundred for CSF expansions built on large orbital sets \cite{CSFG}.
At the same time, the generators allow for {\em a priori} condensations of the CSF space that further adds to the usefulness.
Using \mbox{CSFGs} combined with the associated condensation technique, extremely accurate spectrum calculations with uncertainties of the transition energies down to a few cm$^{-1}$  have been performed for small systems (Be, B, C) based 
on expansions with hundreds of millions of CSFs built from large orbital sets, see for example \cite{Be}. Due to the rapid increase of the number of CSFs as a function of the  number of correlated electron pairs and the size of the orbital set, which basically
acts as a scaling wall, these results are difficult to extend to larger atomic systems (group-VIII elements, lanthanides, and actinides with
open $f$-shell structures). Indeed, for larger systems, the rapid increase of the number of CSFs with the size of the orbital set often restricts the calculations to include 
only valence-valence and core-valence electron correlation. Even so, it can sometimes be difficult to saturate the correlation effects with respect to the orbital set, which impairs 
the accuracy of the computed properties \cite{MgI,Sonetal:2024a,LiI}. 

A way to break the scaling wall, at least partially,  is to rely on non-orthogonal orbitals and orbital sets.  Froese Fischer~\cite{Fro:81a} observed that the convergence of CI expansions is faster if the orbitals associated with different pair-correlations effects are not required to be orthogonal. By extending the underlying Racah algebra, limited non-orthogonalities were included in the non-relativistic multiconfiguration Hartree-Fock (MCHF) program \cite{Hibetal:88a}. Non-orthogonal orbitals were later used to obtain efficient and compact representations of the two-electron valence wave functions in alkaline earths~(see e.g. refs~\cite{FroGod:82a,Vaeetal:88a}).  
In the relativistic context, non-orthogonal orbitals were  explored by Zatsarinny and Froese Fischer in the Breit-Pauli approximation~\cite{ZatFro:00b} and by Desclaux and Indelicato in the MCDHF approach~\cite{MCDFGME}. In these works, the underlying idea of introducing non-orthogonal orbitals is not only to ensure the correctness of the transition matrix element calculation between separately optimized wave functions, but also to speed up convergence by associating non-orthogonal orbitals with different configurations in the many-configuration expansion. In both codes~\cite{ZatFro:00b,MCDFGME}, the wave function expansions are written in terms of Slater determinants, and the determination of the matrix elements between CSFs is reduced to
that of the matrix elements between separate Slater determinants, which, in turn, are reduced to one- or two-electron integrals between spin orbitals. This step is performed
using the cofactor method derived by Löwdin~\cite{Low:55a}. Compared to the {\sc Grasp} code based on Racah algebra techniques, the resulting formulae explicitly include the magnetic quantum numbers and contain additional summations over Slater functions and over all electrons but not shells. As observed in~\cite{ZatFro:00b}, this considerably increases the
number of operations required, making the calculation slow.

A different way to handle non-orthogonal orbital sets, which allows 
Racah algebra to be used for the computation of the matrix elements,   
is provided by a biorthonormal transformation technique developed by Malmqvist \cite{Mal:86a,Olsetal:95a}.  Based on this technique, Verdebout {\em et al.} \cite{Veretal:2013a} devised a divide-and-conquer method in which 
each separate pair-correlation effect is described by CSFs built on their own optimally located, and thus compact, orthonormal orbital set. The orbital sets associated with different pair-correlation effects are not required to be orthonormal, but are transformed to biorthonormal sets, in which case the computation of the Hamiltonian matrix elements proceeds as usual based on ordinary Racah algebra. Whereas the total number of orbitals used to describe correlation may be large, the number of orbitals used to build the different parts of the CSF space is small, reducing the  expansion sizes. 
Similar computationally efficient methods based on biorthonormal expansions of orbitals and CI wave functions have been developed recently in quantum chemistry (see ~\cite{KahOls:2017a} and references therein).

In the present paper, we describe the relevant theory, algorithms and tools used to implement the divide-and-conquer method into {\sc Graspg}, which is the latest version of {\sc Grasp}  based on  generators~\cite{GRASPG}, and establish a framework for treating different electron-correlation effects with compact, separately optimized active spaces. Special attention is paid to the role of the generators, which allow the biorthonormal counter transformations to be organized in a block-diagonal form and thereby provide a basis for an efficient implementation. A number of applications are given, emphasizing various aspects of the divide-and-conquer method.

The present work focuses on the formulation, validation, and convergence properties of the relativistic PCFI method. Since the generator structure has not yet been integrated throughout the complete computational workflow, the present assessment emphasizes the compactness of the required correlation expansions and the convergence behavior of the calculated properties. A fully generator-based implementation and the codes, together with a systematic analysis of computational performance, will be addressed in future work.

\section{Relativistic multiconfiguration methods based on generators}
\subsection{Multiconfiguration methods}
In relativistic multiconfiguration calculations, the atomic state function for a targeted state with quantum label $\Gamma JM_J$ is obtained as an expansion over $jj$-coupled configuration state functions (CSFs)
\begin{equation} \label{eq:jj_ASF}
\Psi(\Gamma JM_J)= \sum_{\alpha=1}^M c_{\alpha} \Phi_{\alpha}(\gamma_{\alpha} JM_J).
\end{equation}
Here $\gamma_{\alpha}$ specifies the orbital occupancy and spin-angular coupling tree quantum numbers of each CSF, whereas $M$ gives the number of CSFs.
The CSFs are constructed from products of the Dirac one-electron orbitals 
\begin{equation}\label{eq:PQ}
\phi_{a} {(\boldsymbol{r})} =  \frac{1}{r}\left(
\begin{array}{c}
P_{nlj}(r)\Omega_{\kappa m}(\theta,\varphi)\\
\mbox{i}~Q_{nlj}(r)\Omega_{-\kappa m}(\theta,\varphi)
\end{array}
\right),
\end{equation}
for which \emph{P} and \emph{Q} are the radial functions and $\Omega$ is the spin-angular function \cite{Review}. 
The label $a$ of the orbital is an abbreviation of $n, \kappa, m$. Requiring the weighted energy functional of the targeted states, together with additional terms
to ensure orthonormality of the orbitals,
to be stationary with respect to small variations of the radial functions gives a set of coupled integro-differential equations for the latter. 
The integro-differential equations are coupled with a  matrix
eigenvalue problem
\begin{equation}
{\bm H}{\bm c} =  E {\bm c},
\end{equation}
where ${\bm H}$ is the Hamiltonian matrix
with elements
$H_{\alpha\alpha'} = \langle \Phi_{\alpha}(\gamma_{\alpha} J) \| {\cal H}\| \Phi_{\alpha'}(\gamma_{\alpha'} J) \rangle$ and ${\bm c} = (c_1,\ldots,c_M)^T$ is the column vector of expansion coefficients.
For simplicity, the Hamiltonian operator, $ {\cal H}$, is often taken as the  Dirac-Coulomb operator.
The integro-differential equations and the eigenvalue problem are solved in a self-consistent field procedure until convergence 
is attained. The above method is referred to as the MCDHF method~\cite{Review,GRASPtheory}. If the radial orbitals are known from a
previous calculation, and only the eigenvalue problem is solved, the method is referred to as the RCI method. At this stage the Breit interaction and quantum electrodynamic (QED) corrections may be added to the Hamiltonian operator \cite{IAN,GRASPtheory,BreitGrant}.
\subsection{Generators}\label{sec:generators}
The computation of Hamiltonian matrix elements between the CSFs in the wave function expansion, a time-consuming part of both MCDHF and RCI calculations, relies on
spin-angular integration. The latter, in its fast implementation based on Racah algebra \cite{G1g,GG2,GRASPGG}, assumes that the CSFs are built on a single orthonormal orbital set. 
As shown in \cite{CSFG}, the time for the spin-angular integration can be cut down substantially for broad classes of CSF expansions 
by applying the so-called generator method.
In this method, as implemented in {\sc Graspg} \cite{GRASPG}, 
the CSF space is divided in a 
labeling space and a correlation space. The CSFs in the labeling space
are obtained by allowing a general class of excitations (SDTQ etc.) from occupied orbitals   
in a set of multireference (MR) configurations to a limited labeling-ordered (LO) set of highly occupied orbitals.\footnote{Instead of labeling ordered such a set is also referred to as spectroscopically ordered.}
The CSFs in the correlation space are 
obtained by SD excitations of occupied orbitals of the MR to a symmetry-ordered (SO) set ($s$, $p$-, $p$, $d$-, $d$ etc.) of correlation orbitals.  
The division of the orbitals into a set of LO orbitals augmented by a set of SO  orbitals allows the correlation space to be split into groups of CSFs, where the CSFs in each group are obtained 
by spin-angular coupling preserving orbital de-excitations within the SO set of occupied orbitals of a generating CSF. The latter we will sometimes simply refer to as a generator, and accordingly we will sometimes talk about CSFs spanned by the generator. The generating CSFs are of four types: generating CSF has one orbital in the SO set, with the $n$ quantum number at its highest value as allowed by the set, generating CSF has two orbitals of different symmetries 
in the SO set, with
the $n$ quantum numbers at their highest values, generating CSF has two orbitals of the same symmetry in the SO set, with
the $n$ quantum numbers at their highest and second-highest values, generating CSF has a doubly occupied orbital 
in the SO set, with the $n$ quantum number at its highest value. Below, as a minimal example, we show a list of CSFs 
in  {\sc Graspg} format for which $\{1s,2s,2p\mbox{-}\}$ are the orbitals in the LO set and
$\{3s,4s,5s,3p\mbox{-},4p\mbox{-},5p\mbox{-}\}$ are the orbitals in the SO set.  There is one CSF in the labeling space, 
built entirely on orbitals in the LO set, and four generating CSFs of types 1, 2, 3, and 4 built on orbitals in the LO and SO sets.

\begin{Verbatim}[fontsize=\footnotesize]
Core subshells:

Peel subshells:
  1s   2s   2p-  3s   4s   5s   3p-  4p-  5p-
CSF(s):
  1s ( 2)  2s ( 1)  2p-( 1)             <-- CSF in labeling space
               1/2      1/2
                           0-
  1s ( 2)  2s ( 1)  5p-( 1)             <-- generating CSF type 1
               1/2      1/2
                           0-													
  1s ( 2)  5s ( 1)  5p-( 1)             <-- generating CSF type 2
               1/2      1/2
                           0-
  2s ( 1)  2p-( 1)  4s ( 1)  5s ( 1)    <-- generating CSF type 3
      1/2      1/2      1/2      1/2 
                    0      1/2      0- 
  2s ( 1)  2p-( 1)  5s ( 2)             <-- generating CSF type 4
      1/2      1/2 
                    0      0-
\end{Verbatim}
\noindent
From the generating CSF of type 1 we have the following group of three CSFs obtained by spin-angular coupling preserving de-excitations of the $5p\mbox{-}$ orbital within the $p\mbox{-}$ symmetry of the SO set
\begin{Verbatim}[fontsize=\footnotesize]
  1s ( 2)  2s ( 1)  3p-( 1)    
               1/2      1/2
                           0-
  1s ( 2)  2s ( 1)  4p-( 1)    
               1/2      1/2
                           0-
  1s ( 2)  2s ( 1)  5p-( 1)    
               1/2      1/2
                           0-
\end{Verbatim}
\noindent
From the generating CSF of type 2 we have the following group of nine CSFs obtained by spin-angular coupling preserving de-excitations of the  $5s$ and $5p\mbox{-}$ orbitals within, respectively, the
$s$ and $p\mbox{-}$ symmetries of the SO sets
\begin{Verbatim}[fontsize=\footnotesize]
  1s ( 2)  3s ( 1)  3p-( 1)    
               1/2      1/2
                           0-
  1s ( 2)  3s ( 1)  4p-( 1)    
               1/2      1/2
                           0-
             ...
                           
  1s ( 2)  5s ( 1)  5p-( 1)    
               1/2      1/2
                           0-
\end{Verbatim}        
\noindent
In the group of CSFs spanned by the generator, the principal quantum number of the rightmost orbital moves more rapidly than the one for the second rightmost orbital.
From the generating CSF of type 3 we have the following group of three CSFs obtained by spin-angular coupling preserving de-excitations of $4s$ and $5s$ within the $s$ symmetry of the SO set
\begin{Verbatim}[fontsize=\footnotesize]
  2s ( 1)  2p-( 1)  3s ( 1)  4s ( 1)   
      1/2      1/2      1/2      1/2 
                    0      1/2      0-
  2s ( 1)  2p-( 1)  3s ( 1)  5s ( 1)   
      1/2      1/2      1/2      1/2 
                    0      1/2      0-
  2s ( 1)  2p-( 1)  4s ( 1)  5s ( 1)   
      1/2      1/2      1/2      1/2 
                    0      1/2      0- 
\end{Verbatim} 
Finally, from the generating CSF of type 4 we have the group below of three CSFs obtained by spin-angular coupling preserving de-excitations of $5s$ 
\begin{Verbatim}[fontsize=\footnotesize]
  2s ( 1)  2p-( 1)  3s ( 2)             
      1/2      1/2 
                    0      0-
  2s ( 1)  2p-( 1)  4s ( 2)             
      1/2      1/2 
                    0      0-
  2s ( 1)  2p-( 1)  5s ( 2)            
      1/2      1/2 
                    0      0-
\end{Verbatim} 
Data obtained from the spin-angular integration between two generating CSFs allow all matrix elements between the CSFs derived from them through orbital de-excitations that preserve the spin-angular coupling to be computed directly by applying the corresponding orbital de-excitations at the integral level. As an example, we consider the matrix elements between the CSFs obtained from the first and second generating CSFs above. 
Spin-angular integration resolves the Dirac-Coulomb matrix element between the two generating CSFs into spin-angular coefficients and radial one-electron integrals $I(a,b)$ and two-electron Slater integrals $R^k(ab;cd)$ as shown below.
\begin{Verbatim}[fontsize=\footnotesize]
    1.000000000 I(2s ,5s )
    1.000000000 R0(2s 5p-,5s 5p-)
   -0.333333333 R1(2s 5s ,5p-5p-)
    2.000000000 R0(1s 2s ,1s 5s )
   -1.000000000 R0(1s 1s ,2s 5s )
\end{Verbatim}
The computer fetches the radial integrals from memory, performs the multiplication, and sums the result. The remaining matrix elements between CSFs for which the symmetry ordered $p\mbox{-}$ orbital for the left-hand CSF is the same as for the right-hand CSF follow directly by 
keeping the spin-angular coefficients, multiplying with radial integrals involving orbitals obtained by the appropriate de-excitations of $5s$ and $5p\mbox{-}$. In conventional generator-based RCI calculations, the reuse of spin-angular data substantially reduces repeated spin-angular integrations and has previously yielded speed-up factors of 100 or more for sufficiently large CSF expansions \cite{GRASPG,CSFG,YTLPRA}.

\section{Electron correlation and divide-and-conquer methods}
\subsection{Electron correlation based on a single orbital set}
\begin{figure}[ht]
\centering
\makebox[\textwidth][c]{%
    \hspace{-5mm}%
    \includegraphics[scale=0.269]{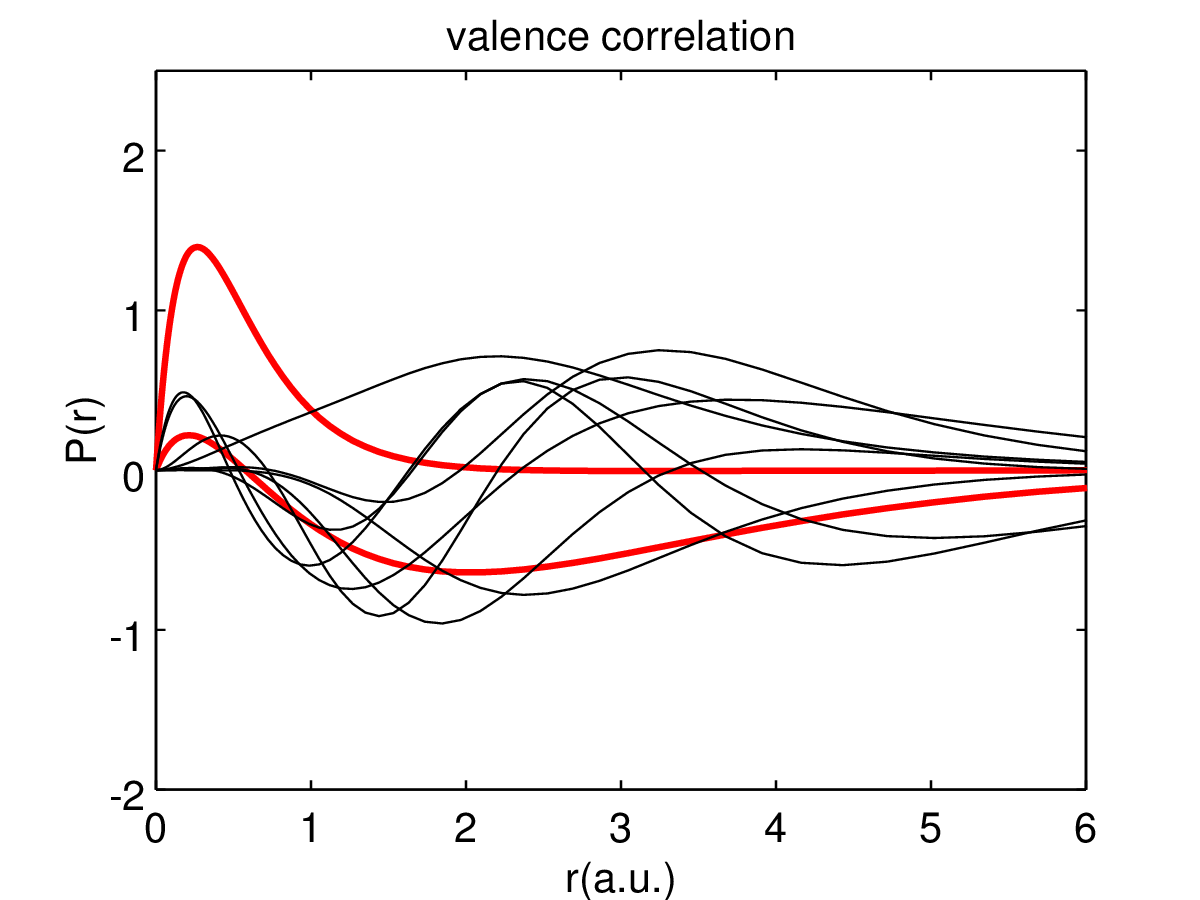}%
    \hspace{-10mm}%
    \includegraphics[scale=0.269]{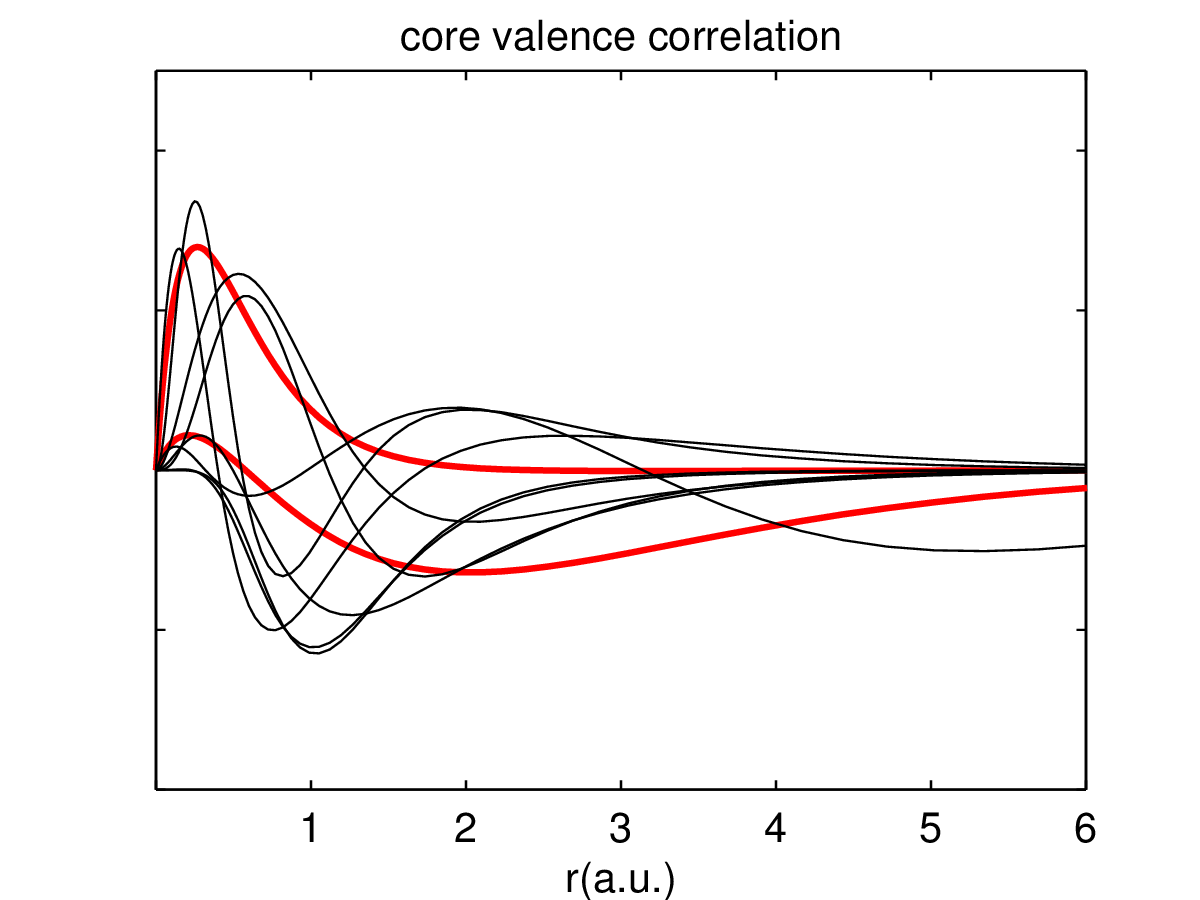}%
    \hspace{-10mm}%
    \includegraphics[scale=0.269]{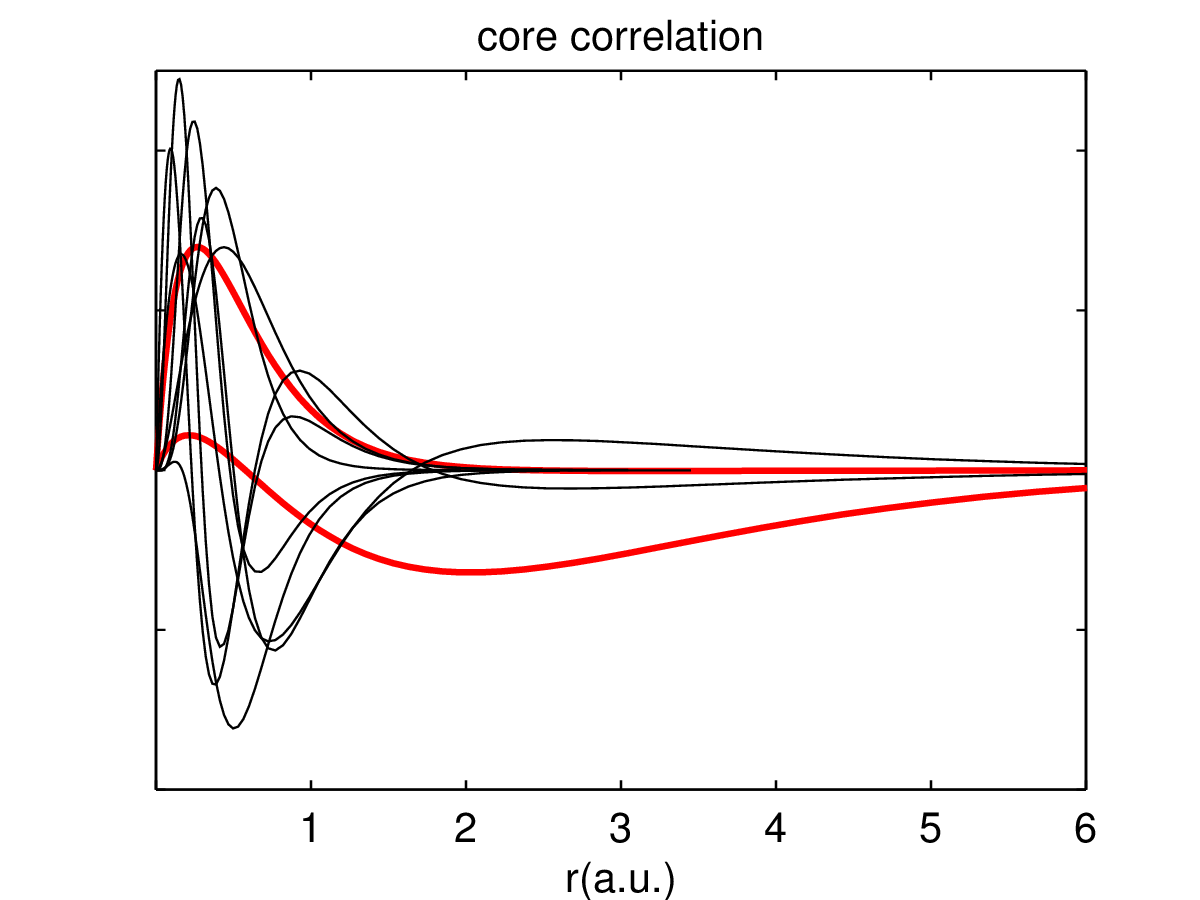}%
}
\caption{Location of the correlation orbitals from VV, CV and
CC correlation MCHF calculations for $1s^2 2s^2~{}^1S$ in Be I.
The two thick red lines correspond to the spectroscopic $1s$
(no node) and $2s$ (one node) orbitals. The other lines represent
the radial distributions of the correlation orbitals of the
$n=4$ active set. Figure adapted from \cite{Veretal:2010a}.}
\label{fig:delta_conv}
\end{figure}

In the generator expansions described above, the CSFs associated with the configurations in the MR account for long-range re-arrangement
of the electron charge distribution that arises from near degeneracies of the Dirac-Hartree-Fock energies. 
The CSFs in the correlation space account for dynamic correlation, a short-range effect that arises from the singularity of the 
 electron-electron interaction at points of coalescence \cite{Review}. Assume, for simplicity, that we have only one configuration in the MR and let $\{a, b, c, ...\}$
denote the occupied orbitals in the LO set and $\{v, v', ...\}$ the orbitals in the SO set of correlation orbitals (active set). 
The function defined by CSFs obtained from single excitations (replacements) $a \rightarrow v$, which we will refer to as a
polarization function (POLF),
accounts for spin- and orbital polarization effects as well as radial correlation. 
The function defined by CSFs from all double excitations $ab \rightarrow vv'$ is called a pair
correlation function (PCF), and it corrects for the cusp in the wave function associated
with the $ab$  electron pair \cite{Review}. The PCFs from all electron pairs correct for the main part of
the dynamic correlation. The PCFs are further classified according to whether the excitations are from valence or core orbitals \cite{Review,Froetal:97b}:
\begin{enumerate}
 \item If $a$ and $b$ are orbitals corresponding to outer electrons, the excitations describe outer, or valence--valence (VV), correlations.
 \index{valence correlation}

 \item If $a$ is a core orbital and $b$ is an outer orbital, the excitations describe core--valence (CV) correlations.
 \index{core-valence correlation}

 \item If both orbitals are core orbitals, the excitations describe core--core (CC) correlations.
 \index{core-core correlation}
\end{enumerate}
As is well known, see for example \cite{Godetal:98a,Veretal:2010a}, the location and shape of the correlation orbitals depend on the energy functional, 
and thus on the CSF expansion used to derive the integro-differential equations. As an illustration, orbitals from non-relativistic MCHF for the $1s^22s^2~^1S$  ground state in Be I based on PCFs describing VV, CV, and CC correlations, respectively, are depicted in Fig.~\ref{fig:delta_conv}. One clearly sees the contraction
of the correlation orbitals when going from a VV to a CC correlation calculation. Obviously, orbitals in the valence region are ill suited to describe correlations in the core region and vice versa.

With this picture in mind, we are now in a position to discuss the main limitations of multiconfiguration methods based on the optimization of a single orthonormal set of orbitals.
As already mentioned in the introduction, the orbital basis must be extended
to large numbers of orbitals of each symmetry, leading to unmanageable CSF expansions, particularly for atoms where many electron-pairs are to be correlated.
In addition, because multiconfiguration calculations are energy-driven, correlation contributions that are less important for the energy may nevertheless be crucial for other properties and therefore not adequately captured through the optimization of a single (truncated) orbital basis \cite{Veretal:2013a,LV,LiI}.

\subsection{Many non-orthogonal orbital sets and divide-and-conquer methods}

As demonstrated by Verdebout {\em et al.} \cite{Veretal:2013a} in the non-relativistic
context, the biorthonormal transformation method developed by Malmqvist \cite{Mal:86a,Olsetal:95a} can be used to relax
the orthogonality constraint of the orbitals such that correlation effects can be described by several 
mutually non-orthogonal sets of correlation orbitals, 
each optimally localized for the correlation effect under consideration. This divide-and-conquer strategy, applied to the fully relativistic MCDHF method, can be summarized as follows:

\begin{enumerate}
 \item Partition the CSFs in the correlation space into subspaces corresponding to different PCFs and POLFs.\footnote{If desired, the POLFs
 can be merged into the PCFs that now will include CSFs obtained by single and double excitations.}
 The CSFs of the PCFs and POLFs are given in terms of generators.
 \item Perform separate MR-MCDHF calculations for each expansion built on the MR together with the CSFs of the PCFs or POLFs.
 If needed, and this is another important quality of the method, the {\tt rmixaccumulate\_csfg} program of {\sc Graspg} \cite{GRASPG} can  be used to
 {\em a priori} extract the most important
 generators and thus CSFs at this early stage of the calculation.
 \item Expand the wave function in CSFs from the MR and CSFs from all PCFs and POLFs. The expansion coefficients are obtained
by computing the Hamiltonian matrix and solving the corresponding eigenvalue problem. The computation of the matrix elements between CSFs
in the different PCFs and POLFs, built on different and mutually non-orthogonal orbital sets, is performed using biorthonormal transformations. 
\end{enumerate}
To specify the different PCFs and POLFs we introduce a curly bracket notation with a superscript S, D or SD to denote the number of excitations from the orbitals inside the bracket. With reference to
$1s^22s^2~^1S_0$ the following 
\begin{equation}
\{vv\}^{\mbox{\scriptsize SD}}, \{1sv\}^{\mbox{\scriptsize D}}, \{1s1s\}^{\mbox{\scriptsize SD}}
\end{equation}
corresponds to one partition accounting for VV correlation including radial correlation, one accounting for CV correlation and one accounting for CC correlation including spin- and orbital polarization.
Another possible partition would be
\begin{equation}
\{vv\}^{\mbox{\scriptsize SD}}, \{1sv\}^{\mbox{\scriptsize D}}, \{1s1s\}^{\mbox{\scriptsize D}},  \{1s\}^{\mbox{\scriptsize S}},
\end{equation}
where now  spin- and orbital polarization of the core is a partition of its own.
It should be noted that
even if the CSFs of the PCF and POLF partitions are built from different orbital sets, they form an orthonormal set. Details can be found in \cite{Veretal:2013a} and its appendix A. Thus, we have an ordinary eigenvalue problem.
Whereas the non-orthogonal divide-and-conquer method comes with many advantages, it also comes with a price: the cost for performing the biorthonormal transformations to build the Hamiltonian matrix. As we will discuss in the next subsection, this price is comparatively low for generator expansions, which results in small groups of CSFs that each supports the transformation. 

\section{Building the Hamiltonian in the divide-and-conquer method}\label{sec:DQ}

In the generator-based divide-and-conquer method, the CSFs in the correlation space
are denoted $\Phi_{\alpha}^{pg}$. The first index, $p$, with $p= 1,\ldots,N$ determines to which PCF or POLF the CSF belongs. Then, for a given $p$, the second index, $g$,  with $g= 1,\ldots,N^p$ determines to which generator the CSF belongs. Finally, for a given $p$ and $g$, the third index $\alpha$, with $\alpha = 1,\ldots, N^{pg}$ labels the CSFs spanned by the generator.
We will now show how to compute the Hamiltonian between these CSFs. 
We start with the submatrix between two groups of CSFs obtained from
two generating CSFs, $g$ and $g'$, belonging to different PCFs or POLFs, $p$ and $p'$.
In the next subsection we will then put everything together and devise an algorithm for computing the full Hamiltonian matrix.

\subsection{Biorthonormal weight matrix method for CSFs in a generator group}\label{sec:BG}

Let 
$\{\Phi^{pg}_{\alpha}\}$ and $\{\Phi^{p'g'}_{\alpha'}\}$  be two groups of CSFs with, respectively, $N^{pg}$ and $N^{p'g'}$ elements obtained from two generating CSFs, $g$ and $g'$. Assume further
that the generating CSFs are parts of two  different PCFs, $p$ and $p'$, built on different and mutually non-orthogonal orbital sets
$\{\phi^p_{n\kappa} \}$ and $\{\phi^{p'}_{n'\kappa'} \}$. Our task is to
compute the $N^{pg} \times N^{p'g'}$ Hamiltonian submatrix 
\begin{equation}
{\bm H}^{pg,p'g'}
\end{equation}
with elements  $H^{pg,p'g'}_{\alpha\alpha'} = 
\langle \Phi^{pg}_{\alpha} \| {\cal H}\| \Phi^{p'g'}_{\alpha'} \rangle$.
For reasons that will be clear from appendix A, we premultiply the Hamiltonian matrix with an $N^{pg} \times N^{pg}$ unit expansion coefficient matrix, ${\bm C}^{pg}$, to the left and with
an $N^{p'g'} \times N^{p'g'}$ unit expansion coefficient matrix, ${\bm C}^{p'g'}$, to the right
\begin{equation}
{\bm H}^{pg,p'g'} \equiv ({\bm C}^{pg})^T{\bm H}^{pg,p'g'}{\bm C}^{p'g'}.
\end{equation}
The CSFs in the two sets are closed under orbital de-excitation (CUD) within each orbital symmetry and, as shown by Malmqvist and collaborators~\cite{Mal:86a,Olsetal:95a},
it is possible to transform the two orbital sets
\begin{equation}
\{\phi^p_{n\kappa} \} \rightarrow \{ \widetilde \phi\,^p_{n\kappa} \}~~~\mbox{and}~~~\{\phi^{p'}_{n'\kappa'} \} \rightarrow \{ \widetilde \phi\,^{p'}_{n'\kappa'} \}
\end{equation}
to become biorthonormal,
$\langle \widetilde \phi^p_{n\kappa} | \widetilde \phi^{p'}_{n'\kappa} \rangle = \delta_{n,n'}$
and at the same time counter transform 
the unit coefficient matrices, for both ${\bm C}^{pg}$ 
\begin{equation}\label{eq:T1}
\underbrace{\left(
\begin{array}{ccccc}
1 & 0 & 0 & \cdots & 0\smallskip\\
0 & 1 & 0 & \cdots & 0  \smallskip\\
0 & 0 & 1 & \cdots & 0  \smallskip\\
\vdots & \vdots & \vdots & \ddots & \vdots\\
0 & 0 & 0 & \cdots & 1 \\
\end{array}
\right)}_{{\bm C}^{pg}} \rightarrow
\underbrace{
\left(
\begin{array}{ccccc}
\widetilde c\,_{11}^{pg} & \widetilde c\,_{12}^{pg} & \widetilde c\,_{13}^{pg} & \cdots & \widetilde c\,_{1N^{pg}}^{pg}\smallskip\\
\widetilde c\,_{21}^{pg} & \widetilde c\,_{22}^{pg} & \widetilde c\,_{23}^{pg} & \cdots & \widetilde c\,_{2N^{pg}}^{pg}\smallskip\\
\widetilde c\,_{31}^{pg} & \widetilde c\,_{32}^{pg} & \widetilde c\,_{33}^{pg} & \cdots & \widetilde c\,_{3N^{pg}}^{pg}\smallskip\\
\vdots & \vdots & \vdots & \ddots & \vdots\\
\widetilde c\,_{N^{pg}1}^{pg} & \widetilde c\,_{N^{pg}2}^{pg} & \widetilde c\,_{N^{pg}3}^{pg} & \cdots & \widetilde c\,_{N^{pg}N^{pg}}^{pg}\\
\end{array}
\right)}_{\widetilde {\bm C}^{pg}}
\end{equation}
and ${\bm C}^{p'g'}$
\begin{equation}\label{eq:T2}
\underbrace{\left(
\begin{array}{ccccc}
1 & 0 & 0 & \cdots & 0\smallskip\\
0 & 1 & 0 & \cdots & 0  \smallskip\\
0 & 0 & 1 & \cdots & 0  \smallskip\\
\vdots & \vdots & \vdots & \ddots & \vdots\\
0 & 0 & 0 & \cdots & 1 \\
\end{array}
\right)}_{{\bm C}^{p'g'}} \rightarrow
\underbrace{
\left(
\begin{array}{ccccc}
\widetilde c\,_{11}^{p'g'} & \widetilde c\,_{12}^{p'g'} & \widetilde c\,_{13}^{p'g'} & \cdots & \widetilde c\,_{1N^{p'g'}}^{p'g'}\smallskip\\
\widetilde c\,_{21}^{p'g'} & \widetilde c\,_{22}^{p'g'} & \widetilde c\,_{23}^{p'g'} & \cdots & \widetilde c\,_{2N^{p'g'}}^{p'g'}\smallskip\\
\widetilde c\,_{31}^{p'g'} & \widetilde c\,_{32}^{p'g'} & \widetilde c\,_{33}^{p'g'} & \cdots & \widetilde c\,_{3N^{p'g'}}^{p'g'}\smallskip\\
\vdots & \vdots & \vdots & \ddots & \vdots\\
\widetilde c\,_{N^{p'g'}1}^{p'g'} & \widetilde c\,_{N^{p'g'}2}^{p'g'} & \widetilde c\,_{N^{p'g'}3}^{p'g'} & \cdots & \widetilde c\,_{N^{p'g'}N^{p'g'}}^{p'g'}\\
\end{array}
\right)}_{\widetilde {\bm C}^{p'g'}}
\end{equation}
such that the Hamiltonian submatrix ${\bm H}^{pg,p'g'}$ is invariant, i.e.
\begin{equation}\label{eq:hamno}
    {\bm H}^{pg,p'g'} \equiv ({\bm C}^{pg})^T{\bm H}^{pg,p'g'}{\bm C}^{p'g'} \equiv (\widetilde {\bm C}\,^{pg})^T 
    \widetilde {\bm H}^{pg,p'g'} \widetilde {\bm C}\,^{p'g'}.
\end{equation}
Here $\widetilde {\bm H}^{pg,p'g'}$ is the reduced matrix with elements
$\widetilde H^{pg,p'g'}_{\alpha\alpha'} = \langle {\widetilde \Phi}^{pg}_{\alpha} \| {\cal H}\| {\widetilde \Phi}^{p'g'}_{\alpha'} \rangle$,
where the CSFs are built on the transformed radial orbitals. Since the transformed radial orbitals are biorthonormal, we can use the
method devised in section~\ref{sec:generators} for the evaluation of $\widetilde H^{pg,p'g'}_{\alpha\alpha'}$. Note however that the orbital indices $a$ and $b$ of the one-electron integrals  $I(a,b)$ now refer to orbitals belonging to 
$\{ \widetilde \phi\,^p_{n\kappa} \}$ and $\{ \widetilde \phi\,^{p'}_{n'\kappa'} \}$, respectively. In a similar way, the orbital indices $(a,b)$ and $(c,d)$ of the two-electron integrals  $R^k(ab;cd)$ refer to orbitals belonging to 
$\{ \widetilde \phi\,^p_{n\kappa} \}$ and $\{ \widetilde \phi\,^{p'}_{n'\kappa'} \}$, respectively. Minor changes are therefore required in the spin-angular part of the reduced matrix element $\widetilde {\bm H}^{pg,p'g'}$ to eliminate the original permutation symmetry properties that only hold when the l.h.s. and r.h.s. are built on a single orbital set.

In terms of data, the biorthogonal transformation requires the radial overlap matrix, ${\bm S}^{\kappa}$,
for each $\kappa$ to compute the radial orbital transformation matrices that will make the two orbital sets biorthonormal.
The overlap matrices will also give the so-called Malmqvist ${\bm T}^{\kappa}$ matrices. The latter,
together with the one-electron spin-angular coefficients
for the interaction between the CSFs in $\{\Phi^{pg}_{\alpha}\}$ and the interaction between the CSFs in $\{\Phi^{p'g'}_{\alpha'}\}$ 
are needed to obtain the expansion coefficient counter transformation matrices ${\bm C}^{pg}$ and ${\bm C}^{p'g'}$, respectively. 
Details of the biorthonormal transformation
are given in \ref{Appendix_A}. Already now, as shown in  \ref{Appendix_B}, we should note that the transformation matrices need not
be computed for each pair of generators $g$ and $g'$, but can be taken from pre-computed templates, thereby reducing repeated computational work.
  
\subsection{Building the full Hamiltonian }\label{sec:algorithm}
We will now build the full Hamiltonian matrix. As a simple illustration
we again look at $1s^22s^2~^1S_0$ in Be I. Denote the set of
CSFs in the MR by $\{\Phi^\textsc{mr}\}$, the set of CSFs obtained by generators in the $\{vv\}^{\mbox{\scriptsize SD}}$ PCF by $\{\Phi^\textsc{vv}\}$, the set of CSFs
obtained from generators 
in the $\{1sv\}^{\mbox{\scriptsize D}}$ PCF by
$\{\Phi^\textsc{cv}\}$ and, finally, the set of CSFs obtained from generators in the $\{1s1s\}^{\mbox{\scriptsize SD}}$ PCF
by $\{\Phi^\textsc{cc}\}$. The resulting Hamiltonian can now schematically be represented as

\renewcommand{\theequation}{\text{\normalsize\arabic{equation}}}
\begin{equation}
\begingroup
\scriptsize
{\bm H} =
\left(
\begin{array}{cccc}
\framebox(71,15)[]{$\langle \{\Phi^\textsc{mr}\} \|{\cal H}\| \{\Phi^\textsc{mr}\} \rangle$} \hspace{-1mm} &  \framebox(71,15)[]{$\langle \{\Phi^\textsc{mr}\} \|{\cal H}\| \{\Phi^\textsc{vv}\} \rangle$} \hspace{-1mm} & 
\framebox(71,15)[]{$\langle \{\Phi^\textsc{mr}\} \|{\cal H}\| \{\Phi^\textsc{cv}\} \rangle$} & \hspace{-1mm}  \framebox(71,15)[]{$\langle \{\Phi^\textsc{mr}\} \|{\cal H}\| \{\Phi^\textsc{cc}\} \rangle$} \\
 & \framebox(71,15)[]{$\langle \{\Phi^\textsc{vv}\} \|{\cal H}\| \{\Phi^\textsc{vv}\} \rangle$} \hspace{-1mm} & \dashbox(71,15)[]{$\langle \{\Phi^\textsc{vv}\} \|{\cal H}\| \{\Phi^\textsc{cv}\} \rangle$} \hspace{-1mm}  & 
 \dashbox(71,15)[]{$\langle \{\Phi^\textsc{vv}\} \|{\cal H}\| \{\Phi^\textsc{cc}\} \rangle$} \\
  &   & \framebox(71,15)[]{$\langle \{\Phi^\textsc{cv}\} \|{\cal H}\| \{\Phi^\textsc{cv}\} \rangle$} \hspace{-1mm}  & \dashbox(71,15)[]{$\langle \{\Phi^\textsc{cv}\} \|{\cal H}\| \{\Phi^\textsc{cc}\} \rangle$} \\
    &   &  & \framebox(71,15)[]{$\langle \{\Phi^\textsc{cc}\} \|{\cal H}\| \{\Phi^\textsc{cc}\} \rangle$} \\
\end{array}
\right)
\endgroup
\end{equation}

\noindent
where all plain-line blocks  involve a single orthonormal orbital set. This holds not only for the diagonal blocks, but also for blocks coupling the PCF and the MR spaces since we do not allow the MR orbitals to vary in the MR-MCDHF calculations for the partitions.
 The building of all other blocks, surrounded by a dashed line, involve non-orthogonal orbitals arising from independent MCDHF calculations and requires, therefore, the use of biorthonormal transformations before the usual methods for spin-angular integration  can be applied. For ease of bookkeeping, all CSFs obtained by excitations to MR-orbitals only, regardless of which PCF they belong to, are transferred to the MR space. In the description
 to come we assume that this has been done.
The interactions between the PCFs and the MR, or more generally a larger space built entirely on LO orbitals, are straightforward to compute using standard methods available in {\sc Grasp}.
We will thus not discuss this part anymore but instead focus on the interaction between the PCFs.
 In the case with $N$ PCFs and POLFs, the algorithm for constructing the Hamiltonian matrix involves loops over PCFs and POLFs as well as over groups and can be expressed as follows:
 \begin{algorithmic}
 \For{$p = 1,\ldots,N$}
       \For{$p' = p,\ldots, N$}
           \If{$p' > p$} \Comment off-diagonal case
               \State perform the biorthonormal transformation to obtain 
               \State $\{ \widetilde \phi\,^{p}_{n\kappa} \}$, $\{ \widetilde \phi\,^{p'}_{n'\kappa'} \}$ and the
               ${\bm T}^{\kappa}$ matrices\smallskip
               \State compute radial integrals based on $\{ \widetilde \phi\,^{p}_{n\kappa} \}$, $\{ \widetilde \phi\,^{p'}_{n'\kappa'} \}$\smallskip
               \State compute ${\bm C}^{pg}$ and ${\bm C}^{p'g'}$ templates\smallskip
               \For{$g = 1,\ldots, N^p$}
               \For{$g' = 1,\ldots, N^{p'}$}
                    \State use generators to compute $\widetilde {\bm H}^{pg,p'g'}$\smallskip
                    \State perform multiplication $(\widetilde {\bm C}\,^{pg})^T 
    \widetilde {\bm H}^{pg,p'g'} \widetilde {\bm C}\,^{p'g'}$ 
                \EndFor
            \EndFor
            \Else \Comment diagonal case
             \State compute radial integrals based on $\{  \phi\,^{p}_{n\kappa} \}$
            \For{$g = 1,\ldots, N^p$}
               \For{$g' = g,\ldots, N^{p}$}
                    \State use generators to compute ${\bm H}^{pg,pg'}$
                \EndFor
            \EndFor
           \EndIf
      \EndFor
 \EndFor
 \end{algorithmic} 
 
\section{Partitioned correlation function interaction}\label{sec:PCFI}

At the price of losing some degrees of variational freedom, the divide-and-conquer method can be modified to reduce 
the size of the Hamiltonian matrix \cite{Veretal:2013a}.  In the generator-based divide-and-conquer method we have wave functions 
\begin{equation}
\Psi^p = \sum_{\alpha=1}^M b^p_{\alpha}   \Phi^{\textsc{mr}}_{\alpha}  + 
\underbrace{\sum^{N^p}_{g=1} \sum^{N^{pg}}_{\alpha=1} c^{pg}_{\alpha}\,  \Phi^{pg}_{\alpha}}_{\Lambda^p},~~~~~p = 1,\ldots,N
\label{eq;wfn1}
\end{equation}
from the separate MR-MCDHF calculations for the $N$ partitions.
As discussed earlier, the sum over CSFs in the MR accounts for static correlation whereas the sum over CSFs in the correlation space
accounts for dynamic electron correlation corresponding to different electron pairs. 
Following Verdebout {\em et al.} \cite{Veretal:2013a} we refer to the latter sum as a partitioned correlation function (PCF)  $\Lambda^p$.
We normalize $\Lambda^p$ to yield  $\overline{\Lambda}^{\,p}$ such that
\begin{equation}
\langle \overline{\Lambda}^{\,p} | \overline{\Lambda}^{\,p} \rangle = 1,~~~~p = 1,\ldots,N.
\end{equation}
Together with the CSFs in the MR, the normalized PCFs
constitute a basis, which leads to the expansion
\begin{equation}
\Psi = \sum_{\alpha=1}^M b^{\textsc{mr}}_{\alpha}\,   \Phi^{\textsc{mr}}_{\alpha} + \sum^N_{p=1} d_p \overline{\Lambda}^{\,p}  \;,
\label{eq;wfn2}
\end{equation}
where the expansion coefficients $\{ b_{\alpha} \} $ and $\{ d_p \}$ are obtained by solving an eigenvalue problem of dimension $M+N$.
Adhering to specific PCF-building rules discussed in appendix A of  \cite{Veretal:2013a}, the basis functions in the expansion are orthonormal and we have an ordinary eigenvalue problem. For cases where we target multiple states with the same parity and $J$, we will have one partitioned correlation function,
$\overline{\Lambda}^{\,ps}$,
for each state $s$ obtained by removing the CSFs of the MR, normalizing each of the CSF expansions corresponding to the different states, and then 
Schmidt orthogonalizing. The wave function expansion is now
\begin{equation}
\Psi = \sum_{\alpha=1}^M b^{\textsc{mr}}_{\alpha}\,   \Phi^{\textsc{mr}}_{\alpha}  + \sum_{s=1}^S\sum^N_{p=1} d_{ps} \overline{\Lambda}^{\,ps}  \;,
\label{eq;wfn3}
\end{equation}
where $S$ is the total number of states with the given parity and $J$.
The resulting eigenvalue problem is now of dimension $M+SN$. We refer to the above as the partitioned correlation function interaction (PCFI) method.
As discussed by Verdebout {\em et al.}, there are many varieties of the PCFI method, and we can, for example, choose
to deconstrain the VV partitioned correlation function to yield an expansion
\begin{equation}
\Psi = \sum_{\alpha=1}^M b^{\textsc{mr}}_{\alpha}\,   \Phi^{\textsc{mr}}_{\alpha}  + \sum_{g = 1}^{N^{\scriptstyle \mathrm{vv}}} \sum_{\alpha=1}^{N^{\scriptstyle \mathrm{vv}g}}b^{\textsc{vv}g}_{\alpha} \,  \Phi^{\textsc{vv}g}_{\alpha} +
\sum_{s=1}^S\sum^{N}_{p=2} d_{ps} \overline{\Lambda}^{\,ps}  \;,
\label{eq;wfn4}
\end{equation}
where now the partitioned correlation functions are used to account for the core-valence and core-core correlation effects. 
In this form, the method resembles a method put forward by Dzuba \cite{Dzuba}. For calculations of hyperfine interaction constants
it could be useful to deconstrain not only the VV partitioned correlation function, but all partitioned functions obtained from single
excitations. The latter contribute to first order to the hyperfine interaction constant.

We now turn to the problem of computing Hamiltonian matrix elements between two partitioned correlation functions that are off-diagonal in $p$.
For simplicity we assume that $S$ is equal to one, and thus we should compute $\langle  \overline{\Lambda}^{\,p}  \| {\cal H}\|  \overline{\Lambda}^{\,p'}  \rangle$.
The partitioned correlation function is
\begin{equation}\label{eq:pcf}
\overline{\Lambda}^{\,p} = \sum^{N^p}_{g=1} \sum^{N^{pg}}_{\alpha=1} \overline{c}^{\,pg}_{\alpha}\,  \Phi^{pg}_{\alpha},
\end{equation}
where $\overline{c}^{\,pg}_{\alpha}$ are the renormalized expansion coefficients.
We start by computing the Hamiltonian matrix ${\bm H}^{p,p'}$ between the CSFs in the two partitioned correlation functions
$ \overline{\Lambda}^{\,p}$ and  $\overline{\Lambda}^{\,p'}$. 
According to the previous section the ${\bm H}^{p,p'}$  matrix is block diagonal with subblocks 
\begin{equation}
    {\bm H}^{pg,p'g'} \equiv ({\bm C}^{pg})^T{\bm H}^{pg,p'g'}{\bm C}^{p'g'} \equiv (\widetilde {\bm C}\,^{pg})^T 
    \widetilde {\bm H}^{pg,p'g'} \widetilde {\bm C}\,^{p'g'}.
\end{equation}
Having computed ${\bm H}^{p,p'}$ we now simply have
\begin{equation}
\langle  \overline{\Lambda}^{\,p}  \| {\cal H}\|  \overline{\Lambda}^{\,p'}  \rangle = (\overline{{\bm C}}^{\,p})^T{\bm H}^{p,p'}\,\overline{{\bm C}}^{\,p'},
\end{equation}
where $\overline{{\bm C}}^{\,p}$ is the column vector with the  renormalized expansion coefficients in eq. (\ref{eq:pcf}). In the general case, targeting $S$ states,
the $S \times S$ Hamiltonian matrix with elements $\langle  \overline{\Lambda}^{\,ps}  \| {\cal H}\|  \overline{\Lambda}^{\,p's'}  \rangle$ is obtained by
\begin{equation}
\langle  \overline{\Lambda}^{\,p}  \| {\cal H}\|  \overline{\Lambda}^{\,p'}  \rangle = (\overline{{\bm C}}^{\,pS})^T{\bm H}^{p,p'}\overline{{\bm C}}^{\,p'S},
\end{equation}
where the $S$ columns of the $\overline{{\bm C}}^{\,pS}$ matrix hold the renormalized expansion coefficients of the $S$ states.
In terms of computation, the algorithm in section \ref{sec:algorithm} needs only to be updated to include the matrix multiplication with the expansion coefficients. 

The gain in computational efficiency compared to the case in section \ref{sec:DQ}, where the expansion is in terms of CSFs of the PCFs, comes from the  smaller dimension of the resulting Hamiltonian matrix making
it possible to perform the diagonalization with all data in the core, substantially speeding up this part of the calculation.
The price to pay is that the expansion coefficients of the CSFs in the partitioned 
correlation functions, $\overline{\Lambda}^{\,ps}$, are locked and there are no relative changes due to the interaction with
other partitioned correlation functions. For energies, the effects of locking the CSFs in the partitioned correlation functions are found to be very small
\cite{Veretal:2013a,Rynkun}. For other expectation values this may not be the case, and it is sensible to expand the wave function in terms of
the CSFs of the MR, the CSFs of the VV PCF and, optionally, CSFs of other PCFs as well, and let the partitioned correlation functions account
for different CC correlations. 
This regains the most important variational freedom, while preserving most of the reduction in the dimension of the final interaction problem.
In these cases, the only change in the computational scheme is that no multiplication by expansion coefficients is required for matrix elements involving CSFs from the deconstrained PCFs.

Overall, the PCFI formulation allows different levels of contraction, ranging from a fully contracted representation, in which the internal CSF coefficients of all PCFs are kept fixed, to a fully deconstrained representation, in which all CSFs are included explicitly in the final interaction calculation. The former reduces the dimension of the final eigenvalue problem, whereas the latter retains the full variational freedom of the underlying CSF expansions. \ref{Appendix_E} provides a numerical validation of the biorthonormal transformation and illustrates the small numerical effect of the contracted representation for the test case considered.

\section{Computation of expectation values}
Consider a wave function
\begin{equation}\label{eq:wavefunc0}
\Psi =  \sum_{\alpha=1}^M b^{\textsc{mr}}_{\alpha}\,   \Phi^{\textsc{mr}}_{\alpha}  + \sum_{p=1}^N \left(\sum_{g,\alpha} c_{\alpha}^{pg}\Phi_{\alpha}^{pg}\right)
\end{equation}
resulting from the final RCI calculation in the divide-and-conquer method based on $N$ PCFs and POLFs. To compute an expectation value
$\langle \Psi \|{\cal O} \| \Psi \rangle$ 
of a tensor operator ${\cal O}$, e.g. the magnetic dipole and electric quadrupole hyperfine operators, we start to compute the operator matrix ${\bm O}$.
The operator matrix is block diagonal, where blocks off-diagonal in $p$ are given by
\begin{equation}\label{eq:operator}
    {\bm O}^{pg,p'g'} \equiv ({\bm C}^{pg})^T{\bm O}^{pg,p'g'}{\bm C}^{p'g'} \equiv (\widetilde {\bm C}\,^{pg})^T 
    \widetilde {\bm O}^{pg,p'g'} \widetilde {\bm C}\,^{p'g'}.
\end{equation}
In the expression above, $\widetilde {\bm O}^{pg,p'g'}$ is the reduced matrix
with elements
$\widetilde O^{pg,p'g'}_{\alpha\alpha'} = \langle {\widetilde \Phi}^{pg}_{\alpha} \| {\cal O}\| {\widetilde \Phi}^{p'g'}_{\alpha'} \rangle$,
with CSFs built on the transformed radial orbitals. The situation is similar to that for the construction of the Hamiltonian matrix and the algorithm in section \ref{sec:algorithm} can be used. Once the operator matrix is available, the expectation value is obtained as
\begin{equation}
\langle \Psi \|{\cal O} \| \Psi \rangle = {\bm C}^T {\bm O}{\bm C},
\end{equation}
where ${\bm C}$ is the column vector of the expansion coefficients in eq. (\ref{eq:wavefunc0}).
The computational procedure has to be slightly changed when considering wave functions built on partitioned correlation functions, see appendix B of \cite{Veretal:2013a}.  

\section{Applications of the PCFI method}
In this section, we apply the PCFI method to three different cases in \mbox{Li I}, Be I and Al I to
illustrate its main features. The performance of the PCFI method is benchmarked against calculations based on a single orbital set. All applications reported in this section employ the fully deconstrained form of PCFI. Thus, the CSFs belonging to the different correlation partitions are included explicitly in the final interaction calculation, and their expansion coefficients are allowed to readjust freely. The results therefore assess the effect of independently optimized, correlation-specific orbital sets without additional constraints from PCF contraction.

\subsection{Computational models}
The calculations were carried out using a common orbital basis optimized for the target states considered in the present work. For each atom, an initial MCDHF calculation was first performed on a MR set in order to generate the spectroscopic orbitals. The MR configurations used for Li I, Be I, and Al I are:
\[
\begin{aligned}
\mathrm{Li~I}: \quad
&1s^2\{2s,2p,3s,3p,3d\},\\
\mathrm{Be~I}: \quad
&1s^2\{2s^2,2s2p,2p^2,2s3s,2s3p,2s3d\},\\
\mathrm{Al~I}: \quad
&1s^2 2s^2 2p^6 3s^2 3p .
\end{aligned}
\]

For Li I and Be I, the MR sets contain configurations of both even and odd parity, and the corresponding parity symmetries were optimized consistently so as to obtain a common set of orbitals for the subsequent correlation calculations. For Al I, a compact single-reference description based on the ground configuration was adopted, and correlation effects were then introduced through systematic enlargements of the active orbital space.

Correlation orbitals were generated by the active space (AS) approach. CSF expansions were produced by allowing single and double excitations from the originally occupied valence subshells to additional excited orbitals in the active set. For
Be I and Al I, the active orbital layers were defined as:
\[
\begin{aligned}
\mathrm{AS}_3 &= \{3s,3p,3d\},\\
\mathrm{AS}_4 &= \mathrm{AS}_3 \cup \{4s,4p,4d,4f\},\\
\mathrm{AS}_5 &= \mathrm{AS}_4 \cup \{5s,5p,5d,5f,5g\},\\
\mathrm{AS}_6 &= \mathrm{AS}_5 \cup \{6s,6p,6d,6f,6g,6h\},\\
\mathrm{AS}_n &= \mathrm{AS}_{n-1} \cup \{ns,np,nd,nf,ng,nh,ni\},
\quad n=7,\ldots,12 .
\end{aligned}
\]

For Li I, the same layer-by-layer strategy was used, but the active set was truncated at the \(h\) symmetry, i.e., no \(i\)-orbitals were included.
The active space was enlarged stepwise in order to monitor the convergence of the computed properties. In this layer-by-layer optimization procedure, only the newly added orbitals were varied at each step, while all
previously optimized orbitals were kept fixed. After the orbital optimization was completed, RCI calculations were carried out with the Breit interaction and leading QED corrections included. For Li I, the active space was extended up to AS$_{10}$, leading to  $34~381$ even-parity CSFs and $17~963$ odd-parity CSFs. For \mbox{Be I}, the largest calculations were performed with AS$_{12}$, giving $252~046$ even-parity CSFs and $174~240$ odd-parity CSFs. For Al I, the active space was extended up to 
AS$_{9}$, and the odd-parity expansion contained $347~756$ CSFs. For each parity, the quoted number denotes the total size of the RCI CSF expansion obtained after summing over the relevant angular-momentum blocks.

Several correlation models were considered. The label VV denotes calculations in which only valence--valence correlation is included during the MCDHF orbital optimization. The model denoted by \(+\mathrm{CV}\) includes core-valence correlation in addition to VV correlation, whereas \(+\mathrm{CC}\) further includes core-core correlation. In order to ensure a consistent comparison among the different orbital sets, the final RCI calculations were performed using the same type of enlarged CSF expansion, including VV, CV, and CC substitutions, together with the Breit interaction and QED corrections.

The CSF expansions described above correspond to the conventional construction, in which a single common orthonormal orbital basis is used for all correlation effects. In addition, calculations based on the PCFI strategy were carried out. In this approach, the correlation-function space is partitioned into several subspaces, each of which is optimized separately to describe a selected correlation effect. For Li I, the partitions were
\begin{equation}
\{v\}^{\mbox{\scriptsize S}} \cup \{1sv\}^{\mbox{\scriptsize D}}, \{1s1s\}^{\mbox{\scriptsize D}}, \{1s\}^{\mbox{\scriptsize S}}.
\end{equation}
The first partition describes the radial correlation of the valence electron together with the CV correlation contribution and the second accounts for CC correlation.
The third accounts for spin- and orbital polarization, important for the hyperfine structure parameters. 
For Be I, the partitions were
\begin{equation}
\{vv\}^{\mbox{\scriptsize SD}}, \{1sv\}^{\mbox{\scriptsize D}}, \{1s1s\}^{\mbox{\scriptsize SD}},
\end{equation}
corresponding to VV, CV, and CC correlations with core-polarization included in CV.  For Al I, a more detailed core-resolved partitioning was adopted given by
\begin{multline}
\{vv\}^{\mbox{\scriptsize SD}}, \{1sv\}^{\mbox{\scriptsize D}},  \{2sv\}^{\mbox{\scriptsize D}},  \{2pv\}^{\mbox{\scriptsize D}},\\
\{1s1s\}^{\mbox{\scriptsize SD}}, \{1s2s\}^{\mbox{\scriptsize D}}, \{1s2p\}^{\mbox{\scriptsize D}}, \{2s2s\}^{\mbox{\scriptsize SD}}, \{2s2p\}^{\mbox{\scriptsize D}},
\{2p2p\}^{\mbox{\scriptsize SD}}.
\end{multline}
 Note that each of the MCDHF calculations for the different partitions is quite small compared to the single orbital +CC MCDHF calculation, a major advantage when considering more complex systems, where the number of CSFs grows very rapidly with the number of orbital layers.

\subsection{Energy levels}

\begin{figure}[h]
    \centering
    \includegraphics[trim={20mm 20mm 23mm 20mm},clip,width=6.2cm]{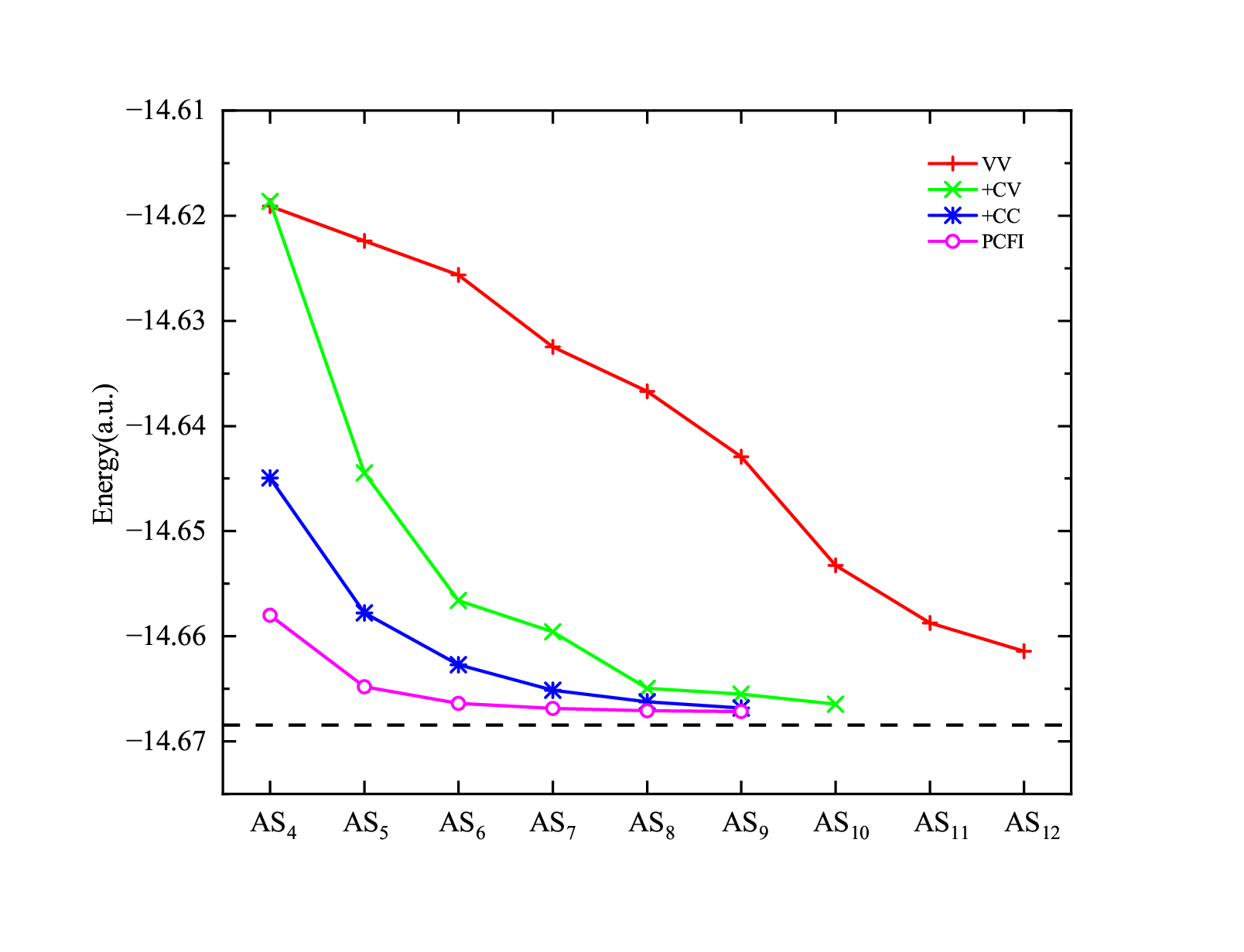}
    \includegraphics[trim={20mm 20mm 23mm 20mm},clip,width=6.2cm]{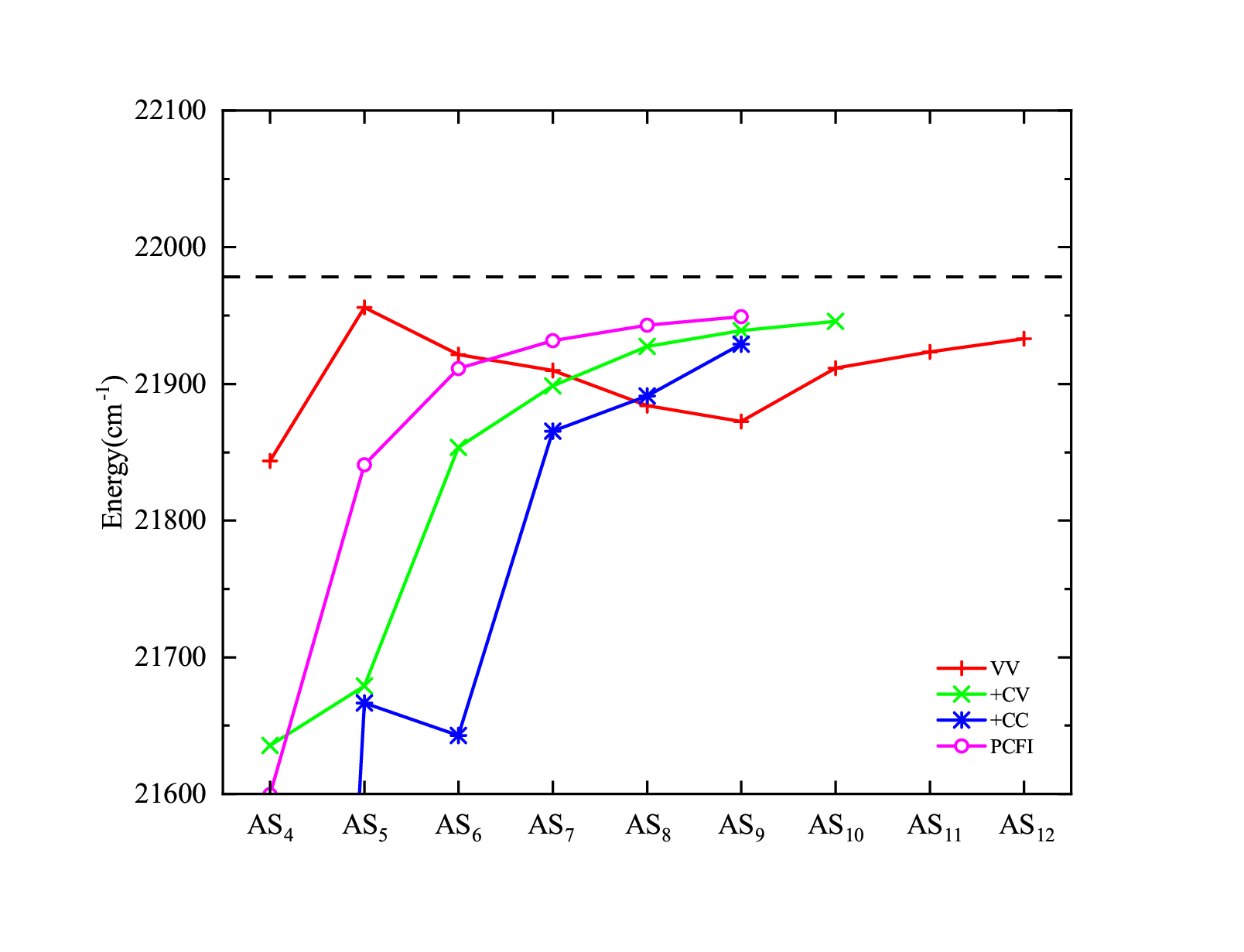}
    \caption{Absolute energies (in a.u.) for the $1s^2 2s^2~^{1}S_{0}$ level and excitation energies (in $\rm{cm^{-1}}$) for the $1s^2 2s 2p~^{3}P_{0}^{\circ}$ level in Be I from RCI calculations accounting for VV, CV and CC correlation for which the orbital bases have been obtained with different computational models, see text. The dashed line indicates the results from NIST.}
    \label{fig:energy}
\end{figure}

As shown in Fig.~\ref{fig:energy} and Table~\ref{tab:energy1}, the RCI calculations accounting for VV, CV and CC correlation including core-polarization for which the orbitals were optimized on VV correlation alone give a rather slow convergence for the absolute energy of the $1s^2 2s^2\,{}^1S_0$ ground state. The energy decreases from $-14.6190696$ a.u. at AS$_4$ to $-14.6614373$ a.u. at AS$_{12}$, but it still remains noticeably above the benchmark explicitly correlated Gaussian functions (ECG) \cite{ECG} and the National Institute of Standards and Technology (NIST) \cite{NIST} values, $-14.668440597(18)$ a.u. and $-14.6684420(18)$ a.u., respectively. This indicates that an orbital basis optimized on VV correlation alone is insufficient for a balanced description of the four-electron system. When CV and CC correlations are included in the orbital optimization, the convergence is substantially accelerated. In particular, the +CC models already produce much lower total energies at relatively small active spaces, demonstrating the importance of CC correlation.

The PCFI results show the most rapid convergence pattern among the models considered. For the ground state, the PCFI energy reaches $-14.6579869$ a.u. already at AS$_4$ and improves to $-14.6671793$ a.u. at AS$_9$. Although this value is still above the ECG and NIST benchmarks, the remaining deviation is much smaller than that from RCI calculation with orbitals optimized on the basis of the VV model at much larger active spaces. This behavior confirms that separately optimized PCFs can describe the dominant short-range dynamical correlation more efficiently than a single orthonormal orbital basis. The improvement is not simply due to increasing the number of orbitals, but rather to the better radial adaptation of different orbital sets to the physical nature of each correlation effect.

A similar trend is observed for the excitation energies of the \(1s^2 2s2p\,{}^3P^{\circ}_{0,1}\) levels in Table~\ref{tab:energy2}. Since excitation energies involve energy differences between states, their convergence is more sensitive to the balance of correlation effects in the lower and upper levels. For the \(1s^2 2s2p\,{}^3P^{\circ}_0\) level, the excitation energy from RCI calculations with orbital optimized on VV correlation increases from \(21843.61~\mathrm{cm}^{-1}\) at AS$_{4}$ to \(21933.04~\mathrm{cm}^{-1}\) at AS$_{12}$, remaining below the NIST value of \(21978.31~\mathrm{cm}^{-1}\). In comparison, the PCFI result reaches \(21949.29~\mathrm{cm}^{-1}\) already at AS$_{9}$. For the \(1s^2 2s2p\,{}^3P^{\circ}_1\) level, the same behavior is found: the PCFI excitation energy increases from \(21598.18~\mathrm{cm}^{-1}\) at AS$_{4}$ to \(21949.94~\mathrm{cm}^{-1}\) at AS$_{9}$, approaching the NIST value of \(21978.925~\mathrm{cm}^{-1}\) more rapidly than the VV sequence. The remaining deviations indicate that further correlation and higher-order effects are still present, but the smooth convergence pattern compared to the models based on single orbital sets shows that PCFI provides a more balanced treatment of the two states involved in the transition.

The same conclusion is supported by the Al I calculations, although the system is more complex because of the presence of several closed-core subshells. As shown in Table~\ref{tab:energy3}, the PCFI energies are consistently lower than the corresponding VV, \(+\mathrm{CV}\), and \(+\mathrm{CC}\) results for both fine-structure levels of the ground configuration \(1s^2 2s^2 2p^6 3s^2 3p\,{}^2P^{\circ}_{1/2,3/2}\). For the \(J=1/2\) level, the PCFI energy improves from \(-242.3961078\) a.u. at AS$_3$ to \(-242.6970732\) a.u. at AS$_{9}$, compared with \(-242.6907465\) a.u. from the \(+\mathrm{CC}\) model at the same active space. For the \(J=3/2\) level, PCFI similarly reaches \(-242.6965409\) a.u. at AS$_{9}$, lower than the corresponding \(+\mathrm{CC}\) value of \(-242.6902138\) a.u.. These results indicate that the advantage observed in Be I is not restricted to a four-electron atom. Also for Al I, where the correlation space is divided into several core-resolved partitions, the PCFI construction remains effective in capturing the dominant correlation contributions with separately optimized orbital sets.

\subsection{Mass shifts}
\begin{figure}[h]
    \centering
    \includegraphics[trim={20mm 20mm 23mm 20mm},clip,width=6.2cm]{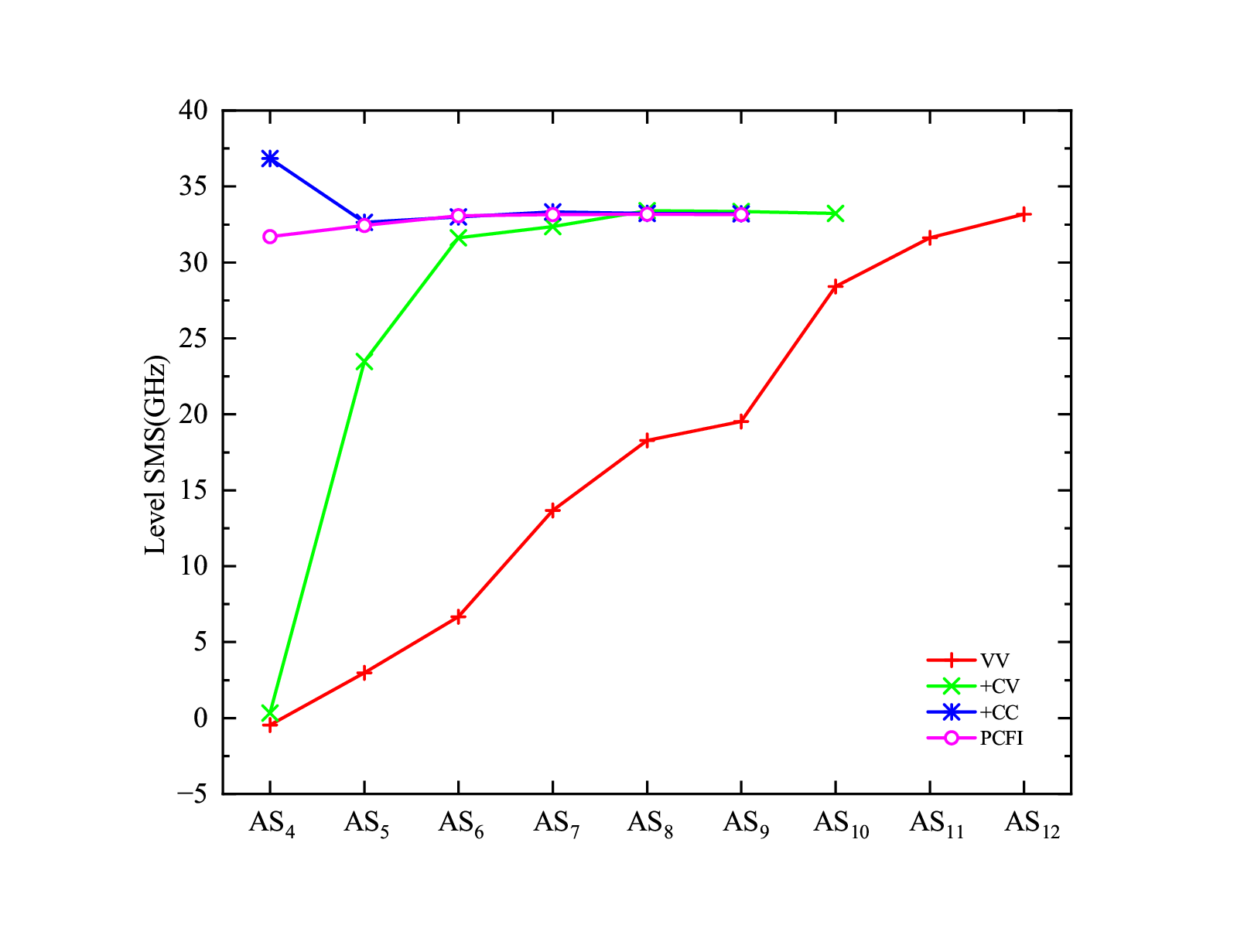}
    \includegraphics[trim={20mm 20mm 23mm 20mm},clip,width=6.2cm]{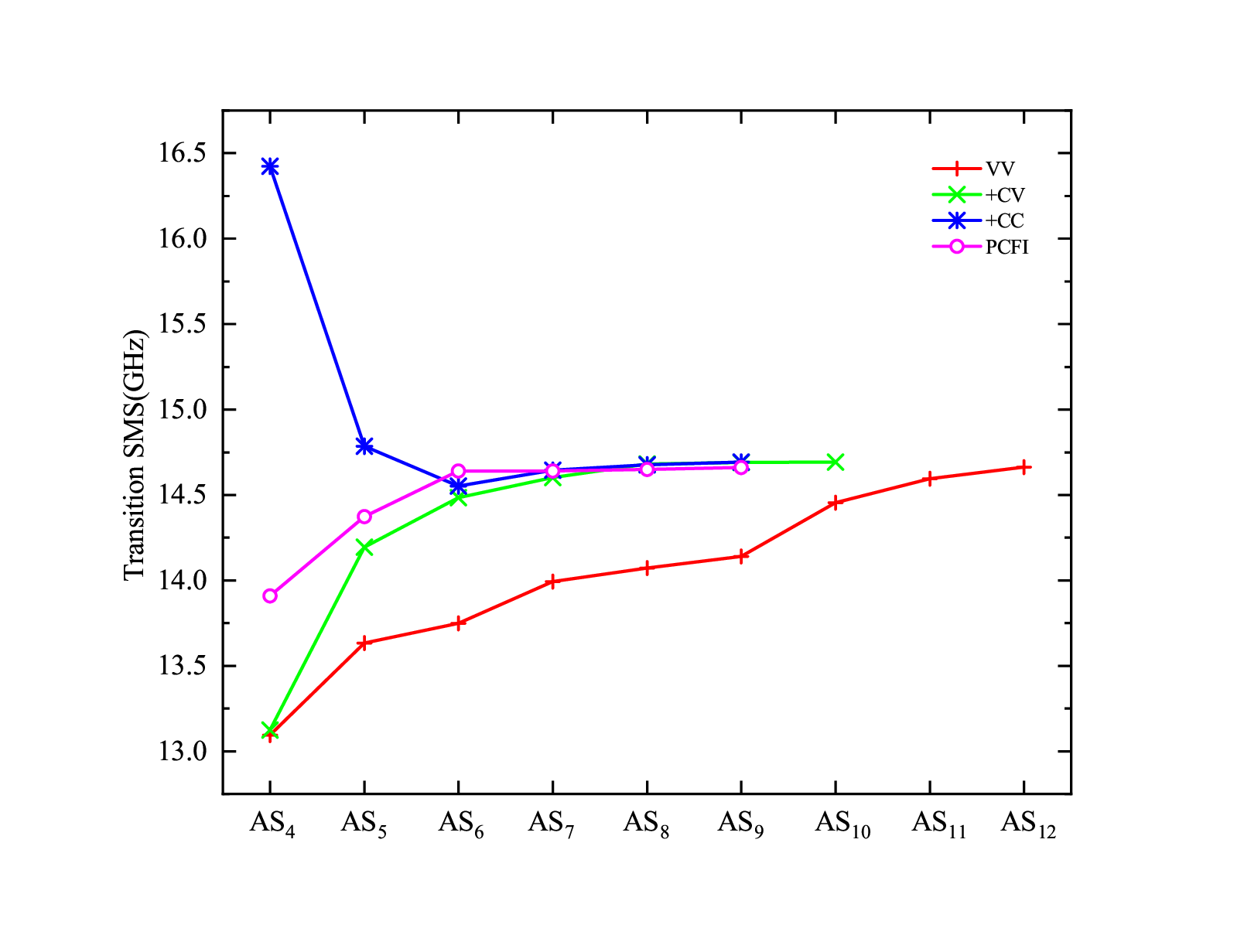}
    \caption{Level specific mass shift (in GHz) for the $1s^2 2s^2~^{1}S_{0}$ level, and transition specific mass shift (in GHz) for $1s^2 2s^2~^{1}S_{0}$ - $1s^2 2s 2p~^{3}P_{0}^{\circ}$ in Be I from RCI calculations accounting for VV, CV and CC correlation for which the orbital bases have been obtained with different computational models, see text.}
    \label{fig:ms}
\end{figure}

The mass shifts for Be I obtained with different computational models are presented in Fig.~\ref{fig:ms} and Table~\ref{tab:nms1}. For \(1s^2 2s^2\,{}^1S_0\) , the PCFI level normal mass shift (NMS) factor changes from \(1070.08\) GHz at AS$_{4}$ to \(1071.23\) GHz at AS$_{9}$. For the \(1s^2 2s2p\,{}^3P^{\rm o}_1\) level, the corresponding value vary from \(1062.69\) GHz to \(1063.81\) GHz over the same range of active spaces. The variation is already small from the first active layers, indicating that PCFI provides a stable description of the one-body mass-polarization contribution. In comparison, the VV, \(+\mathrm{CV}\), and \(+\mathrm{CC}\) sequences display more visible changes at small and intermediate active spaces. For example, the \(S_{\rm NMS}^{+\mathrm{CC}}\) value of the ground state decreases from \(1076.22\) GHz at AS$_{4}$ to \(1071.03\) GHz at AS$_{9}$, while the PCFI sequence remains confined within about \(1.2\) GHz over the same range.

The level specific mass shift (SMS) factors show a stronger dependence on the correlation model. This is expected because the SMS operator involves two-electron momentum-correlation terms and is therefore more sensitive to the details of the correlated wave function. For the \(1s^2 2s^2\,{}^1S_0\) level, the VV sequence changes from \(-0.47\) GHz at AS$_{4}$ to \(33.17\) GHz at AS$_{12}$, showing that 
RCI calculations based on orbitals optimized on VV correlation alone requires a much larger active space to reach the stable region. The \(+\mathrm{CV}\) sequence also changes rapidly in the first few layers, increasing from \(0.31\) GHz at AS$_{4}$ to \(31.61\) GHz at AS$_{6}$. In contrast, the PCFI value is already \(31.70\) GHz at AS$_{4}$, and then varies only slightly, reaching \(33.16\) GHz at AS$_{9}$. The same behavior is found for the \(1s^2 2s2p\,{}^3P^{\rm o}_1\) level. The results
from the RCI calculations with orbitals optimized on VV correlation evolves from \(-13.57\) GHz to \(18.51\) GHz over the full active-space sequence, whereas the PCFI values remain in the narrow range \(17.79\)--\(18.50\) GHz. Thus, for the level SMS factors, the dominant correlation contribution is captured much earlier in the PCFI sequence.

For the transition NMS factor of \(1s^2 2s^2\,{}^1S_0 - 1s^2 2s2p\,{}^3P^{\rm o}_1\), all models show a more regular convergence pattern, as listed in Table~\ref{tab:nms2}. The PCFI values remain close to \(7.4\) GHz, changing only from \(7.39\) GHz at AS$_{4}$ to \(7.42\) GHz at AS$_{9}$. The VV and \(+\mathrm{CV}\) sequences also approach the same region at larger active spaces, and the \(+\mathrm{CC}\) values become consistent with them from AS$_{6}$ onward. Compared with the individual level NMS factors, the transition NMS factor is less affected by the residual variations in each level, because the lower and upper level contributions partially cancel in the transition quantity. In contrast, the transition SMS factor retains a clearer sensitivity to the correlation model as shown in Fig.~\ref{fig:ms}. For the \(1s^2 2s^2\,{}^1S_0 - 1s^2 2s2p\,{}^3P^{\rm o}_1\) transition, \(S_{\rm SMS}^{\rm PCFI}\) increases from \(13.91\) GHz at AS$_{4}$ to \(14.66\) GHz at AS$_{9}$, and is already essentially stable from AS$_{6}$ onward. The \(+\mathrm{CV}\) and \(+\mathrm{CC}\) models approach a similar limiting value, around \(14.69\) GHz, but their early-layer behavior is less uniform. The VV sequence also approaches the same value, but only after the active space has been extended to the largest layers. The PCFI sequence therefore provides a faster and smoother stabilization of the transition SMS factor, while retaining a balanced description of the two levels involved in the transition.

\subsection{Hyperfine structure}
The hyperfine structure constants are strongly affected by correlation effects that are not necessarily dominant in the energy functional. In particular, the magnetic-dipole constant \(A\) is sensitive to the spin polarization of the closed \(1s\) shell, while the electric-quadrupole constant \(B\) is also affected by orbital polarization. Following the analysis of core-polarization effects in Li I \cite{LiI}, the CSFs generated by single excitations from the \(1s\) orbital were treated as a separate partition in the present PCFI calculations. This partition was optimized independently, so that the corresponding orbitals could be better adapted to the spatial region responsible for the hyperfine interaction, instead of being driven mainly by the lowering of the total energy.

\begin{figure}[h]
    \centering
    \includegraphics[trim={20mm 10mm 23mm 10mm},clip,width=6.2cm]{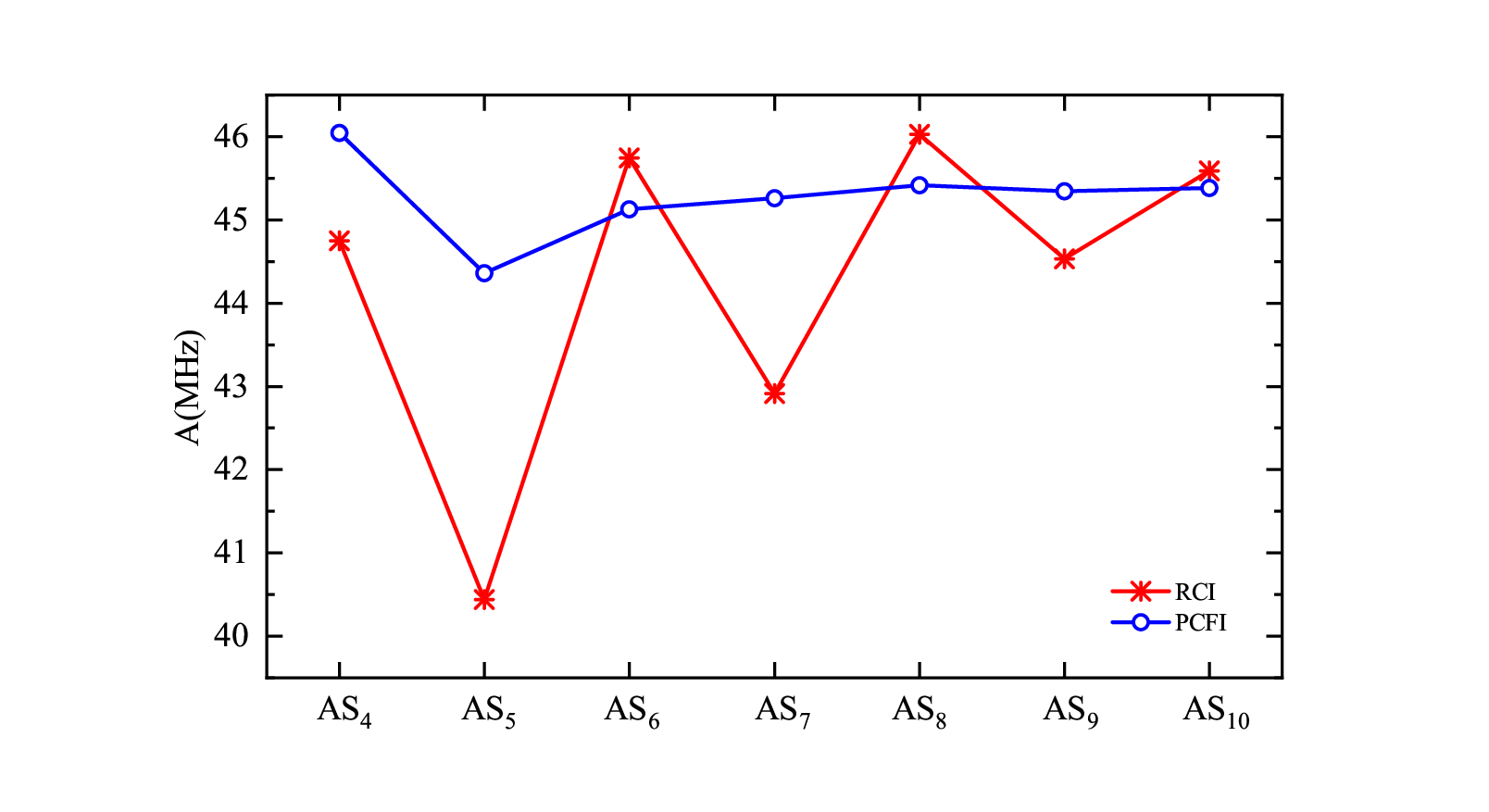}
    \includegraphics[trim={20mm 10mm 23mm 10mm},clip,width=6.2cm]{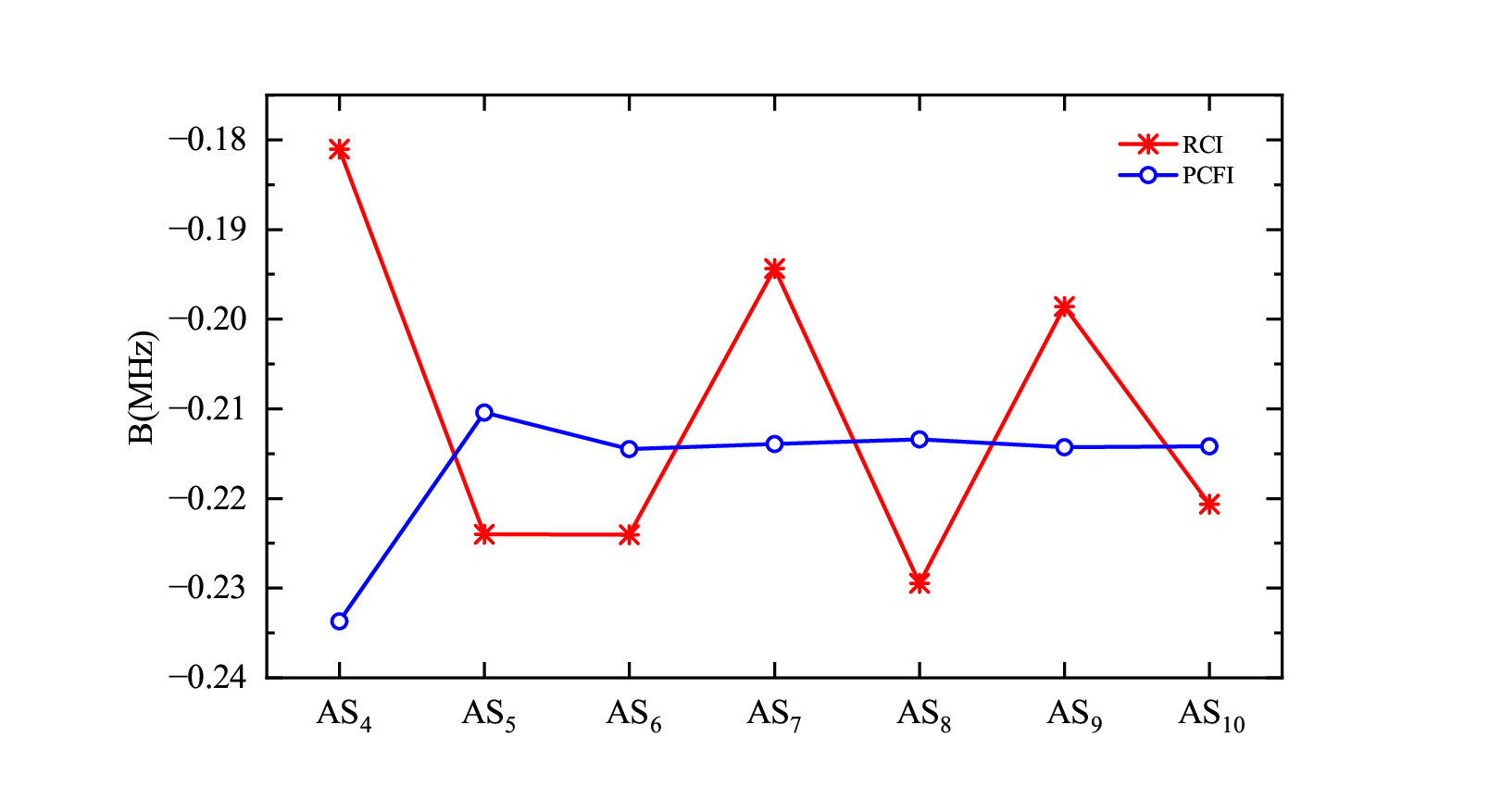}
    \caption{Convergence of the $A$ hyperfine structure constants (in MHz) of the $1s^22p~^2P_{1/2}^{\circ}$ level and $B$ hyperfine structure constants (in MHz) of the $1s^22p~^2P_{3/2}^{\circ}$ level in Li I from RCI calculations based on a single orbital set. PCFI denotes calculations based on three orbital
sets separately optimized to account for CV and CC correlation as well as core-polarization effects, see text.}
    \label{fig:hfs}
\end{figure}

The effect of this treatment is clearly seen in Fig.~\ref{fig:hfs} and Table~\ref{tab:hfs}. In the conventional \(+\mathrm{CC}\) sequence, the hyperfine constants show pronounced irregularities as the active space is increased. For example, the \(A\) constant of the \(1s^2 2s\,{}^2S_{1/2}\) level decreases from \(388.94\) MHz at AS$_{4}$ to \(374.47\) MHz at
AS$_{5}$, and then increases again towards \(398.78\) MHz at AS$_{10}$. A similar oscillatory behavior is observed for the $A$ constant of $1s^2 2p\,{}^2P^{\rm o}_{1/2}$, which varies between \(40.44\) and \(46.03\) MHz in the intermediate active spaces. These oscillations indicate that the orbitals generated in the ordinary RCI sequence do not provide a uniformly stable description of the polarization contribution to the hyperfine operator. In contrast, the PCFI results are much more regular. For \(1s^2 2s\,{}^2S_{1/2}\), the $A$ constant increases smoothly from \(377.43\) MHz at AS$_{4}$ to \(399.48\) MHz at AS$_{10}$, with only small changes after AS$_{6}$. For \(1s^2 2p\,{}^2P^{\rm o}_{1/2}\), the PCFI values remain in a narrow interval, from \(44.36\) to \(45.41\) MHz for AS$_{5}$--AS$_{10}$, whereas the corresponding \(+\mathrm{CC}\) values still display visible fluctuations. The same improvement is found for the \(1s^2 2p\,{}^2P^{\rm o}_{3/2}\) level. The \(+\mathrm{CC}\) sequence for \(A\) alternates between \(-3.55\), \(-1.21\), \(-3.32\), \(-2.05\), \(-3.31\), \(-2.72\), and \(-3.02\) MHz from AS$_{4}$ to AS$_{10}$, while the PCFI sequence is rapidly damped and stabilizes around \(-3\) MHz. For the quadrupole constant \(B\), the PCFI values are nearly constant, approximately \(-0.21\) MHz, from AS$_{5}$ onward. This is in stark contrast to the calculations based on a single orbital set that show a weakly damped oscillatory behavior.

The improved convergence can be attributed to the separate optimization of the \(1s\) single excitation partition. These CSFs describe the polarization of the closed \(1s\) core induced by the valence electron, which gives an important contribution to the hyperfine structure constants but is weakly connected to the total-energy minimization. By isolating this contribution in a dedicated PCF and coupling it with the other correlation partitions in the final PCFI calculation, the wave function gains the flexibility needed for the hyperfine operator. As a result, the oscillatory behavior observed in the conventional RCI sequence is largely removed, and stable hyperfine structure constants can be obtained with moderate active spaces.
\section{Conclusions and outlook}

In this work, the PCFI method was applied to relativistic atomic structure calculations of Li I, Be I, and Al I. The central idea of this approach is to divide the correlation space into physically motivated subspaces, such as VV, CV, CC, or more detailed core-resolved sectors, and to optimize the corresponding correlation functions separately. The use of independent orbital sets allows each partition to adapt to the radial region most relevant for the targeted correlation effect. The different partitions are then coupled in the final interaction calculation through biorthonormal transformations. In this way, PCFI provides a compact and flexible alternative to conventional RCI calculations based on a single common orthonormal orbital basis.

The numerical results demonstrate the advantage of this construction for atomic properties. For Be I, the PCFI calculations give a more rapid and smoother convergence of total energies and excitation energies than the conventional VV, \(+\mathrm{CV}\), and \(+\mathrm{CC}\) sequences. The same behavior is found for the isotope shift parameters, especially for the specific mass shift factors, which are more sensitive to electron correlation. For Li I, the single-excitation contribution from the \(1s\) shell was optimized as a separate partition. This treatment improves the description of core-polarization effects and strongly suppresses the oscillatory convergence behavior observed in the hyperfine structure constants from the conventional RCI sequence. For Al I, the core-resolved partitioning also leads to lower and more stable energies, indicating that the method remains effective for systems with a more complicated closed-core structure. 

We note that the contributions to total energies, energy separations, and various expectation values from different partitions converge at different rates as the AS increases. Consequently, for a given target uncertainty, one can independently minimize the AS size for each partition. Although this strategy was not applied in the present calculations, it can be utilized in future work and further highlights the advantages of the PCFI method.

The present results suggest several directions for future work. First, the choice of partitioning should be further optimized according to the property under consideration, since energy levels, transition rates, hyperfine structure constants and isotope shifts probe different parts of the wave function \cite{LV,NITROGEN}. Second, the PCFI strategy should be extended to heavier atoms and ions, where the rapid growth of CSF expansions makes the inclusion of CC correlation increasingly difficult in conventional calculations. 

The present implementation was developed primarily to establish and validate the relativistic PCFI formulation and to assess its convergence properties. Since the generator structure is not yet used throughout the complete workflow, no direct timing comparison with conventional RCI calculations is attempted here.
Extending the current partial use of CSF generators toward a fully generator-based implementation would make the PCFI approach a promising route toward accurate and computationally feasible calculations of correlation-sensitive atomic properties in larger systems.

Finally, we would like to emphasize that beyond the bound-state applications considered here, the same general matrix-element machinery is not restricted to discrete orbital sets. The developed methodology for computing general matrix elements based on the biorthonormal transformation is also applicable to continuum wave functions. For continuum wave functions generated by the {\sc Graspc} program \cite{GRASPC}, a {\sc Grasp} module has recently been developed for the calculation of Auger rates that includes orbital relaxation effects \cite{jorgen}.

\clearpage

\clearpage
\begin{table}[h]
\centering
\caption{Total energies, $E$ (in a.u.), of $1s^22s^2~^{1}S_0$, $1s^22s2p~^3P_0^{\circ}$ and $1s^22s2p~^3P_1^{\circ}$ in Be I
from RCI calculations accounting for VV, CV and CC correlation. The orbital basis were obtained with different computational models: VV orbitals optimized on VV expansions, +CV orbitals optimized on VV and CV expansions, +CC orbitals optimized on VV, CV and CC expansions. PCFI denotes
RCI calculations based on three orbital sets separately optimized to account for VV, CV and CC correlation, respectively.}
\label{tab:energy1}
\begin{tabular}{p{1.6cm}|p{3.2cm}p{2.8cm}p{2.8cm}p{2.8cm}}
\toprule
\multicolumn{5}{c}{$1s^22s^2~^1S_0$} \\
\midrule
AS & $E_\text{VV}$ & $E_\text{+CV}$ & $E_\text{+CC}$ & $E_\text{PCFI}$ \\
\midrule

$\mathrm{AS}_4$    & $-14.6190696$ & $-14.6186402$ & $-14.6449620$ & $-14.6579869$ \\
$\mathrm{AS}_5$    & $-14.6224045$ & $-14.6444823$ & $-14.6577698$ & $-14.6648174$ \\
$\mathrm{AS}_6$    & $-14.6256184$ & $-14.6566207$ & $-14.6627083$ & $-14.6663979$ \\
$\mathrm{AS}_7$    & $-14.6324619$ & $-14.6596023$ & $-14.6651234$ & $-14.6668752$ \\
$\mathrm{AS}_8$    & $-14.6367209$ & $-14.6649620$ & $-14.6662620$ & $-14.6670817$ \\
$\mathrm{AS}_9$    & $-14.6429158$ & $-14.6654942$ & $-14.6668263$ & $-14.6671793$ \\
$\mathrm{AS}_{10}$ & $-14.6532584$ & $-14.6664722$ & $\dots$       & $\dots$     \\
$\mathrm{AS}_{11}$ & $-14.6587462$ & $\dots$       & $\dots$       & $\dots$     \\
$\mathrm{AS}_{12}$ & $-14.6614373$ & $\dots$       & $\dots$       & $\dots$     \\
\addlinespace[6pt]
$E^{\mathrm{ECG}}$\cite{ECG} & $-14.668440597(18)$ &  &  &  \\
$E^{\mathrm{NIST}}$\cite{NIST}  & $-14.6684420(18)$ &  &  &  \\
\midrule
\multicolumn{5}{c}{$1s^22s2p~^3P_0^{\circ}$} \\
\midrule
$AS$ & $E_\text{VV}$ & $E_\text{+CV}$ & $E_\text{+CC}$ & $E_\text{PCFI}$ \\
\midrule

$\mathrm{AS}_{4}$   & $-14.5195428$ & $-14.5200625$ & $-14.5520147$ & $-14.5595728$ \\
$\mathrm{AS}_{5}$   & $-14.5223659$ & $-14.5457056$ & $-14.5590502$ & $-14.5653038$ \\
$\mathrm{AS}_{6}$   & $-14.5257364$ & $-14.5570484$ & $-14.5640962$ & $-14.5665624$ \\
$\mathrm{AS}_{7}$   & $-14.5326334$ & $-14.5598257$ & $-14.5654971$ & $-14.5669477$ \\
$\mathrm{AS}_{8}$   & $-14.5370091$ & $-14.5650535$ & $-14.5665189$ & $-14.5671019$ \\
$\mathrm{AS}_{9}$   & $-14.5432580$ & $-14.5655328$ & $-14.5669096$ & $-14.5671710$ \\
$\mathrm{AS}_{10}$  & $-14.5534219$ & $-14.5664803$ & $\dots$       & $\dots$     \\
$\mathrm{AS}_{11}$  & $-14.5588560$ & $\dots$       & $\dots$       & $\dots$     \\
$\mathrm{AS}_{12}$  & $-14.5615030$ & $\dots$       & $\dots$       & $\dots$     \\
\midrule
\multicolumn{5}{c}{$1s^22s2p~^3P_1^{\circ}$} \\
\midrule
AS & $E_\text{VV}$ & $E_\text{+CV}$ & $E_\text{+CC}$ & $E_\text{PCFI}$ \\
\midrule

$\mathrm{AS}_{4}$   & $-14.5195486$ & $-14.5200651$ & $-14.5520230$ & $-14.5595784$ \\
$\mathrm{AS}_{5}$   & $-14.5223666$ & $-14.5457009$ & $-14.5590460$ & $-14.5653024$ \\
$\mathrm{AS}_{6}$   & $-14.5257357$ & $-14.5570441$ & $-14.5640907$ & $-14.5665594$ \\
$\mathrm{AS}_{7}$   & $-14.5326309$ & $-14.5598231$ & $-14.5654929$ & $-14.5669447$ \\
$\mathrm{AS}_{8}$   & $-14.5370068$ & $-14.5650507$ & $-14.5665146$ & $-14.5670989$ \\
$\mathrm{AS}_{9}$   & $-14.5432559$ & $-14.5655299$ & $-14.5669066$ & $-14.5671680$ \\
$\mathrm{AS}_{10}$  & $-14.5534202$ & $-14.5664773$ & $\dots$       & $\dots$     \\
$\mathrm{AS}_{11}$  & $-14.5588533$ & $\dots$       & $\dots$       & $\dots$     \\
$\mathrm{AS}_{12}$  & $-14.5615002$ & $\dots$       & $\dots$       & $\dots$     \\
\bottomrule
\end{tabular}
\end{table}

\begin{table}[h]
\centering
\caption{Excitation energies, $E$ (in $\rm{cm^{-1}}$), of $1s^2~2s2p~^3P_0^{\circ}$ and $1s^2~2s2p~^3P_1^{\circ}$ in Be I 
from RCI calculations accounting for VV, CV and CC correlation. The orbital basis were obtained with different computational models: VV orbitals optimized on VV expansions, +CV orbitals optimized on VV and CV expansions, +CC orbitals optimized on VV, CV and CC expansions. PCFI denotes
RCI calculations based on three orbital sets separately optimized to account for VV, CV and CC correlation, respectively.}
\label{tab:energy2}
\begin{tabular}{p{1.6cm}|p{2.8cm}p{2.8cm}p{2.8cm}p{2.8cm}}
\toprule
\multicolumn{5}{c}{$1s^22s2p~^3P_0^{\circ}$} \\
\midrule
AS & $E_\text{VV}$ & $E_\text{+CV}$ & $E_\text{+CC}$ & $E_\text{PCFI}$ \\
\midrule

$\mathrm{AS}_{4}$   & 21~843.61    & 21~635.30    & 20~399.55    & 21~599.39    \\
$\mathrm{AS}_{5}$   & 21~955.94    & 21~679.00    & 21~666.46    & 21~840.71    \\
$\mathrm{AS}_{6}$   & 21~921.57    & 21~853.59    & 21~642.84    & 21~911.37    \\
$\mathrm{AS}_{7}$   & 21~909.82    & 21~898.44    & 21~865.45    & 21~931.55    \\
$\mathrm{AS}_{8}$   & 21~884.21    & 21~927.39    & 21~891.09    & 21~943.03    \\
$\mathrm{AS}_{9}$   & 21~872.37    & 21~938.98    & 21~929.19    & 21~949.29    \\
$\mathrm{AS}_{10}$  & 21~911.57    & 21~945.70    & $\dots$      & $\dots$      \\
$\mathrm{AS}_{11}$  & 21~923.35    & $\dots$      & $\dots$      & $\dots$      \\
$\mathrm{AS}_{12}$  & 21~933.04    & $\dots$      & $\dots$      & $\dots$      \\
\addlinespace[6pt] 
$E^{\mathrm{NIST}}$\cite{NIST}  & 21~978.31 &  &  &  \\
\midrule
\multicolumn{5}{c}{$1s^22s2p~^3P_1^{\circ}$} \\
\midrule
AS & $E_\text{VV}$ & $E_\text{+CV}$ & $E_\text{+CC}$ & $E_\text{PCFI}$ \\
\midrule

$\mathrm{AS}_{4}$   & 21~842.34    & 21~634.75    & 20~397.74    & 21~598.18  \\
$\mathrm{AS}_{5}$   & 21~955.78    & 21~680.02    & 21~667.39    & 21~841.03  \\
$\mathrm{AS}_{6}$   & 21~921.72    & 21~854.53    & 21~644.06    & 21~912.02  \\
$\mathrm{AS}_{7}$   & 21~910.37    & 21~899.02    & 21~866.37    & 21~932.21  \\
$\mathrm{AS}_{8}$   & 21~884.71    & 21~927.99    & 21~892.01    & 21~943.70  \\
$\mathrm{AS}_{9}$   & 21~872.82    & 21~939.62    & 21~929.85    & 21~949.94  \\
$\mathrm{AS}_{10}$  & 21~911.95    & 21~946.34    & $\dots$      & $\dots$    \\
$\mathrm{AS}_{11}$  & 21~923.96    & $\dots$      & $\dots$      & $\dots$    \\
$\mathrm{AS}_{12}$  & 21~933.65    & $\dots$      & $\dots$      & $\dots$    \\
\addlinespace[6pt]
$E^{\mathrm{NIST}}$\cite{NIST}  & 21~978.925 &  &  &  \\
\bottomrule
\end{tabular}
\end{table}

\begin{table}[h]
\centering
\caption{Total energies, $E$ (in a.u.), of $1s^22s^22p^63s^23p~^{2}P_{1/2}^{\circ}$ and $^{2}P_{3/2}^{\circ}$ in Al I
obtained with different computational models, see text.  PCFI denotes RCI calculations based on ten orbital
sets separately optimized to account for VV, CV as well as different CC correlation effects.}
\label{tab:energy3}
\begin{tabular}{p{1.6cm}|p{3.2cm}p{2.8cm}p{2.8cm}p{2.8cm}}
\toprule
\multicolumn{5}{c}{$1s^22s^22p^63s^23p~^{2}P_{1/2}^{\circ}$} \\
\midrule
AS & $E_\text{VV}$ & $E_\text{+CV}$ & $E_\text{+CC}$ & $E_\text{PCFI}$ \\
\midrule

$\mathrm{AS}_{3}$   & $-242.2981182$ & $-242.2981719$ & $-242.2981671$ & $-242.3961078$ \\
$\mathrm{AS}_{4}$   & $-242.3202079$ & $-242.3771563$ & $-242.5394736$ & $-242.6305146$ \\
$\mathrm{AS}_{5}$   & $-242.3381245$ & $-242.5002587$ & $-242.6231888$ & $-242.6767577$ \\
$\mathrm{AS}_{6}$   & $-242.3814440$ & $-242.5956562$ & $-242.6563264$ & $-242.6907817$ \\
$\mathrm{AS}_{7}$   & $-242.5118960$ & $-242.6307266$ & $-242.6750239$ & $-242.6948993$ \\
$\mathrm{AS}_{8}$   & $-242.5361343$ & $-242.6471015$ & $-242.6863100$ & $-242.6964123$ \\
$\mathrm{AS}_{9}$   & $-242.5864766$ & $-242.6767827$ & $-242.6907465$ & $-242.6970732$ \\
\addlinespace[6pt]
$E^{\mathrm{NIST}}$\cite{NIST}  & $-242.727(8)$ &  &  &  \\
\midrule
\multicolumn{5}{c}{$1s^22s^22p^63s^23p~^{2}P_{3/2}^{\circ}$} \\
\midrule
AS & $E_\text{VV}$ & $E_\text{+CV}$ & $E_\text{+CC}$ & $E_\text{PCFI}$ \\
\midrule

$\mathrm{AS}_{3}$   & $-242.2976296$ & $-242.2976747$ & $-242.2976666$ & $-242.3955859$ \\
$\mathrm{AS}_{4}$   & $-242.3197179$ & $-242.3766411$ & $-242.5389692$ & $-242.6299859$ \\
$\mathrm{AS}_{5}$   & $-242.3376130$ & $-242.4997245$ & $-242.6226797$ & $-242.6762246$ \\
$\mathrm{AS}_{6}$   & $-242.3809146$ & $-242.5951263$ & $-242.6558056$ & $-242.6902497$ \\
$\mathrm{AS}_{7}$   & $-242.5113370$ & $-242.6301965$ & $-242.6744938$ & $-242.6943674$ \\
$\mathrm{AS}_{8}$   & $-242.5356024$ & $-242.6465711$ & $-242.6857774$ & $-242.6958802$ \\
$\mathrm{AS}_{9}$   & $-242.5859494$ & $-242.6762513$ & $-242.6902138$ & $-242.6965409$ \\
\bottomrule
\end{tabular}
\end{table}

\begin{table}[h]
\centering
\caption{Level normal mass shift, $S_{\text{NMS}}$ (in GHz), and specific mass shift, $S_{\text{SMS}}$ (in GHz), of $1s^22s^2~^1S_0$ and $1s^22s2p~^3P_1^{\circ}$ in Be I obtained with different computational models, see text.}
\label{tab:nms1}
\begin{tabular}{p{1cm}|p{1.3cm}p{1.3cm}p{1.3cm}p{1.3cm}|p{1.2cm}p{1.2cm}p{1.2cm}p{1.2cm}}
\toprule
\multicolumn{9}{c}{$1s^22s^2~^1S_0$} \\
\midrule
AS & $S^\text{VV}_\text{NMS}$ & $S^\text{+CV}_\text{NMS}$ & $S^\text{+CC}_\text{NMS}$ & $S^\text{PCFI}_\text{NMS}$ & $S^\text{VV}_\text{SMS}$ & $S^\text{+CV}_\text{SMS}$ & $S^\text{+CC}_\text{SMS}$ & $S^\text{PCFI}_\text{SMS}$ \\
\midrule

$\mathrm{AS}_{4}$   & 1~067.42      & 1~066.43      & 1~076.22      & 1~070.08     & $-0.47$   & 0.31    & 36.84   &  31.70   \\
$\mathrm{AS}_{5}$   & 1~066.77      & 1~065.14      & 1~069.54      & 1~070.69     & 2.98    & 23.46   & 32.65   &  32.45   \\
$\mathrm{AS}_{6}$   & 1~066.21      & 1~068.53      & 1~070.27      & 1~071.10     & 6.66    & 31.61   & 32.99   &  33.07   \\
$\mathrm{AS}_{7}$   & 1~065.13      & 1~068.83      & 1~070.71      & 1~071.20     & 13.67   & 32.37   & 33.32   &  33.14   \\
$\mathrm{AS}_{8}$   & 1~065.10      & 1~070.50      & 1~071.11      & 1~071.21     & 18.28   & 33.40   & 33.24   &  33.17   \\
$\mathrm{AS}_{9}$   & 1~065.16      & 1~070.62      & 1~071.03      & 1~071.23     & 19.53   & 33.36   & 33.21   &  33.16   \\
$\mathrm{AS}_{10}$  & 1~067.16      & 1~070.87      & $\dots$       & $\dots$      & 28.41   & 33.23   & $\dots$ &  $\dots$ \\
$\mathrm{AS}_{11}$  & 1~068.70      & $\dots$       & $\dots$       & $\dots$      & 31.61   & $\dots$ & $\dots$ &  $\dots$ \\
$\mathrm{AS}_{12}$  & 1~069.42      & $\dots$       & $\dots$       & $\dots$      & 33.17   & $\dots$ & $\dots$ &  $\dots$ \\
\midrule
\multicolumn{9}{c}{$1s^22s2p~^3P_1^{\circ}$} \\
\midrule
AS & $S^\text{VV}_\text{NMS}$ & $S^\text{+CV}_\text{NMS}$ & $S^\text{+CC}_\text{NMS}$ & $S^\text{PCFI}_\text{NMS}$ & $S^\text{VV}_\text{SMS}$ & $S^\text{+CV}_\text{SMS}$ & $S^\text{+CC}_\text{SMS}$ & $S^\text{PCFI}_\text{SMS}$ \\
\midrule

$\mathrm{AS}_{4}$   & 1~059.94       & 1~059.35       & 1~065.56       & 1~062.69      & $-13.57$  & $-12.81$  & 20.41   &  17.79   \\
$\mathrm{AS}_{5}$   & 1~059.19       & 1~057.64       & 1~062.47       & 1~063.53      & $-10.66$  & 9.27    & 17.86   &  18.07   \\
$\mathrm{AS}_{6}$   & 1~058.47       & 1~061.10       & 1~063.14       & 1~063.74      & $-7.09$   & 17.13   & 18.44   &  18.43   \\
$\mathrm{AS}_{7}$   & 1~057.38       & 1~061.52       & 1~063.27       & 1~063.79      & $-0.32$   & 17.76   & 18.67   &  18.50   \\
$\mathrm{AS}_{8}$   & 1~057.41       & 1~063.10       & 1~063.55       & 1~063.78      & 4.21    & 18.72   & 18.56   &  18.52   \\
$\mathrm{AS}_{9}$   & 1~057.71       & 1~063.22       & 1~063.63       & 1~063.81      & 5.39    & 18.67   & 18.51   &  18.50   \\
$\mathrm{AS}_{10}$  & 1~059.79       & 1~063.46       & $\dots$        & $\dots$       & 13.95   & 18.53   & $\dots$ &  $\dots$ \\
$\mathrm{AS}_{11}$  & 1~061.34       & $\dots$        & $\dots$        & $\dots$       & 17.02   & $\dots$ & $\dots$ &  $\dots$ \\
$\mathrm{AS}_{12}$  & 1~062.04       & $\dots$        & $\dots$        & $\dots$       & 18.51   & $\dots$ & $\dots$ &  $\dots$ \\
\bottomrule
\end{tabular}
\end{table}

\begin{table}[h]
\centering
\caption{Transition normal mass shift, $S_{\text{NMS}}$ (in GHz), and specific mass shift, $S_{\text{SMS}}$ (in GHz), of $1s^2~2s^2~^1S_0$ - $1s^2~2s2p~^3P_1^{\circ}$ in Be obtained with different computational models, see text.}
\label{tab:nms2}
\begin{tabular}{p{1cm}|p{1.3cm}p{1.3cm}p{1.3cm}p{1.3cm}|p{1.2cm}p{1.2cm}p{1.2cm}p{1.2cm}}
\toprule
AS & $S^\text{VV}_\text{NMS}$ & $S^\text{+CV}_\text{NMS}$ & $S^\text{+CC}_\text{NMS}$ & $S^\text{PCFI}_\text{NMS}$ & $S^\text{VV}_\text{SMS}$ & $S^\text{+CV}_\text{SMS}$ & $S^\text{+CC}_\text{SMS}$ & $S^\text{PCFI}_\text{SMS}$ \\
\midrule

$\mathrm{AS}_{4}$   & 7.48       & 7.09       & 10.66       & 7.39       & 13.09    & 13.12   & 16.42   &  13.91   \\
$\mathrm{AS}_{5}$   & 7.58       & 7.50       & 7.07        & 7.16       & 13.63    & 14.20   & 14.79   &  14.37   \\
$\mathrm{AS}_{6}$   & 7.74       & 7.43       & 7.13        & 7.37       & 13.75    & 14.48   & 14.55   &  14.64   \\
$\mathrm{AS}_{7}$   & 7.75       & 7.31       & 7.44        & 7.41       & 13.99    & 14.60   & 14.64   &  14.64   \\
$\mathrm{AS}_{8}$   & 7.69       & 7.40       & 7.56        & 7.43       & 14.07    & 14.68   & 14.68   &  14.65   \\
$\mathrm{AS}_{9}$   & 7.46       & 7.40       & 7.41        & 7.42       & 14.14    & 14.69   & 14.69   &  14.66   \\
$\mathrm{AS}_{10}$  & 7.37       & 7.41       & $\dots$     & $\dots$    & 14.45    & 14.69   & $\dots$ &  $\dots$ \\
$\mathrm{AS}_{11}$  & 7.36       & $\dots$    & $\dots$     & $\dots$    & 14.60    & $\dots$ & $\dots$ &  $\dots$ \\
$\mathrm{AS}_{12}$  & 7.38       & $\dots$    & $\dots$     & $\dots$    & 14.66    & $\dots$ & $\dots$ &  $\dots$ \\
\bottomrule
\end{tabular}
\end{table}

\begin{table}
\centering
\caption{Hyperfine interaction constants, $A$ and $B$ (in MHz), in Li I obtained with different computational models.
RCI denotes calculations based on a single orbital sets. PCFI denotes calculations based on three orbital
sets separately optimized to account for CV and CC correlation as well as core-polarization effects.}
\label{tab:hfs}
\begin{tabular}{l| ll| ll| llll}
\toprule
\multirow{2}{*}{AS}
& \multicolumn{2}{c}{$1s^2 2s\,{}^2S_{1/2}$}
& \multicolumn{2}{c}{$1s^2 2p\,{}^2P^{\circ}_{1/2}$}
& \multicolumn{4}{c}{$1s^2 2p\,{}^2P^{\circ}_{3/2}$} \\
\cmidrule(lr){2-3}
\cmidrule(lr){4-5}
\cmidrule(lr){6-9}
& $A_{\rm{RCI}}$ & $A_{\rm{PCFI}}$ & $A_{\rm{RCI}}$ & $A_{\rm{PCFI}}$ & $A_{\rm{RCI}}$ & $A_{\rm{PCFI}}$ & $B_{\rm{RCI}}$ & $B_{\rm{PCFI}}$ \\
\midrule
$\mathrm{AS}_{4}$   & 388.94 & 377.43 & 44.75 & 46.04 & $-3.55$ & $-4.46$ & $-0.181$ & $-0.234$ \\
$\mathrm{AS}_{5}$   & 374.47 & 395.19 & 40.44 & 44.36 & $-1.21$ & $-2.81$ & $-0.224$ & $-0.210$ \\
$\mathrm{AS}_{6}$   & 391.11 & 398.41 & 45.74 & 45.13 & $-3.32$ & $-2.87$ & $-0.224$ & $-0.214$ \\
$\mathrm{AS}_{7}$   & 392.82 & 398.86 & 42.91 & 45.26 & $-2.05$ & $-2.90$ & $-0.194$ & $-0.214$ \\
$\mathrm{AS}_{8}$   & 396.70 & 399.23 & 46.03 & 45.41 & $-3.31$ & $-3.16$ & $-0.229$ & $-0.213$ \\
$\mathrm{AS}_{9}$   & 398.30 & 399.30 & 44.53 & 45.34 & $-2.72$ & $-2.87$ & $-0.199$ & $-0.214$ \\
$\mathrm{AS}_{10}$  & 398.78 & 399.48 & 45.59 & 45.39 & $-3.02$ & $-2.96$ & $-0.221$ & $-0.214$ \\
\bottomrule
\end{tabular}
\end{table}

\clearpage
\section{Acknowledgment}
SW, RS and CYC acknowledge support from National Key Research and Development Project of China (No.~2022YFA1602500 and No.~2022YFA1602303) and the National Natural Science Foundation of China (No.~12674316 and No.~12393824).  PJ acknowledges support from the Swedish research council under contract 2023-05367. MG acknowledges support from the FWO and F.R.S.-FNRS under the Excellence of Science (EOS) program (No.~O022818F). 

\clearpage

\appendix
\section{The biorthonormal transformation method}
\label{Appendix_A}
\renewcommand{\theequation}{A.\arabic{equation}}
\setcounter{equation}{0}
\renewcommand{\thetable}{A.\arabic{table}}
\setcounter{table}{0}

In this appendix we describe the biorthonormal transformation method as presented in \cite{Olsetal:95a}.
Let 
\begin{equation}
\Psi_l = \sum_{\alpha = 1}^{N_l} c_{\alpha}^{\,l}\Phi^l_{\alpha}  \mbox{~~~~~and~~~~~} \Psi_r = \sum_{\alpha' = 1}^{N_r} c_{\alpha'}^{\,r}\Phi^r_{\alpha'} 
\end{equation}
be two CSF expansions, where now $l$ and $r$ stands for left and right, built on different and mutually non-orthogonal orbital sets.  
Under the assumption that both CSF expansions are closed under orbital de-excitation (CUD) within each orbital symmetry $\kappa$, it is possible to transform the orbital sets 
\begin{equation}\label{eq:orbtransformation}
\{\phi^l_{n\kappa} \} \rightarrow \{ \widetilde \phi\,^l_{n\kappa} \}~~~\mbox{and}~~~\{\phi^{r}_{n'\kappa'} \} \rightarrow \{ \widetilde \phi\,^{r}_{n'\kappa'} \}
\end{equation}
to become biorthonormal $\langle \widetilde \phi^l_{n\kappa} | \widetilde \phi^r_{n'\kappa} \rangle = \delta_{n,n'}$,
where tilde denotes the transformed orbitals, and at the same time counter transform 
the expansion coefficients 
\begin{equation}\label{eq:transformc}
\underbrace{\left(
\begin{array}{c}
c^{\,l}_1    \smallskip\\
c^{\,l}_{2}  \smallskip\\
  \vdots         \\
c^{\,l}_{N^l} \\
\end{array}
\right)}_{{\bm c}^{\,l}} \rightarrow
\underbrace{\left(
\begin{array}{c}
\widetilde c^{\,l}_1  \smallskip\\
\widetilde c^{\,l}_{2} \smallskip\\
  \vdots    \\
\widetilde c^{\,l}_{N^{l}}\\
\end{array}
\right)}_{\widetilde {\bm c}^{\,l}} 
~~~\mbox{and}~~~
\underbrace{\left(
\begin{array}{c}
c^{\,r}_1    \smallskip\\
c^{\,r}_{2}  \smallskip\\
  \vdots         \\
c^{\,r}_{N^l} \\
\end{array}
\right)}_{{\bm c}^{\,r}} \rightarrow
\underbrace{\left(
\begin{array}{c}
\widetilde c^{\,r}_1  \smallskip\\
\widetilde c^{\,r}_{2} \smallskip\\
  \vdots    \\
\widetilde c^{\,r}_{N^{r}}\\
\end{array}
\right)}_{\widetilde {\bm c}^{\,r}}
\end{equation}
to leave the
total wave functions invariant 
\begin{equation}
\Psi_l  \equiv \sum_{\alpha = 1}^{N_l} \widetilde c_{\alpha}^{\,l} \widetilde \Phi^l_{\alpha} \mbox{~~~~~and~~~~~} \Psi_r \equiv \sum_{\alpha' = 1}^{N_r} 
\widetilde c_{\alpha'}^{\,r} \widetilde \Phi^r_{\alpha'}.
\end{equation}
In the transformed representation, an expectation value of a general tensor operator ${\cal O}$ can be written
\begin{equation}\label{eq:op}
\langle \Psi_l  \|{\cal O}\| \Psi_r \rangle = (\widetilde {\bm c}^{\,l})^T \widetilde {\bm O}^{\,lr} \widetilde {\bm c}^{\,r},
\end{equation}
where 
$\widetilde {\bm O}^{\,lr}$ is the matrix
with elements
$\widetilde O^{\,lr}_{\alpha\alpha'} = \langle {\widetilde \Phi}^{l}_{\alpha} \| {\cal O}\| {\widetilde \Phi}^{r}_{\alpha'} \rangle$.
Since the CSFs are built on the biorthonormal orbitals, the matrix elements can be obtained 
as a sum of radial integrals weighted by spin-angular coefficients obtained 
from standard Racah algebra. To compute the Hamiltonian matrix
\begin{equation}
{\bm H}^{lr}
\end{equation}
with elements
$\langle \Phi^l_{\alpha}  \|{ \cal H}\| \Phi^r_{\alpha'} \rangle$ 
we premultiply the Hamiltonian matrix with an $N^{l} \times N^{l}$ unit expansion coefficient matrix, ${\bm C}^{\,l}$, to the left and with
an $N^{r} \times N^{r}$ unit expansion coefficient matrix, ${\bm C}^{\,r}$, to the right
\begin{equation}
{\bm H}^{lr} \equiv ({\bm C}^{\,l})^T{\bm H}^{lr}{\bm C}^{\,r}.
\end{equation}
Transforming the coefficients of each column in the unit  expansion coefficient matrices ${\bm C}^{\,l}$ and ${\bm C}^{\,r}$ according to \ref{eq:transformc} and utilizing \ref{eq:op} we have
\begin{equation}\label{eq:A8}
{\bm H}^{lr}  \equiv ({\bm C}^{\,l})^T{\bm H}^{lr}{\bm C}^{\,r} \equiv (\widetilde {\bm C}^{\,l})^T \widetilde {\bm H}^{lr} \widetilde {\bm C}^{\,r},
\end{equation}
where 
$\widetilde {\bm H}^{lr}$ is the matrix
with elements
$\widetilde H^{\,lr}_{\alpha\alpha'} = \langle {\widetilde \Phi}^{l}_{\alpha} \| {\cal H}\| {\widetilde \Phi}^{r}_{\alpha'} \rangle$ which can now be computed using ordinary Racah algebra.

\subsection{Biorthonormal orbital transformation}\label{sec:BT1}

With due adaptations to adhere to the relativistic formalism, the algorithm in \cite{Olsetal:95a} for computing the orbital transformation in eq. \ref{eq:orbtransformation} is as follows:
 \begin{algorithmic}\label{alg:orb}
 \For{each $\kappa$}
         \State compute radial overlap matrix $S^{lr}_{ij}=\langle \phi^l_i | \phi^r_j \rangle$, where
         \State orbitals are indexed by their order numbers $i$, $j$\smallskip
         \State UL-decompose the inverse such that ${\bm U}{\bm L} = ({\bm S\,}^{lr})^ {-1}$\smallskip
         \State obtain transformation matrices ${\bm C}_{\scriptstyle \mathrm{orb}}^{\,l} = {\bm L}^T$, ${\bm C}_{\scriptstyle \mathrm{orb}}^{\,r} = {\bm U}$\smallskip
         \State transform orbitals:
          \[(\widetilde \phi^l_{1},\ldots,\widetilde \phi^l_{m_l}) =
          (\phi^l_{1},\ldots,\phi^l_{m_l}) {\bm C}_{\scriptstyle \mathrm{orb}}^{\,l}\]
          \[(\widetilde \phi^r_{1},\ldots,\widetilde \phi^r_{m_r}) =
          (\phi^r_{1},\ldots,\phi^r_{m_r}) {\bm C}_{\scriptstyle \mathrm{orb}}^{\,r}
           \]    
\EndFor
\end{algorithmic}

\noindent
The algorithm is implemented in the {\tt rbiotransform} program of {\sc Grasp} \cite{GRASP2018}. For convenience, {\sc Matlab} functions for the different parts are available in appendix C.
\subsection{Expansion coefficient counter transformation}\label{sec:BT2}
As shown in \cite{Olsetal:95a} the expansion coefficient counter transformation \ref{eq:transformc}, where the left, $l$, and right, $r$,  superscripts have been suppressed,
 can be inferred from the relation 
\begin{equation}
\label{eq:wct}
\sum_{\alpha =1}^N c_{\alpha}  \Phi_{\alpha}  = \sum_{\alpha = 1}^N c_{\alpha} ~\prod_{n \kappa} 
\left[
\sum_{k=0}^{2|\kappa|} \frac{1}{k!} {\hat S}^k_{n\kappa} \right] 
t^{{\hat N}_{n \kappa}^{\alpha}}_{n \kappa, n \kappa}  {\widetilde \Phi}_{\alpha} ,
\end{equation}
where 
\begin{equation}
\label{eq:S_nkappa}
{\hat S}_{n\kappa} = \sum_{n'\kappa < n \kappa} \frac{t_{n' \kappa,n \kappa}}{t_{n \kappa, n \kappa}}
\left[ a^{(j)}_{\left( n' \kappa \right)}\hspace{-0.13cm}\times\hspace{-0.08cm}\tilde{a}^{(j)}_{\left(n \kappa \right)} \right]^{(0)}.
\end{equation}
Here ${\hat N}_{n\kappa}^{\alpha}$ is the occupation number of the $n\kappa$ subshell in  ${\widetilde \Phi}_{\alpha}$, $t_{n'\kappa,n\kappa}$ are elements of the ${\bm T}$ matrix.
For the left hand side, the lower $(l)$ and upper $(u)$ triangular parts, ${\bm T}^{(l)}$ and ${\bm T}^{(u)}$, respectively, 
of  ${\bm T}$ can be inferred from the relations
\begin{equation}
{\bm L} = {\bm I} - {\bm T}^{(l)}~~~~\mbox{and}~~~~{\bm U} = \left({\bm T}^{(u)}\right)^{-1},
\end{equation}
where ${\bm L}{\bm U} = {\bm C}_{\scriptstyle \mathrm{orb}}^{\,l}$.
For the right-hand side,  the lower and upper triangular parts, ${\bm T}^{(l)}$ and ${\bm T}^{(u)}$, respectively, 
of  ${\bm T}$ can be inferred from the similar relations based on ${\bm L}{\bm U} = {\bm C}_{\scriptstyle \mathrm{orb}}^{\,r}$.
The de-excitation operator\footnote{We use the same phase system convention and the same notation for second quantization operators, their tensors, and their tensor products as defined in \cite{Gaigalas:26c}. Therefore, the de-excitation operator 
$\left[ a^{(j)}_{\left( n' \kappa \right)}\hspace{-0.13cm}\times\hspace{-0.08cm}\tilde{a}^{(j)}_{\left(n \kappa \right)} \right]^{(0)}$ 
corresponds to the notation
$\left( {\bf a}^{\dagger}_{n^{\prime}l} {\hat {\bf a}}_{nl} \right)^{(00)}$
(in non-relativistic atomic theory) as defined in \cite{Boretal:2010a}. Further details and a discussion of the various conventions and notations in atomic theory can be found in \cite{Gaigalas:26c}.
},
$\left[ a^{(j)}_{\left( n' \kappa \right)}\hspace{-0.13cm}\times\hspace{-0.08cm}\tilde{a}^{(j)}_{\left(n \kappa \right)} \right]^{(0)}$, maps CSFs, which we index $b$, appearing to the right of the operator to other CSFs, which we index $a$, and multiplies them with a factor. The mappings for the specified $n' \kappa, n \kappa$, i.e., the connected CSF$_a$ and CSF$_b$ and the multiplicative factor, are obtained from the spin-angular coefficients of the one-electron integrals $I(n'\kappa,n\kappa)$ of the one-electron scalar matrix elements between the CSFs.

\section{Matrix based counter transformations and templates}
\label{Appendix_B}
\renewcommand{\theequation}{B.\arabic{equation}}
\setcounter{equation}{0}
\renewcommand{\thetable}{B.\arabic{table}}
\setcounter{table}{0}
We now consider the expansion coefficient counter transformation of a unit matrix, to the left or the right hand side, needed for the construction of the Hamiltonian matrix in eq.~(\ref{eq:A8}). Instead of applying the
relation in eq.~(\ref{eq:wct}) to transform each of the columns in the unit matrix, 
we use the matrix representations of the de-excitation operators
$\left[ a^{(j)}_{\left( n' \kappa \right)}\hspace{-0.13cm}\times\hspace{-0.08cm}\tilde{a}^{(j)}_{\left(n \kappa \right)} \right]^{(0)}$
to build a matrix representation of the expansion coefficient counter transformation \cite{Simon}.
For computational efficiency, the matrices should be in sparse format. 
\subsection*{Expansion coefficient counter transformation -- type 2}
As a concrete example of the transformation for CSFs spanned by a generator of type 2 we consider  the list
\begin{Verbatim}[fontsize=\footnotesize]
  1s ( 2)  3s ( 1)  3p-( 1)    
               1/2      1/2
                           0-
  1s ( 2)  3s ( 1)  4p-( 1)    
               1/2      1/2
                           0-
  1s ( 2)  4s ( 1)  3p-( 1)    
               1/2      1/2
                           0-
  1s ( 2)  4s ( 1)  4p-( 1)    
               1/2      1/2
                           0-
  1s ( 2)  5s ( 1)  3p-( 1)    
               1/2      1/2
                           0-
  1s ( 2)  5s ( 1)  4p-( 1)    
               1/2      1/2
                           0-
\end{Verbatim}
from section \ref{sec:generators}, where $1s$ and $2s$ belong to the LO set. 
The expansion is closed under de-excitation of the $s$ and $p$- orbitals in the SO set and supports the biorthonormal transformation. We start by computing
the spin-angular coefficients of the one-electron integrals between off-diagonal CSFs by calls to the {\tt onescalar} routine in {\sc Graspg}.  Sorted by
integral we obtain the following result
\begin{Verbatim}[fontsize=\footnotesize]   
    1    3    1.000000000 I(3s ,4s )   
    2    4    1.000000000 I(3s ,4s )  
    1    5    1.000000000 I(3s ,5s )
    2    6    1.000000000 I(3s ,5s ) 
    3    5    1.000000000 I(4s ,5s )   
    4    6    1.000000000 I(4s ,5s )  
    1    2    1.000000000 I(3p-,4p-)   
    3    4    1.000000000 I(3p-,4p-)   
    5    6    1.000000000 I(3p-,4p-)   
\end{Verbatim}
In terms of mapping this means that the de-excitation $4s\mbox{} \rightarrow 3s\mbox{}$ maps CSF 3 onto CSF 1 and CSF 4 onto CSF 2 
with a coefficient 1, the de-excitation $5s\mbox{} \rightarrow 3s\mbox{}$ maps CSF 5 onto CSF 1 and CSF 6 onto CSF 2 with a coefficient 1 etc.
From this we have the following matrix representations of the de-excitation operators

\renewcommand{\theequation}{\text{\normalsize B.\arabic{equation}}}
\begin{eqnarray}
\begingroup
\scriptsize
\left[ a^{(1/2)}_{\left( 3 s \right)}\hspace{-0.13cm}\times\hspace{-0.08cm}\tilde{a}^{(1/2)}_{\left(4 s \right)} \right]^{(0)}
\hspace{-0.12cm} = \hspace{-0.1cm}
\left( \hspace{-0.1cm}
\begin{array}{cccccc}
 0 & 0 & 1 & 0 & 0 & 0\\
 0 & 0 & 0 & 1 & 0 & 0\\
 0 & 0 & 0 & 0 & 0 & 0\\
 0 & 0 & 0 & 0 & 0 & 0\\
 0 & 0 & 0 & 0 & 0 & 0\\
 0 & 0 & 0 & 0 & 0 & 0\\
\end{array}
\hspace{-0.1cm} \right),~~~
\left[ a^{(1/2)}_{\left( 3 s \right)}\hspace{-0.13cm}\times\hspace{-0.08cm}\tilde{a}^{(1/2)}_{\left(5 s \right)} \right]^{(0)}
\hspace{-0.12cm} = \hspace{-0.1cm}
\left( \hspace{-0.1cm}
\begin{array}{cccccc}
 0 & 0 & 0 & 0 & 1 & 0\\
 0 & 0 & 0 & 0 & 0 & 1\\
 0 & 0 & 0 & 0 & 0 & 0\\
 0 & 0 & 0 & 0 & 0 & 0\\
 0 & 0 & 0 & 0 & 0 & 0\\
 0 & 0 & 0 & 0 & 0 & 0\medskip\\
\end{array}
\hspace{-0.1cm} \right) 
\endgroup
\\
\begingroup
\scriptsize
\left[ a^{(1/2)}_{\left( 4 s \right)}\hspace{-0.13cm}\times\hspace{-0.08cm}\tilde{a}^{(1/2)}_{\left(5 s \right)} \right]^{(0)}
\hspace{-0.21cm} = \hspace{-0.1cm}
\left( \hspace{-0.1cm}
\begin{array}{cccccc}
 0 & 0 & 0 & 0 & 0 & 0\\
 0 & 0 & 0 & 0 & 0 & 0\\
 0 & 0 & 0 & 0 & 1 & 0\\
 0 & 0 & 0 & 0 & 0 & 1\\
 0 & 0 & 0 & 0 & 0 & 0\\
 0 & 0 & 0 & 0 & 0 & 0\\  
\end{array}
\hspace{-0.1cm} \right),~
\left[ a^{(1/2)}_{\left( 3p\mbox{-} \right)}\hspace{-0.13cm}\times\hspace{-0.08cm}\tilde{a}^{(1/2)}_{\left(4p\mbox{-} \right)} \right]^{(0)}
\hspace{-0.12cm} = \hspace{-0.1cm}
\left( \hspace{-0.1cm}
\begin{array}{cccccc}
 0 & 1 & 0 & 0 & 0 & 0\\
 0 & 0 & 0 & 0 & 0 & 0\\
 0 & 0 & 0 & 1 & 0 & 0\\
 0 & 0 & 0 & 0 & 0 & 0\\
 0 & 0 & 0 & 0 & 0 & 1\\
 0 & 0 & 0 & 0 & 0 & 0\\   
\end{array}
\hspace{-0.1cm} \right). 
\nonumber 
\endgroup
\end{eqnarray}
We now construct the different counter transformation operators.
For $3s$ there is no de-excitation and 
we only construct the diagonal part $t_{3s\mbox{\scriptsize }3s\mbox{\scriptsize }}^{{\hat N}_{3s\mbox{\scriptsize }}}$
\renewcommand{\theequation}{\text{\normalsize B.\arabic{equation}}}
\begin{equation}
\begingroup
\footnotesize
{\bm t}_{3s\mbox{\scriptsize }3s\mbox{\scriptsize }}^{{\hat N}_{3s\mbox{\scriptsize }}} = 
\left(
\begin{array}{cccccc}
 {t}_{3s\mbox{\scriptsize }3s\mbox{\scriptsize }} & 0 & 0 & 0 & 0 & 0\\
 0 &  {t}_{3s\mbox{\scriptsize }3s\mbox{\scriptsize }} & 0 & 0 & 0 & 0\\
 0 & 0 & 1 & 0 & 0 & 0\\
 0 & 0 & 0 & 1 & 0 & 0\\
 0 & 0 & 0 & 0 & 1 & 0\\
 0 & 0 & 0 & 0 & 0 & 1\\   
\end{array}
\right).
\endgroup
\end{equation}
The counter transformation for $3s$ is $\widetilde {\bm C}_{3s\mbox{\scriptsize }} = {\bm t}_{3s\mbox{\scriptsize }3s\mbox{\scriptsize }}^{{\hat N}_{3s\mbox{\scriptsize }}}$.
For $4s$ the diagonal part is
\renewcommand{\theequation}{\text{\normalsize B.\arabic{equation}}}
\begin{equation}
\begingroup
\footnotesize
{\bm t}_{4s\mbox{\scriptsize }4s\mbox{\scriptsize }}^{{\hat N}_{4s\mbox{\scriptsize }}} = 
\left(
\begin{array}{cccccc}
 1 & 0 & 0 & 0 & 0 & 0\\
 0 & 1 & 0 & 0 & 0 & 0\\
 0 & 0 &  {t}_{4s\mbox{\scriptsize }4s\mbox{\scriptsize }} & 0 & 0 & 0\\
 0 & 0 & 0 &  {t}_{4s\mbox{\scriptsize }4s\mbox{\scriptsize }} & 0 & 0\\
 0 & 0 & 0 & 0 & 1 & 0\\
 0 & 0 & 0 & 0 & 0 & 1\\   
\end{array}
\right).
\endgroup
\end{equation}
There is one de-excitation and 
\begin{equation}
{\hat {\bm S}}_{4s\mbox{\scriptsize }} =
 \frac{t_{3s\mbox{\scriptsize }4s\mbox{\scriptsize }} }{t_{4s\mbox{\scriptsize }4s\mbox{\scriptsize }} }
 \left[ a^{(1/2)}_{\left( 3 s \right)}\hspace{-0.13cm}\times\hspace{-0.08cm}\tilde{a}^{(1/2)}_{\left(4 s \right)} \right]^{(0)}.
\end{equation}
The counter transformation for $4s$ is 
\begin{equation}
\widetilde {\bm C}_{4s\mbox{\scriptsize }}  = \left({\bm I} + {\hat {\bm S}}_{4s\mbox{\scriptsize }} + \frac{1}{2} {\hat {\bm S}}_{4s\mbox{\scriptsize }}^ 2 \right) {\bm t}_{4s\mbox{\scriptsize }4s\mbox{\scriptsize }}^{{\hat N}_{4s\mbox{\scriptsize }}},
\end{equation}
where ${\bm I}$ is the unit matrix.
For $5s$ the diagonal part is
\renewcommand{\theequation}{\text{\normalsize B.\arabic{equation}}}
\begin{equation}
\begingroup
\footnotesize
{\bm t}_{5s\mbox{\scriptsize }5s\mbox{\scriptsize }}^{{\hat N}_{5s\mbox{\scriptsize -}}} = 
\left(
\begin{array}{cccccc}
1 & 0 & 0 & 0 & 0 & 0\\
 0 & 1 & 0 & 0 & 0 & 0\\
 0 & 0 & 1 & 0 & 0 & 0\\
 0 & 0 & 0 & 1 & 0 & 0\\
 0 & 0 & 0 & 0 & {t}_{5s\mbox{\scriptsize }5s\mbox{\scriptsize }} & 0\\
 0 & 0 & 0 & 0 & 0 & {t}_{5s\mbox{\scriptsize }5s\mbox{\scriptsize }}\\  
\end{array}
\right).
\endgroup
\end{equation}
There are two de-excitations and 
\begin{equation}
{\hat {\bm S}}_{5s\mbox{\scriptsize }} =
 \frac{t_{3s\mbox{\scriptsize }5s\mbox{\scriptsize }} }{t_{5s\mbox{\scriptsize }5s\mbox{\scriptsize }} }
 \left[ a^{(1/2)}_{\left( 3 s \right)}\hspace{-0.13cm}\times\hspace{-0.08cm}\tilde{a}^{(1/2)}_{\left(5 s \right)} \right]^{(0)}
+ \frac{t_{4s\mbox{\scriptsize }5s\mbox{\scriptsize }} }{t_{5s\mbox{\scriptsize }5s\mbox{\scriptsize }} }
\left[ a^{(1/2)}_{\left( 4 s \right)}\hspace{-0.13cm}\times\hspace{-0.08cm}\tilde{a}^{(1/2)}_{\left(5 s \right)} \right]^{(0)}.
\end{equation}
The counter transformation for $5s$ is 
\begin{equation}
\widetilde {\bm C}_{5s\mbox{\scriptsize }}  = \left({\bm I} + {\hat {\bm S}}_{5s\mbox{\scriptsize }} + \frac{1}{2} {\hat {\bm S}}_{5s\mbox{\scriptsize }}^ 2 \right) {\bm t}_{5s\mbox{\scriptsize }5s\mbox{\scriptsize }}^{{\hat N}_{5s\mbox{\scriptsize }}}.
\end{equation}
For $3p$- there is no de-excitation.  
We only construct the diagonal part ${\bm t}_{3p\mbox{\scriptsize -}3p\mbox{\scriptsize -}}^{{\hat N}_{3p\mbox{\scriptsize -}}}$
\renewcommand{\theequation}{\text{\normalsize B.\arabic{equation}}}
\begin{equation}
\begingroup
\footnotesize
{\bm t}_{3p\mbox{\scriptsize -}3p\mbox{\scriptsize -}}^{{\hat N}_{3p\mbox{\scriptsize -}}} = 
\left(
\begin{array}{cccccc}
 {t}_{3p\mbox{\scriptsize -}3p\mbox{\scriptsize -}} & 0 & 0 & 0 & 0 & 0\\
 0 & 1 & 0 & 0 & 0 & 0\\
 0 & 0 & {t}_{3p\mbox{\scriptsize -}3p\mbox{\scriptsize -}} & 0 & 0 & 0\\
 0 & 0 & 0 & 1 & 0 & 0\\
 0 & 0 & 0 & 0 & {t}_{3p\mbox{\scriptsize -}3p\mbox{\scriptsize -}} & 0\\
 0 & 0 & 0 & 0 & 0 & 1\\   
\end{array}
\right).
\endgroup
\end{equation}
The counter transformation for $3p$- is $\widetilde {\bm C}_{3p\mbox{\scriptsize -}} = {\bm t}_{3p\mbox{\scriptsize -}3p\mbox{\scriptsize -}}^{{\hat N}_{3p\mbox{\scriptsize -}}}$.
For $4p$- the diagonal part is
\renewcommand{\theequation}{\text{\normalsize B.\arabic{equation}}}
\begin{equation}
\begingroup
\footnotesize
{\bm t}_{4p\mbox{\scriptsize -}4p\mbox{\scriptsize -}}^{{\hat N}_{4p\mbox{\scriptsize -}}} = 
\left(
\begin{array}{cccccc}
 1 & 0 & 0 & 0 & 0 & 0\\
 0 & {t}_{4p\mbox{\scriptsize -}4p\mbox{\scriptsize -}} & 0 & 0 & 0 & 0\\
 0 & 0 & 1 & 0 & 0 & 0\\
 0 & 0 & 0 & {t}_{4p\mbox{\scriptsize -}4p\mbox{\scriptsize -}} & 0 & 0\\
 0 & 0 & 0 & 0 & 1 & 0\\
 0 & 0 & 0 & 0 & 0 & {t}_{4p\mbox{\scriptsize -}4p\mbox{\scriptsize -}}\\   
\end{array}
\right).
\endgroup
\end{equation}
There is one de-excitation and 
\begin{equation}
{\hat {\bm S}}_{4p\mbox{\scriptsize -}} =
 \frac{t_{3p\mbox{\scriptsize -}4p\mbox{\scriptsize -}} }{t_{4p\mbox{\scriptsize -}4p\mbox{\scriptsize -}} }
\left[ a^{(1/2)}_{\left( 3p\mbox{-} \right)}\hspace{-0.13cm}\times\hspace{-0.08cm}\tilde{a}^{(1/2)}_{\left(4p\mbox{-} \right)} \right]^{(0)}.
\end{equation}
The counter transformation for $4p$- is 
\begin{equation}
\widetilde {\bm C}_{4p\mbox{\scriptsize -}}  = \left({\bm I} + {\hat {\bm S}}_{4p\mbox{\scriptsize -}} + \frac{1}{2} {\hat {\bm S}}_{4p\mbox{\scriptsize -}}^ 2 \right) {\bm t}_{4p\mbox{\scriptsize -}4p\mbox{\scriptsize -}}^{{\hat N}_{4p\mbox{\scriptsize -}}}.
\end{equation}
The total counter transformation matrix is
\begin{equation}
\widetilde {\bm C} = \widetilde {\bm C}_{5s} \widetilde {\bm C}_{4s} \widetilde {\bm C}_{3s} \widetilde {\bm C}_{4p\mbox{\scriptsize -}}\widetilde {\bm C}_{3p\mbox{\scriptsize -}}.
\end{equation}
Due to the ordering of the CSFs in the list spanned by the generator, de-excitations always map CSFs to CSFs that occur earlier in the list and the   total transformation matrix is thus upper triangular.  
\subsection*{Expansion coefficient counter transformation -- type 3 and 4}
As a concrete example of the transformation for CSFs spanned by generators of type 3 and 4 that needs to be combined for CUD we consider  
\begin{Verbatim}[fontsize=\footnotesize]
  2s ( 1)  2p-( 1)  3s ( 1)  4s ( 1)   
      1/2      1/2      1/2      1/2 
                    0      1/2      0-
  2s ( 1)  2p-( 1)  3s ( 1)  5s ( 1)   
      1/2      1/2      1/2      1/2 
                    0      1/2      0-
  2s ( 1)  2p-( 1)  4s ( 1)  5s ( 1)   
      1/2      1/2      1/2      1/2 
                    0      1/2      0- 
  2s ( 1)  2p-( 1)  3s ( 2)             
      1/2      1/2 
                    0      0-
  2s ( 1)  2p-( 1)  4s ( 2)             
      1/2      1/2 
                    0      0-
  2s ( 1)  2p-( 1)  5s ( 2)            
      1/2      1/2 
                    0      0-
\end{Verbatim} 
from section \ref{sec:generators}, where $1s$ and $2s$ again belong to the LO set. 
The combined expansion is closed under de-excitation of the $s$  orbitals in the SO set and supports the biorthonormal transformation.
We start by computing
the spin-angular coefficients of the one-electron integrals between off-diagonal CSFs, again by calls to the {\tt onescalar} routine. Sorted by de-excitation we obtain the following result
\begin{Verbatim}[fontsize=\footnotesize]   
    2    3    1.000000000 I(3s ,4s )
    4    1    1.414213562 I(3s ,4s )
    1    5    1.414213562 I(3s ,4s ) 
    2    6    1.414213562 I(3s ,5s ) 
    1    3    1.000000000 I(3s ,5s )
    4    2    1.414213562 I(3s ,5s )
    1    2    1.000000000 I(4s ,5s )
    5    3    1.414213562 I(4s ,5s )
    3    6    1.414213562 I(4s ,5s ) 
\end{Verbatim}
In terms of mapping this means that the de-excitation $4s\mbox{} \rightarrow 3s\mbox{}$ maps CSF 3 onto CSF 2 with coefficient 1, CSF 1 onto CSF 4 and CSF 5 onto CSF 1, 
each with a coefficient $\sqrt{2}$ etc.
From this we have the following matrix representations
\renewcommand{\theequation}{\text{\normalsize B.\arabic{equation}}}
\begin{eqnarray}
\begingroup
\scriptsize
\left[ a^{(1/2)}_{\left( 3 s \right)}\hspace{-0.13cm}\times\hspace{-0.08cm}\tilde{a}^{(1/2)}_{\left(4 s \right)} \right]^{(0)}
\hspace{-0.12cm} = \hspace{-0.1cm} \left( \hspace{-0.2cm}
\begin{array}{cccccc}
 0 & 0 & 0 & 0 & \sqrt{2} & 0\\
 0 & 0 & 1 & 0 & 0 & 0\\
 0 & 0 & 0 & 0 & 0 & 0\\
 \sqrt{2} & 0 & 0 & 0 & 0 & 0\\
 0 & 0 & 0 & 0 & 0 & 0\\
 0 & 0 & 0 & 0 & 0 & 0\\
\end{array}
\hspace{-0.1cm} \right),~~~
 \left[ a^{(1/2)}_{\left( 3 s \right)}\hspace{-0.13cm}\times\hspace{-0.08cm}\tilde{a}^{(1/2)}_{\left(5 s \right)} \right]^{(0)}
\hspace{-0.12cm} = \hspace{-0.1cm} \left( \hspace{-0.1cm}
\begin{array}{cccccc}
 0 & 0 & 1 & 0 & 0 & 0\\
 0 & 0 & 0 & 0 & 0 & \sqrt{2}\\
 0 & 0 & 0 & 0 & 0 & 0\\
 0 & \sqrt{2} & 0 & 0 & 0 & 0\\
 0 & 0 & 0 & 0 & 0 & 0\\
 0 & 0 & 0 & 0 & 0 & 0\medskip\\
\end{array}
\hspace{-0.12cm} \right)
\endgroup
\\
\begingroup
\scriptsize
\left[ a^{(1/2)}_{\left( 4 s \right)}\hspace{-0.13cm}\times\hspace{-0.08cm}\tilde{a}^{(1/2)}_{\left(5 s \right)} \right]^{(0)}
\hspace{-0.12cm} = \hspace{-0.1cm} \left( \hspace{-0.1cm}
\begin{array}{cccccc}
 0 & 1 & 0 & 0 & 0 & 0\\
 0 & 0 & 0 & 0 & 0 & 0\\
 0 & 0 & 0 & 0 & 0 & \sqrt{2}\\
 0 & 0 & 0 & 0 & 0 & 0\\
 0 & 0 & \sqrt{2} & 0 & 0 & 0\\
 0 & 0 & 0 & 0 & 0 & 0\\  
\end{array}
\hspace{-0.12cm} \right).~
\textcolor{white}{
 \left[ a^{(1/2)}_{\left( 4 s \right)}\hspace{-0.13cm}\times\hspace{-0.08cm}\tilde{a}^{(1/2)}_{\left(5 s \right)} \right]^{(0)}
\hspace{-0.1cm} = \hspace{-0.1cm}
\left(
\begin{array}{cccccc}
 0 & 1 & 0 & 0 & 0 & 0\\
 0 & 0 & 0 & 0 & 0 & 0\\
 0 & 0 & 0 & 0 & 0 & \sqrt{2}\\
 0 & 0 & 0 & 0 & 0 & 0\\
 0 & 0 & \sqrt{2} & 0 & 0 & 0\\
 0 & 0 & 0 & 0 & 0 & 0\\  
\end{array}
\right)
} \nonumber 
\endgroup
\end{eqnarray}
We now construct the different counter transformation operators.
For $3s$ there is no de-excitation and 
we only construct the diagonal part ${\bm t}_{3s\mbox{\scriptsize }3s\mbox{\scriptsize }}^{{\hat N}_{3s\mbox{\scriptsize }}}$
\renewcommand{\theequation}{\text{\normalsize B.\arabic{equation}}}
\begin{equation}
\begingroup
\footnotesize
{\bm t}_{3s\mbox{\scriptsize }3s\mbox{\scriptsize }}^{{\hat N}_{3s\mbox{\scriptsize }}} = 
\left(
\begin{array}{cccccc}
 {t}_{3s\mbox{\scriptsize }3s\mbox{\scriptsize }} & 0 & 0 & 0 & 0 & 0\\
 0 &  {t}_{3s\mbox{\scriptsize }3s\mbox{\scriptsize }} & 0 & 0 & 0 & 0\\
 0 & 0 & 1 & 0 & 0 & 0\\
 0 & 0 & 0 & {t}_{3s\mbox{\scriptsize }3s\mbox{\scriptsize }}^2 & 0 & 0\\
 0 & 0 & 0 & 0 & 1 & 0\\
 0 & 0 & 0 & 0 & 0 & 1\\   
\end{array}
\right).
\endgroup
\end{equation}
The counter transformation for $3s$ is $\widetilde {\bm C}_{3s\mbox{\scriptsize }} = {\bm t}_{3s\mbox{\scriptsize }3s\mbox{\scriptsize }}^{{\hat N}_{3s\mbox{\scriptsize }}}$.
For $4s$ the diagonal part is
\renewcommand{\theequation}{\text{\normalsize B.\arabic{equation}}}
\begin{equation}
\begingroup
\footnotesize
{\bm t}_{4s\mbox{\scriptsize }4s\mbox{\scriptsize }}^{{\hat N}_{4s\mbox{\scriptsize }}} = 
\left(
\begin{array}{cccccc}
 {t}_{4s\mbox{\scriptsize }4s\mbox{\scriptsize }} & 0 & 0 & 0 & 0 & 0\\
 0 & 1 & 0 & 0 & 0 & 0\\
 0 & 0 &  {t}_{4s\mbox{\scriptsize }4s\mbox{\scriptsize }} & 0 & 0 & 0\\
 0 & 0 & 0 &  1 & 0 & 0\\
 0 & 0 & 0 & 0 & {t}_{4s\mbox{\scriptsize }4s\mbox{\scriptsize }}^2 & 0\\
 0 & 0 & 0 & 0 & 0 & 1\\   
\end{array}
\right).
\endgroup
\end{equation}
There is one de-excitation and 
\begin{equation}
{\hat {\bm S}}_{4s\mbox{\scriptsize }} =
\frac{t_{3s\mbox{\scriptsize }4s\mbox{\scriptsize }} }{t_{4s\mbox{\scriptsize }4s\mbox{\scriptsize }} }
 \left[ a^{(1/2)}_{\left( 3 s \right)}\hspace{-0.13cm}\times\hspace{-0.08cm}\tilde{a}^{(1/2)}_{\left(4 s \right)} \right]^{(0)}.
\end{equation}
The counter transformation for $4s$ is 
\begin{equation}
\widetilde {\bm C}_{4s\mbox{\scriptsize }}  = \left({\bm I} + {\hat {\bm S}}_{4s\mbox{\scriptsize }} + \frac{1}{2} {\hat {\bm S}}_{4s\mbox{\scriptsize }}^ 2 \right) {\bm t}_{4s\mbox{\scriptsize }4s\mbox{\scriptsize }}^{{\hat N}_{4s\mbox{\scriptsize }}}.
\end{equation}
For $5s$ the diagonal part is
\renewcommand{\theequation}{\text{\normalsize B.\arabic{equation}}}
\begin{equation}
\begingroup
\footnotesize
{\bm t}_{5s\mbox{\scriptsize }5s\mbox{\scriptsize }}^{{\hat N}_{5s\mbox{\scriptsize -}}} = 
\left(
\begin{array}{cccccc}
1 & 0 & 0 & 0 & 0 & 0\\
 0 & {t}_{5s\mbox{\scriptsize }5s\mbox{\scriptsize }} & 0 & 0 & 0 & 0\\
 0 & 0 & {t}_{5s\mbox{\scriptsize }5s\mbox{\scriptsize }} & 0 & 0 & 0\\
 0 & 0 & 0 & 1 & 0 & 0\\
 0 & 0 & 0 & 0 & 1 & 0\\
 0 & 0 & 0 & 0 & 0 & {t}_{5s\mbox{\scriptsize }5s\mbox{\scriptsize }}^2\\  
\end{array}
\right).
\endgroup
\end{equation}
There are two de-excitations and 
\begin{equation}
{\hat {\bm S}}_{5s\mbox{\scriptsize }} =
 \frac{t_{3s\mbox{\scriptsize }5s\mbox{\scriptsize }} }{t_{5s\mbox{\scriptsize }5s\mbox{\scriptsize }} }
  \left[ a^{(1/2)}_{\left( 3 s \right)}\hspace{-0.13cm}\times\hspace{-0.08cm}\tilde{a}^{(1/2)}_{\left(5 s \right)} \right]^{(0)}
+ \frac{t_{4s\mbox{\scriptsize }5s\mbox{\scriptsize }} }{t_{5s\mbox{\scriptsize }5s\mbox{\scriptsize }} }
 \left[ a^{(1/2)}_{\left( 4 s \right)}\hspace{-0.13cm}\times\hspace{-0.08cm}\tilde{a}^{(1/2)}_{\left(5 s \right)} \right]^{(0)}.
\end{equation}
The counter transformation for $5s$ is 
\begin{equation}
\widetilde {\bm C}_{5s\mbox{\scriptsize }}  = \left({\bm I} + {\hat {\bm S}}_{5s\mbox{\scriptsize }} + \frac{1}{2} {\hat {\bm S}}_{5s\mbox{\scriptsize }}^ 2 \right) {\bm t}_{5s\mbox{\scriptsize }5s\mbox{\scriptsize }}^{{\hat N}_{5s\mbox{\scriptsize }}}.
\end{equation}
The total counter transformation matrix is
\begin{equation}
\widetilde {\bm C} = \widetilde {\bm C}_{5s} \widetilde {\bm C}_{4s} \widetilde {\bm C}_{3s}.
\end{equation}
In this case the matrix is no longer upper triangular.
\subsection*{Redundancies of  counter transformation matrices -- templates}\label{sec:redundant}
A very important result in the context of counter transformations, rigorously proven in \ref{Appendix_F}, is that the spin-angular coefficients for the off-diagonal one-particle scalar interaction between CSFs spanned by a generator of type 1 and CSFs spanned by a generator type 2 are always 1. This means that the expansion coefficient counter transformation matrices are the same for any two generators of type 1 with the same $\kappa$ symmetry of the SO orbitals and that the expansion coefficient counter transformation matrices are the same for any two generators of type 2 with the same $\kappa, \kappa'$ symmetry of the SO orbitals.
For CSF spanned by generators of type 3 that need not be combined with CSFs spanned by a generator of type 4 for CUD, the
spin-angular coefficients can take on the values 1 or $-1$, and again the expansion coefficient counter transformation matrices depend only on the $\kappa$ symmetry.
We refer to the above matrices as $\kappa$ dependent expansion coefficient counter transformation templates.
For convenience, {\sc Matlab} functions for computing these templates can be found in \ref{AppendixD}.
For generator groups of type 3 and 4, that need to be combined for CUD, the spin-angular coefficients between the CSFs differ from one and depend on the spin-angular coupling tree of the CSFs.  In practical terms, the matrix representations of the de-excitation operators
$\left[ a^{(j)}_{\left( n' \kappa \right)}\hspace{-0.13cm}\times\hspace{-0.08cm}\tilde{a}^{(j)}_{\left(n \kappa \right)} \right]^{(0)}$ are obtained by calls to the {\tt onescalar} routine \cite{GRASPGG} of {\sc Graspg} and
then the computation of the templates proceeds as in the previous section.
 Having computed the templates, that are saved in core, the only overhead from the biorthonormal transformation, according to the algorithm in section \ref{sec:algorithm}, for computing the Hamiltonian matrix compared to the ordinary case based on a single orthonormal
orbital set, comes from the multiplication of the Hamiltonian submatrices $\widetilde {\bm H}^{pg,p'g'}$ with $(\widetilde {\bm C}\,^{pg})^T$ and $ \widetilde {\bm C}\,^{p'g'}$.


\section{Closure under de-excitation}
\label{Appendix_C}
\renewcommand{\theequation}{C.\arabic{equation}}
\setcounter{equation}{0}
\renewcommand{\thetable}{C.\arabic{table}}
\setcounter{table}{0}
The requirement for the biorthonormal transformation to hold is that the application of 
\begin{equation}\label{eq:BO}
\prod_{n \kappa} 
\left[
\sum_{k=0}^{2|\kappa|} \frac{1}{k!} {\hat S}^k_{n \kappa} \right] 
t^{{\hat N}_{n \kappa}}_{n \kappa, n \kappa}~~~\mbox{where}~~~{\hat S}_{n\kappa} =  \sum_{n'\kappa < n \kappa} \frac{t_{n' \kappa,n \kappa}}{t_{n \kappa, n \kappa}}
\left[ a^{(j)}_{\left( n' \kappa \right)}\hspace{-0.13cm}\times\hspace{-0.08cm}\tilde{a}^{(j)}_{\left(n \kappa \right)} \right]^{(0)}
\end{equation}
to a set of CSFs does not lead outside the set. Such a set of CSFs is said to be closed under de-excitation (CUD). 
For generators of types 1 and 2 it is clear from 
the construction that the scalar 
de-excitation $\left[ a^{(j)}_{\left( n' \kappa \right)}\hspace{-0.13cm}\times\hspace{-0.08cm}\tilde{a}^{(j)}_{\left(n \kappa \right)} \right]^{(0)}$ from an $n\kappa$ in the SO set to another $n'\kappa$ in the same set
does not lead outside the set of CSFs. 
CSFs corresponding to types 3 and 4 sometimes have to be combined to ensure CUD. In the example in section \ref{sec:generators} the CSFs spanned by 
\begin{Verbatim}[fontsize=\footnotesize]
  2s ( 1)  2p-( 1)  4s ( 1)  5s ( 1)
      1/2      1/2      1/2      1/2 
                    1      1/2      0-
\end{Verbatim}
is CUD by itself. The CSFs spanned by
\begin{Verbatim}[fontsize=\footnotesize]
  2s ( 1)  2p-( 1)  4s ( 1)  5s ( 1)
      1/2      1/2      1/2      1/2 
                    0      1/2      0-
\end{Verbatim}
have, however, to be combined by the CSFs spanned by 
\begin{Verbatim}[fontsize=\footnotesize]
  2s ( 1)  2p-( 1)  5s ( 2)
      1/2      1/2 
                    0      0-
\end{Verbatim} 
to ensure the CUD property. Which CSFs corresponding to types 3 and 4 that need to be combined is determined by the coupling of the orbitals in the LO set.

Now we must also prove that the generator CSFs are CUD with respect to de-excitation from orbitals $n \kappa$ in the SO set 
to orbitals $n' \kappa$ in the LO set and additionally with respect to de-excitation from orbitals $n \kappa$ in the LO set to other orbitals 
$n' \kappa$ in the same set. To this end we note that the de-excitation $n\kappa \rightarrow n'\kappa$ connects to other CSFs only when $t_{n'\kappa, n\kappa} \ne 0$. Thus
the question if a set of CSFs is CUD is also governed by the Malmqvist ${\bm T}$ matrix.
In our divide-and-conquer approach, the orbitals in the LO set are the same for all PCFs, and it is only the orbitals in the SO set that need to be biorthogonalized
against SO orbitals in another PCF.  If, for example, $1s,2s,3s$ are in the LO set and $4s, 5s$ belong to the SO set, 
then the radial overlap matrix ${\bm S}$ for this kappa has the form
\begin{equation}
{\bm S} = \left(
\begin{array}{ll}
{\bm I}  & {\bm O} \\
{\bm O}^T & {\bm S}'  \\
\end{array}
\right),
\end{equation}
where ${\bm I}$ is a $3 \times 3$ unit matrix, ${\bm O}$ a $3 \times 2$ zero matrix and ${\bm S}'$ the $2 \times 2$ overlap matrix 
\begin{equation}
{\bm S}' = \left(
\begin{array}{ll}
\langle \phi_{4s} | \phi_{4s} \rangle  & \langle \phi_{4s} | \phi_{5s} \rangle \\
\langle \phi_{5s} | \phi_{4s} \rangle  & \langle \phi_{5s} | \phi_{5s} \rangle \\
\end{array}
\right)
\end{equation}
with non-zero off-diagonal elements. Now, the ${\bm T}$
matrix is obtained by a sequence of operations involving inversion, UL and LU-decompositions.
These operations are all block diagonal from which it follows that 
\begin{equation}
{\bm T} = \left(
\begin{array}{ll}
{\bm I}  & {\bm O} \\
{\bm O}^T & {\bm T}'  \\
\end{array}
\right),
\end{equation}
where ${\bm T}'$ is the ${\bm T}$ matrix resulting from ${\bm S}'$. This form of the ${\bm T}$ matrix implies that de-excitations from $4s$ and $5s$ to $1s, 2s, 3s$ will not connect to any new CSFs. Thus the CSF list needs only to be CUD with respect to de-excitations within the SO orbital set $4s,5s$. This proof extends to the general case.

\section{ {\sc Matlab} functions for transformation matrices}\label{AppendixD}
In this appendix, we present {\sc Matlab} functions for computing the orbital transformation matrices $ {\bm C}_{\scriptstyle \mathrm{orb}}^{\,l}$ and $ {\bm C}_{\scriptstyle \mathrm{orb}}^{\,r}$ from a radial
overlap matrix ${\bm S}^{lr}$ as well as functions for computing the ${\bm T}^{\,l}$ and ${\bm T}^{\,r}$ matrices. We also include functions for computing 
expansion coefficient counter transformation templates for generators of type 1 and type 2.\medskip\\
Function for computing orbital transformation matrices  ${\bm C}_{\scriptstyle \mathrm{orb}}^{\,l}$ and ${\bm C}_{\scriptstyle \mathrm{orb}}^{\,r}$ for a given $\kappa$ from the
orbital overlap matrix ${\bm S}^{lr}$, see  \ref{sec:BT1}.
\begin{Verbatim}[fontsize=\footnotesize]
function [CL,CR] = orbtrans(S)
% Computes orbital transformation matrices CL and CR 
% given the overlap matrix S for a specified kappa.
% Special care taken when S is not quadratic, see
% appendix C of Olsen et al. Phys Rev E, 52, 4499, (1995).

% Number of rows and columns
[nr,nc] = size(S); 

% Minimal and maximal dimensions
nmin = min([nr nc]);
nmax = max([nr nc]);
ndiff = nmax - nmin;

% Invert quadratic part of S
T = S(1:nmin,1:nmin);
Sinv = inv(T);

% UL-decompose Sinv
[U,L] = ulla(Sinv);

% Obtain the transformation matrices 
CL = L';
CR = U;

% Biorthogonalize remaining part to get the full transformation
% matrices, see Appendix C of Olsen et al.
if nr < nc
   Z = S(:,nmin+1:nmax);
   CRBIO = -inv(CL'*T)*CL'*Z;
   CR = [CR CRBIO];
   CR = [CR 
         zeros(ndiff,nmin) eye(ndiff)];
elseif nr > nc
   Z = S(nmin+1:nmax,:);
   CLBIO = -inv(CR'*T')*CR'*Z';
   CL = [CL CLBIO];
   CL = [CL 
         zeros(ndiff,nmin) eye(ndiff)];
end
\end{Verbatim}
Function for computing  ${\bm T}^{\,l}$ and ${\bm T}^{\,r}$ from the orbital transformation matrices ${\bm C}_{\scriptstyle \mathrm{orb}}^{\,l}$ and ${\bm C}_{\scriptstyle \mathrm{orb}}^{\,r}$, see \ref{sec:BT2}
\begin{Verbatim}[fontsize=\footnotesize]
function [TL,TR] = tmatrix(CL,CR)
% Computes TL and TR matrices given orbital
% transformation matrices CL and CR. 

% T matrix for left hand side
[L,U] = lulu(CL);
TL = inv(U);
n = size(CL,1);
for i = 1:n
   for j = 1:i-1
      TL(i,j) = - L(i,j);
   end 
end

% T matrix for right hand side
[L,U] = lulu(CR);
TR = inv(U);
n = size(CR,1);
for i = 1:n
   for j = 1:i-1
      TR(i,j) = - L(i,j);
   end 
end
\end{Verbatim}

\noindent
Ancillary function for LU-decomposition without pivoting 
\begin{Verbatim}[fontsize=\footnotesize]
function [L,U] = lulu(A)
% LU-decomposition of a square matrix A without pivoting.

n = size(A,1);  % Obtain number of rows (should equal number of columns)
L = eye(n);     % Start L off as identity and populate the lower triangular half slowly
for k = 1:n
   % For each row k, access columns from k+1 to the end and divide by
   % the diagonal coefficient at A(k,k)
   L(k+1:n,k) = A(k+1:n,k)/A(k,k);

   % For each row k+1 to the end, perform Gaussian elimination
   % In the end, A will contain U
   for l = k+1:n
      A(l,:) = A(l,:) - L(l,k)*A(k,:);
   end
end
U = A;
\end{Verbatim}
Ancillary function for UL-decomposition
\begin{Verbatim}[fontsize=\footnotesize]
function [U,L] = ulla(A)
% UL decomposition of A
% In order to change into UL form introduce the orthogonal matrix
%    P(i,j) = delta(i,n-j+1)
% and rewrite
% A = P P A P P  = P L U P = PLP PUP
% where LU is an LU decomposition of PAP, since PLP is upper
% tringular and PUP is lower traingular we have obtained the goal

n = size(A, 1); % Obtain number of rows (equal to number of columns) 
P = zeros(n,n);
for i = 1:n
   P(i,n-i+1) = 1;
end

% LU decomposition, without pivoting, of PAP
% LS, US scratch matrices
[LS,US] = lulu(P*A*P); 
U = P*LS*P;
L = P*US*P;
\end{Verbatim}
Function for expansion coefficient counter transformation templates for CSFs spanned by generators of type 1 based
on orbitals in the SO set of $\kappa$ symmetry. 
\begin{Verbatim}[fontsize=\footnotesize]
function Ctot = template1(kappa,norb,T)
% Computes the expansion coefficient counter transformation template 
% Ctot for CSFs spanned by a generator of type 1 
% norb         = number of orbitals in SO set with kappa symmetry
% T(norb,norb) = T matrix for given kappa symmetry

% Number of spanned CSFs by the generator
ncsf = norb;

Ctot = eye(ncsf,ncsf);
for k = 1:norb       % loop over orbitals in SO set of given kappa symmetry
  S = zeros(ncsf,ncsf);
  if k == 1          % no de-excitation 
    C = eye(ncsf,ncsf);
  else            
    for l = 1:k-1    % loop over orbital de-excitations
      % Construct AdA matrix
      AdA = zeros(ncsf,ncsf);
      ncol = k;
      nrow = l;
      AdA(nrow,ncol) = 1;
      % Form S matrix
      S = T(l,k)*AdA/T(k,k) + S;
    end
    C = zeros(ncsf,ncsf);
    % Compute exp(S) by a terminating Taylor expansion
    for l = 0:2*abs(kappa)
      C = S^l/factorial(l) + C;
    end
  endif
  % Construct diagonal Tn
  Tn = eye(ncsf,ncsf);
  Tn(k,k) = T(k,k);

  % C matrix for orbital k
  C = C*Tn;

  % Accumulate C matrix for all orbitals in SO set
  Ctot = C*Ctot;
end
\end{Verbatim}
Function for expansion coefficient counter transformation templates for CSFs spanned by generators of type 2 based
on orbitals in the SO set with $\kappa_1$ and $\kappa_2$  symmetry. 
\begin{Verbatim}[fontsize=\footnotesize]
function Ctot = template2(kappa1,norb1,T1,kappa2,norb2,T2)
% Computes the expansion coefficient counter transformation template 
% Ctot for CSFs spanned by a generator of type 2 
% norb1           = number of orbitals in SO set with kappa1 symmetry 
% T1(norb1,norb1) = T matrix for given kappa1 symmetry
% norb2           = number of orbitals in SO set with kappa2 symmetry 
% T2(norb2,norb2) = T matrix for given kappa2 symmetry

% Number of spanned CSFs by the generator
ncsf = norb1*norb2;

% Construct C matrix for kappa1 symmetry
C1 = eye(ncsf,ncsf);
for k = 1:norb1      % loop over orbitals in SO set with kappa1 symmetry
  S = zeros(ncsf,ncsf);
  if k == 1          % no de-excitation
    C = eye(ncsf,ncsf);
  else            
    for l = 1:k-1    % loop over orbital de-excitations
      % Construct AdA matrix
      AdA = zeros(ncsf,ncsf);
      for m = 1:norb2
        ncol = m + norb2*(k-1);
        nrow = ncol - (k-l)*norb2;
        AdA(nrow,ncol) = 1;
      end
      % Form the S matrix
      S = T1(l,k)*AdA/T1(k,k) + S;
    end
    C = zeros(ncsf,ncsf);
    % Compute exp(S) by a terminating Taylor expansion
    for l = 0:2*abs(kappa1)
      C = S^l/factorial(l) + C;
    end
  end
  % Construct diagonal Tn
  Tn = eye(ncsf,ncsf);
  for l = 1:norb2
    Tn((k-1)*norb2+l,(k-1)*norb2+l) = T1(k,k);
  end
  % Final C matrix for orbital k
  C = C*Tn;
  % Accumulate C matrix for all orbitals
  C1 = C*C1;
end

% Construct C matrix for last kappa2 symmetry
% Loop over de-excitation pairs
C2 = eye(ncsf,ncsf);
for k = 1:norb2      % loop over orbitals in SO set with kappa2 symmetry
  S = zeros(ncsf,ncsf);
  if k == 1          % no de-excitation
    C = eye(ncsf,ncsf);
  else            
    for l = 1:k-1    % loop over orbital de-excitations
      % Construct AdA matrix
      AdA = zeros(ncsf,ncsf);
      for m = 1:norb1
        ncol = k + (m-1)*norb2;
        nrow = ncol - (k-l);
        AdA(nrow,ncol) = 1;
      end
      % Form the S matrix
      S = T2(l,k)*AdA/T2(k,k) + S;
    end
    C = zeros(ncsf,ncsf);
    % Compute exp(S) by a terminating Taylor expansion
    for l = 0:2*abs(kappa2)
      C = S^l/factorial(l) + C;
    end
  end
  % Construct diagonal Tn
  Tn = eye(ncsf,ncsf);
  for l = 1:norb1
    Tn((l-1)*norb2+k,(l-1)*norb2+k) = T2(k,k);
  end
  % Final C matrix for orbital k
  C = C*Tn;
  % Accumulate C2 matrix for all orbitals
  C2 = C*C2;
end
% Total weight transformation given by the C matrices of the two symmetries
Ctot = C1*C2;
\end{Verbatim}

\section{Validation}\label{Appendix_E}
\renewcommand{\theequation}{E.\arabic{equation}}
\setcounter{equation}{0}
The biorthogonal transformation can be validated by orbital rotations. To this end consider a CSF expansion, build on a single orbital basis, that can be divided in two parts $\{\Phi^1_{\alpha}\}$ and $\{\Phi^2_{\alpha'}\}$ 
that are both CUD. 
The Hamiltonian matrix can now be written
\begin{equation}\label{eq:eq1}
{\bm H} = \left(
\begin{array}{cc}
{\bm H}^{11} & {\bm H}^{12} \\
{\bm H}^{21} & {\bm H}^{22} \\
\end{array}
\right),
\end{equation}
where the ${\bm H}^{ik} = \langle \{\Phi^i_{\alpha}\} \|{\cal H}\| \{\Phi^k_{\alpha'} \} \rangle$. Diagonalize the total Hamiltonian matrix to get the energy eigenvalue and wave function of the lowest state. Now, for validation, build $\{\Phi^1_{\alpha}\}$ on the original orbitals and  $\{\Phi^2_{\alpha'}\}$ on an orbital set where we original orbitals have been rotated. Due to the fact that the orbital sets
are CUD, the expansion spans the same functional space as in the original case. Whereas the diagonal parts ${\bm H}^{11}$ and ${\bm H}^{22}$ can be constructed in the normal way based on the original orbital set and the rotated orbital set, respectively, we must apply the biorthonormal transformation for the off-diagonal part that is obtained as
\begin{equation}\label{eq:eq2}
{\bm H}^{12} =(\widetilde {\bm C}^{\,l})^T \widetilde {\bm H}^{12} \widetilde {\bm C}^{\,r},
\end{equation}
where $\widetilde {\bm H}^{12} = \langle \{\widetilde \Phi^1_{\alpha}\} \|{\cal H}\| \{\widetilde \Phi^2_{\alpha'} \} \rangle$ with tilde indicating that the CSFs are built on the biorthonormal orbitals. Diagonalizing the Hamiltonian matrix obtained in this way should give exactly the same energy and wave function as in the original case with a single orbital set.
This serves as a strict validation that the procedure with the biorthonormal transformation is correctly implemented. 

As a minimal example we consider the following CSF expansion
\begin{Verbatim}[fontsize=\footnotesize]
  1s ( 2)  2s ( 2)

                  0+
  1s ( 2)  3s ( 2)

                  0+
  1s ( 2)  3s ( 1)  4s ( 1)
               1/2      1/2
                           0+
  1s ( 2)  4s ( 2)

                  0+
  1s ( 1)  2s ( 1)  3s ( 2)
      1/2      1/2
                    0      0+
  1s ( 1)  2s ( 1)  3s ( 1)  4s ( 1)
      1/2      1/2      1/2      1/2
                    0      1/2      0+
  1s ( 1)  2s ( 1)  4s ( 2)
      1/2      1/2
                    0      0+
\end{Verbatim}
We divide the expansion in two parts
\begin{Verbatim}[fontsize=\footnotesize]
  1s ( 2)  2s ( 2)

                  0+
  1s ( 2)  3s ( 2)

                  0+
  1s ( 2)  3s ( 1)  4s ( 1)
               1/2      1/2
                           0+
  1s ( 2)  4s ( 2)

                  0+
\end{Verbatim}
and
\begin{Verbatim}[fontsize=\footnotesize]
  1s ( 1)  2s ( 1)  3s ( 2)
      1/2      1/2
                    0      0+
  1s ( 1)  2s ( 1)  3s ( 1)  4s ( 1)
      1/2      1/2      1/2      1/2
                    0      1/2      0+
  1s ( 1)  2s ( 1)  4s ( 2)
      1/2      1/2
                    0      0+
\end{Verbatim} 
Based on a single orthonormal orbital set from an MCDHF calculation targeted on the ground state of Be I we have the
following diagonal submatrices  (we show only upper triangular half and restrict the number of digits to save space)
\renewcommand{\theequation}{\text{\normalsize E.\arabic{equation}}}
\begin{equation}
\begingroup
\footnotesize
{\bm H}^{11} = \left(
\begin{array}{rrrr}
  -14.575891 &   0.054385 &  0.018142  & 0.029237  \\
             & -13.698394 &  0.331830  & 0.075088  \\
             &            & -7.930535  & 0.668493   \\
             &            &            & -1.510215  \\
\end{array}
\right),
\endgroup
\end{equation}
and
\renewcommand{\theequation}{\text{\normalsize E.\arabic{equation}}}
\begin{equation}\label{eq:D4}
\begingroup
\footnotesize
{\bm H}^{22} = \left(
\begin{array}{rrr}
  -9.516465 &  -0.044698 &   0.075088 \\
            & -4.646772  &  0.291963  \\
            &            & 0.875381   \\
\end{array}
\right).
\endgroup
\end{equation}
The off-diagonal $4 \times 3$ matrix is
\renewcommand{\theequation}{\text{\normalsize E.\arabic{equation}}}
\begin{equation}\label{eq:D5}
\begingroup
\footnotesize
{\bm H}^{12} = \left(
\begin{array}{rrr}
 -0.014487  & -0.052053  & -0.108875 \\
   0.010711  &  0.044548  &  0.000000 \\
   0.036940  &  0.137218  &  0.044548 \\
   0.000000 &   0.036940 &   0.263725 \\
\end{array}
\right).
\endgroup
\end{equation}
Diagonalizing the total Hamiltonian matrix in eq. (\ref{eq:eq1})  gives the
energy \mbox{$E = -14.5803913 $ a.u.}. For validation we apply rotations of the $3s$ and $4s$ orbitals in the SO set for the second part
\begin{eqnarray}
\phi_{3s} \leftarrow \frac{1}{\sqrt{2}}\left(\phi_{3s} + \phi_{4s}\right),~~~~\phi_{4s} \leftarrow \frac{1}{\sqrt{2}}\left(\phi_{4s} - \phi_{3s}\right).
\end{eqnarray}
The first diagonal block, ${\bm H}^{11}$, remains the same as above.
The second diagonal block, ${\bm H}^{22}$,
computed in the rotated orbital basis, is
\renewcommand{\theequation}{\text{\normalsize E.\arabic{equation}}}
\begin{equation}\label{eq:D7}
\begingroup
\footnotesize
{\bm H}^{22} = \left(
\begin{array}{rrr}
   -4.271270 &  3.842404   &  0.200659 \\
             &  -4.395630  &  3.505741 \\
             &             & -4.620955 \\
\end{array}
\right).
\endgroup
\end{equation}
To compute the off-diagonal block we perform the biorthonormal transformation and compute
$\widetilde {\bm H}^{12}  = \langle \{\widetilde \Phi^1_{\alpha}\} \|{\cal H}\| \{\widetilde \Phi^2_{\alpha'} \} \rangle$, where now tilde indicates that the CSFs are built on the biorthonormal orbitals.
We have
\renewcommand{\theequation}{\text{\normalsize E.\arabic{equation}}}
\begin{equation}
\begingroup
\footnotesize
\widetilde {\bm H}^{12} = \left(
\begin{array}{rrr}
   -0.098489  & -0.206026  & -0.217750  \\
    0.073711  &  0.089096  & 0.000000 \\
    0.085650  &  0.137218  & 0.089096 \\
    0.000000  &  0.085650  & 0.200724 \\
\end{array}
\right).
\endgroup
\end{equation}
The left and right hand side expansion coefficient transformation matrices, $\widetilde {\bm C}^l$ and $\widetilde {\bm C}^r$, respectively,   are
\renewcommand{\theequation}{\text{\normalsize E.\arabic{equation}}}
\begin{equation}
\begingroup
\footnotesize
\widetilde {\bm C}^{l} = \left(
\begin{array}{rrrr}
   1.000000  & 0.000000  & 0.000000  & 0.000000 \\
   0.000000  & 0.500000  & 0.707106  & 0.500000 \\
   0.000000  & 0.000000  & 1.000000  & 1.414213 \\
   0.000000  & 0.000000  & 0.000000  & 2.000000 \\
\end{array}
\right),
\endgroup
\end{equation}
and
\renewcommand{\theequation}{\text{\normalsize E.\arabic{equation}}}
\begin{equation}
\begingroup
\footnotesize
\widetilde {\bm C}^{r} = \left(
\begin{array}{rrr}
   1.000000  & -1.414213  &  1.000000 \\
   0.000000  &  1.000000  & -1.414213 \\
   0.000000  &  0.000000  &  1.000000 \\
\end{array}
\right).
\endgroup
\end{equation}
The expression ${\bm H}^{12} =(\widetilde {\bm C}^{\,l})^T \widetilde {\bm H}^{12} \widetilde {\bm C}^{\,r}$ gives
\renewcommand{\theequation}{\text{\normalsize E.\arabic{equation}}}
\begin{equation}\label{eq:D11}
\begingroup
\footnotesize
{\bm H}^{12} = \left(
\begin{array}{rrr}
  -0.098489  & -0.066742  & -0.024873 \\
   0.036855  & -0.007573  & -0.026144 \\
   0.137772  &  0.005379  & -0.056283 \\
   0.157983  &  0.186481  & 0.105741  \\
\end{array}
\right).
\endgroup
\end{equation}

Diagonalizing the total Hamiltonian obtained in the above way yields the energy $E = -14.5803913$ a.u., which is the same as in the original case with a single orbital basis. Thus, we have validated the method
of biorthonormal transformation.

We will now validate the partitioned correlation function interaction in section 5.
Again consider the minimalist example divided in two parts. Adding the reference CSF to the CSFs in the last part and performing an RCI calculation in the radial orbital basis from
the previous calculation we get the following expansion
coefficients 
\begin{Verbatim}[fontsize=\footnotesize]
  1s ( 2)  2s ( 2)                         0.99995925

                  0+
  1s ( 1)  2s ( 1)  3s ( 2)                0.00280443
      1/2      1/2
                    0      0+
  1s ( 1)  2s ( 1)  3s ( 1)  4s ( 1)       0.00505044
      1/2      1/2      1/2      1/2
                    0      1/2      0+
  1s ( 1)  2s ( 1)  4s ( 2)                0.00693654
      1/2      1/2
                    0      0+
\end{Verbatim}
We can view the three last CSFs as correlation corrections to the MR. By normalizing we get the partitioned correlation function 
\begin{Verbatim}[fontsize=\footnotesize]
  1s ( 1)  2s ( 1)  3s ( 2)                0.310670
      1/2      1/2
                    0      0+
  1s ( 1)  2s ( 1)  3s ( 1)  4s ( 1)       0.559480
      1/2      1/2      1/2      1/2
                    0      1/2      0+
  1s ( 1)  2s ( 1)  4s ( 2)                0.768418
      1/2      1/2
                    0      0+
\end{Verbatim}
The four CSFs from the first part plus the partitioned correlation function make up 5 orthonormal basis functions.
The first diagonal Hamiltonian matrix block, ${\bm H}^{11}$. is the same as the one in the previous cases. 
The $4 \times 1$ off-diagonal matrix between the four first CSFs and the partitioned correlation function is obtained by the matrix vector product
${\bm H}^{12}\, {\bm C}$. With ${\bm H}^{12}$ from eq. (\ref{eq:D5}) and the vector of expansion coefficients  ${\bm C}^T = (0.310670,\,0.559480,\,0.768418)$ from above we have
\renewcommand{\theequation}{\text{\normalsize E.\arabic{equation}}}
\begin{equation}
\begingroup
\footnotesize
 {\bm H}^{12}\,{\bm C} =     \left(
\begin{array}{r}
  -0.117285  \\
   0.028251  \\
   0.122479  \\
   0.223319  \\
\end{array}
\right).
\endgroup
\end{equation}
The diagonal matrix element of the partitioned correlation function is given by
${\bm C}^T\,{\bm H}^{22}\,{\bm C}$. Using ${\bm H}^{22}$ from eq. (\ref{eq:D4}) and performing the matrix vector multiplications gives
\renewcommand{\theequation}{\text{\normalsize E.\arabic{equation}}}
\begin{equation}
\begingroup
\footnotesize
{\bm C}^T\,{\bm H}^{22}\,{\bm C}  = 
-1.584776. 
\endgroup
\end{equation}
Diagonalizing the resulting $5 \times 5$ matrix yields the energy \mbox{$E = -14.5803907$ a.u.} to be compared  
with original $7 \times 7$ result $E = -14.5803913$ a.u..  
We see that the constraining has a very small influence on the total energy. Now for the validation.
Redoing the RCI calculation in the rotated orbital basis gives  
\begin{Verbatim}[fontsize=\footnotesize]
  1s ( 2)  2s ( 2)                         0.99995925

                  0+
  1s ( 1)  2s ( 1)  3s ( 2)                0.00844169
      1/2      1/2
                    0      0+
  1s ( 1)  2s ( 1)  3s ( 1)  4s ( 1)       0.00292184
      1/2      1/2      1/2      1/2
                    0      1/2      0+
  1s ( 1)  2s ( 1)  4s ( 2)                0.00129928
      1/2      1/2
                    0      0+
\end{Verbatim}
By normalizing we get the partitioned correlation function
\begin{Verbatim}[fontsize=\footnotesize]
  1s ( 1)  2s ( 1)  3s ( 2)                0.935156 
      1/2      1/2
                    0      0+
  1s ( 1)  2s ( 1)  3s ( 1)  4s ( 1)       0.323677
      1/2      1/2      1/2      1/2
                    0      1/2      0+
  1s ( 1)  2s ( 1)  4s ( 2)                0.143932
      1/2      1/2
                    0      0+
\end{Verbatim}
The four CSFs from the first part plus the constrained correlation function make up 5 orthonormal basis functions.
The first diagonal block, ${\bm H}^{11}$, is the same as in the previous cases. 
The $4 \times 1$ off-diagonal matrix between the four first CSFs and the partitioned correlation function in the rotated orbital basis 
is given by the matrix vector product
${\bm H}^{12}\, {\bm C}$ where now  ${\bm H}^{12}$ is taken from eq. (\ref{eq:D11}) and the vector of expansion coefficients ${\bm C}^T = ( 0.935156,\,0.323677,\,0.143932)$. Performing the multiplication we regain the values from 
the unrotated case, i.e.,
\renewcommand{\theequation}{\text{\normalsize E.\arabic{equation}}}
\begin{equation}
\begingroup
\footnotesize
{\bm H}^{12}\,{\bm C} =     
\left(
\begin{array}{r}
  -0.117285  \\
   0.028251  \\
   0.122479  \\
   0.223319  \\
\end{array}
\right).
\endgroup
\end{equation}
The diagonal matrix element of the partitioned correlation function is given by
${\bm C}^T\,{\bm H}^{22}\,{\bm C}$. Using ${\bm H}^{22}$ in the rotated orbital basis from eq. (\ref{eq:D7}) and performing the matrix vector multiplications gives the same value as in the unrotated basis
\renewcommand{\theequation}{\text{\normalsize E.\arabic{equation}}}
\begin{equation}
\begingroup
\footnotesize
{\bm C}^T\,{\bm H}^{22}\,{\bm C}  = 
-1.584776. 
\endgroup
\end{equation}
We have the same energy $E = -14.5803907$ a.u. as in the unrotated case and we have validated the biorthonormal transformation also in this case.

\section{Reduced matrix elements of one-particle scalar operator}
\label{Appendix_F}
\renewcommand{\theequation}{F.\arabic{equation}}
\setcounter{equation}{0}
\renewcommand{\thetable}{F.\arabic{table}}
\setcounter{table}{0}

In this appendix, we will discuss cases in which the spin-angular coefficients of reduced matrix elements take very simple values, even though their original expressions are quite complex and consist of products of coefficients of fractional parentage, unit tensors, and recoupling matrices. We will also identify cases where the spin-angular coefficients of reduced matrix elements between CSFs belonging to generator types 1–3 are equal to $1$. Obtaining expressions in the simplest possible form is important for the implementation of the PCFI method, since such expressions facilitate the calculations and reduce the number of algebraic operations required.

The reduced matrix elements of a one-particle scalar operator $\widehat{F}^{(0)}$ between configuration state functions with any number of open subshells, according to Eqs. (7) and (8) of \cite{GRASPGG}, can be expressed as a sum over one-electron contributions

\begin{eqnarray}
\label{eq:one-a}
    \redmem{\gamma_{\alpha} J}{\, \widehat{F}^{(0)} \,}{\gamma_{\beta} J^{\prime} } =
    {\sum_{n_i \ell _i j_{i}}} \; {\sum_{n_j \ell_j j_{j}}}
		\redmem{\gamma_{\alpha} J}{\, \widehat{F} ( n_i \ell _i j_{i}, n_j \ell_j j_{j}) \,}{\gamma_{\beta} J^{\prime }},
\end{eqnarray}
where

\begin{eqnarray}
\label{eq:one-b}
   \redmem{\gamma_{\alpha} J}{\, \widehat{F} ( n_i \ell _i j_{i}, n_j \ell_j j_{j} ) \,}{\gamma_{\beta} J^{\prime }}
\hspace{6.8cm}
      \nonumber  \\[1ex]
=
( -1)^{\Delta +1} \sqrt{2j_i +1} \; 
   R\left( j_i, j_j,\Lambda
  ^{bra},\Lambda ^{ket} \right) \;
	\redmem{(n_i \ell_i)\, j_{i}}{\, f^{(0)} \,}{(n_j\ell_j)\, j_{j}}
\hspace{0.5cm} 
    \nonumber  \\[1ex]
   \times  \left\{ \delta \left( i, j \right)
	 \redmem{(n_i\ell_i)j_{i}^{w_i} \;\alpha _i Q_i J_i}
	 {\, \left[  a^{( q \hspace{0.1cm} j_i )}_{\hspace{0.1cm}\frac{1}{2}} \times  a^{\hspace{0.05cm}( \hspace{0.05cm}q \hspace{0.1cm} j_i )}_{-\frac{1}{2}} \right] ^{\left( 0 \right)} \,}
	 {(n_i\ell_i)j_{i}^{w_{i}} \;\alpha _i Q_i  J_i }
\hspace{-0.2cm}
   \nonumber \right. \\[1ex]
   \left. + (1-\delta ( i , j )) \;
	 \redmem{(n_i\ell_i)j_{i}^{w_i}\,\alpha _i Q_i J_i}{\, a^{( q \hspace{0.1cm} j_i )}_{\hspace{0.1cm}\frac{1}{2}} \,}{(n_i\ell_i)j_i^{w_{i}^{\prime }} \,\alpha _i ^{\prime} Q_i ^{\prime} J_i ^{\prime}}
\hspace{1.0cm}
   \nonumber \right. \\[1ex]
   \times \left.
	 \redmem{(n_j\ell_j)j_j^{w_j}\;\alpha _j Q_j J_j}{\, a^{\hspace{0.05cm}( \hspace{0.05cm}q \hspace{0.1cm} j_j )}_{-\frac{1}{2}} \,}{(n_j\ell_j)j_j^{w_{j}^{\prime }} \,\alpha_j ^{\prime} Q_j ^{\prime} J_j ^{\prime}}
   \right\}.
\hspace{1.0cm}
\end{eqnarray} 
For the case
\begin{eqnarray}
\label{eq:one-c}
\bram{\gamma_{\alpha} J} = \bram{\mathcal{A}_{\alpha}}
\hspace{9.0cm}
      \nonumber  \\[1ex]
 \equiv
\bram{\gamma_{\alpha_1} J_{\alpha_1} \,\, (n_1\ell)j^{w_1}\,\alpha _1 Q_1 J_1 \, (J_{\alpha_1 1}) \, (n_2\ell)j^{w_2} \,\alpha _2 Q_2 J_2 \, \,\,  J}
\hspace{1.8cm}
\end{eqnarray} 
and 
\begin{eqnarray}
\label{eq:one-d}
\ketm{\gamma_{\beta} J} = \ketm{\mathcal{A}_{\beta}}
\hspace{9.0cm}
      \nonumber  \\[1ex]
 \equiv
\ketm{\gamma_{\alpha_1} J_{\alpha_1} \,\, (n_1\ell)j^{w_1-1}\,\alpha' _1 Q'_1 J'_1 \, (J'_{\alpha_1 1}) \, (n_2\ell)j^{w_2+1} \,\alpha' _2 Q'_2 J'_2 \, \,\, J}
\hspace{1.0cm}
\end{eqnarray} 
the recoupling matrix in (\ref{eq:one-b}) according to (11) from \cite{GRASPGG} has the following form
\begin{eqnarray}
\label{eq:one-e}
   R\left( j_1, j_1,\Lambda
  ^{bra},\Lambda ^{ket} \right)
\hspace{6cm}
        \nonumber  \\[1ex]
\hspace{-1.5cm} =
( -1)^{1+J_{\alpha_1}+J'_1+J_2+J_{\alpha_1 1}+J'_{\alpha_1 1}+J} \sqrt{\frac{\left[J_{\alpha_1 1},J'_{\alpha_1 1}\right]}{\left[j\right]}} 
   \nonumber \\[1ex]
\hspace{-1.0cm}
\times
\left\{
   \begin{array}{lll}
     j            & J'_{1}          & J_{1} \\
     J_{\alpha_1} & J_{\alpha_1 1} & J'_{\alpha_1 1}
   \end{array}
\hspace{-0.1cm} \right\}
\,
\left\{
   \begin{array}{lll}
     j & J'_{2}          & J_{2} \\
     J & J_{\alpha_1 1} & J'_{\alpha_1 1}
   \end{array}
\hspace{-0.1cm} \right\}
\end{eqnarray} 
the phase factor $\Delta$ according to (15) from \cite{GRASPGG} is given by
\begin{equation}
\label{eq:one-g}
   \Delta = w_1+1.
\hspace{8cm}
\end{equation}

For the orbital biorthonormal transformation, the one-electron reduced matrix element in (\ref{eq:one-b}) is as follows
\begin{eqnarray}
\label{eq:one-f}
       \redmem{(n_1 \ell)\, j}{\, f^{(0)} \,}{(n_2 \ell)\, j}\; = \; 1.
\hspace{5cm}
\end{eqnarray}

Then reduced matrix element of one-particle scalar operator in Eq. (\ref{eq:one-b}) between configuration state functions Eqs. (\ref{eq:one-c}) and (\ref{eq:one-d}) is then given by
\begin{eqnarray}
\label{eq:one-h}
   \redmem{\mathcal{A}_{\alpha}}{\, \widehat{F} ( n_1 \ell  j, n_2 \ell j ) \,}{\mathcal{A}_{\beta}}
\hspace{6cm}
      \nonumber  \\[1ex]
 =
(-1)^{w_1+1+J_{\alpha_1}+J'_1+J_2+J_{\alpha_1 1}+J'_{\alpha_1 1}+J}
\sqrt{\left[J_{\alpha_1 1},J'_{\alpha_1 1}\right]}
\hspace{1.0cm}
   \nonumber \\[1ex]
\times
\left\{
   \begin{array}{lll}
     j            & J'_{1}          & J_{1} \\
     J_{\alpha_1} & J_{\alpha_1 1} & J'_{\alpha_1 1}
   \end{array}
\hspace{-0.1cm} \right\}
\,
\left\{
   \begin{array}{lll}
     j & J'_{2}          & J_{2} \\
     J & J_{\alpha_1 1} & J'_{\alpha_1 1}
   \end{array}
\hspace{-0.1cm} \right\}
\hspace{1.1cm}
   \nonumber \\[1ex]
   \times 
	 \redmem{(n_1\ell)j^{w_1}\,\alpha _1 Q_1 J_1}{\, a^{( q \hspace{0.1cm} j )}_{\hspace{0.1cm}\frac{1}{2}} \,}{(n_1\ell)j^{w_{1}-1} \,\alpha _1 ^{\prime} Q_1 ^{\prime} J_1 ^{\prime}}
\hspace{0.2cm}
   \nonumber \\[1ex]
\hspace{-1.0cm}
   \times 
	 \redmem{(n_2\ell)j^{w_2}\;\alpha _2 Q_2 J_2}{\, a^{\hspace{0.05cm}( \hspace{0.05cm}q \hspace{0.1cm} j )}_{-\frac{1}{2}} \,}{(n_2\ell)j^{w_{2}+1} \,\alpha_2 ^{\prime} Q_2 ^{\prime} J_2 ^{\prime}} .
\end{eqnarray}

The reduced matrix elements of operators $a^{( q \hspace{0.15cm} j )}_{\hspace{0.1cm}\frac{1}{2} \, m_j}\equiv a\dn_{m_j}^{( j )} \,$ ($a\dn_{m_j}^{( j )}$ is an electron annihilation operator) and $a^{\hspace{0.05cm}( \hspace{0.05cm}q \hspace{0.15cm} j )}_{-\frac{1}{2} \, m_j}\equiv \tilde a\dn_{\hspace{0.05cm}m_j}^{\left( j \right)}\equiv \left( -1 \right)^{j-m_j }a\dn_{-m_j } ^{\dagger \, ( j )  \,}$ ($a\dn_{-m_j } ^{\dagger \, ( j ) }$ is an electron creation operator) of second quantization in (\ref{eq:one-h}) according to (6) of \cite{Gaietal:2000a} and (5), (16) of \cite{Gaietal:2000a} can be expressed, respectively, via the coefficient of fractional parentage 
\begin{eqnarray}
\label{eq:one-ag}
\redmem{(n\ell)j^{w}\,\alpha Q J}{\, a^{( q \hspace{0.1cm} j )}_{\hspace{0.1cm}\frac{1}{2}} \,}{(n\ell)j^{w-1} \,\alpha ^{\prime} Q^{\prime} J^{\prime}}
\hspace{2cm}
      \nonumber  \\[1ex]
= (-1)^{w} \sqrt{w \left[J\right]}
\left(j^{w-1} \: \alpha^{\prime} Q^{\prime} J^{\prime} \: |\} \:j^w \: \alpha QJ \right),
\end{eqnarray}
\begin{eqnarray}
\label{eq:one-am}
\redmem{(n\ell)j^{w-1}\;\alpha Q J}{\, a^{\hspace{0.05cm}( \hspace{0.05cm}q \hspace{0.1cm} j )}_{-\frac{1}{2}} \,}{(n\ell)j^{w} \,\alpha ^{\prime} Q^{\prime} J^{\prime}}
\hspace{2cm}
      \nonumber  \\[1ex]
= (-1)^{w+J-J'+j+1} \sqrt{w' \left[J'\right]}
\left(j^{w-1} \: \alpha Q J \: |\} \:j^w \: \alpha^{\prime} Q^{\prime} J^{\prime} \right).
\hspace{-1.0cm}
\end{eqnarray}

Note that the notation $\gamma_{\alpha_1} J_{\alpha_1}$ in (\ref{eq:one-c}) and (\ref{eq:one-d}) defines the part of the configuration state functions $\bram{\mathcal{A}_{\alpha}}$ and $\ketm{\mathcal{A}_{\beta}}$ that is the same in both the bra and ket configuration state functions and consists of any number of subshells. The resultant angular momentum of this part in $j$ space is $J_{\alpha_1}$. We will use the same notation below to define other types of configuration state functions.

For the case
\begin{eqnarray}
\label{eq:one-i}
\bram{\gamma_{\alpha} J} = \bram{\mathcal{B}_{\alpha}}
\hspace{9.0cm}
      \nonumber  \\[1ex]
 \equiv
\bram{\gamma_{\alpha_1} J_{\alpha_1} \,\, (n_1\ell)j^{1}\,\alpha _1 Q_1 J_1 \hspace{0.05cm}\shorteq\hspace{0.05cm} j \, (J_{\alpha_1 1}) \, (n_2\ell)j^{0} \,\alpha _2 Q_2 J_2 \hspace{0.05cm}\shorteq\hspace{0.05cm} 0 \, \, \, J \hspace{0.05cm}\shorteq\hspace{0.05cm} J_{\alpha_1 1}}, 
\hspace{0.6cm}
\end{eqnarray} 
\begin{eqnarray}
\label{eq:one-j}
\ketm{\gamma_{\beta} J} = \ketm{\mathcal{B}_{\beta}}
\hspace{9.0cm}
      \nonumber  \\[1ex]
 \equiv
\ketm{\gamma_{\alpha_1} J_{\alpha_1} \,\, (n_1\ell)j^{0}\,\alpha' _1 Q'_1 J'_1 \hspace{0.05cm}\shorteq\hspace{0.05cm} 0 \, (J'_{\alpha_1 1} \hspace{0.05cm}\shorteq\hspace{0.05cm} J_{\alpha_1}) \, (n_2\ell)j^{1} \,\alpha' _2 Q'_2 J'_2 \hspace{0.05cm}\shorteq\hspace{0.05cm} j \, \, \, J \hspace{0.05cm}\shorteq\hspace{0.05cm} J_{\alpha_1 1} }
\end{eqnarray} 
and taking into account (6.8.39) from \cite{IAN}
\begin{eqnarray}
\label{eq:one-a0}
\left(j^{0} \: \alpha^{\prime} Q^{\prime} J^{\prime} \shorteq\hspace{0.05cm} 0 \: |\} \:j^1 \: \alpha QJ \shorteq\hspace{0.05cm} j \right) = 1
\end{eqnarray}
the reduced matrix element (\ref{eq:one-h}) becomes
\begin{eqnarray}
\label{eq:one-k}
   \redmem{\mathcal{B}_{\alpha}}{\, \widehat{F} ( n_1 \ell  j, n_2 \ell j ) \,}{\mathcal{B}_{\beta}} = 1.
\end{eqnarray}

Meanwhile, for the following configuration state functions
\begin{eqnarray}
\label{eq:one-l}
\bram{\gamma_{\alpha} J} = \bram{\mathcal{C}_{\alpha}}
\hspace{8.5cm}
      \nonumber  \\[1ex]
 \equiv
\bram{\gamma_{\alpha_1} J_{\alpha_1} \,\, (n_1\ell)j^{2}\,\alpha _1 Q_1 J_1 \, (J_{\alpha_1 1}) \, (n_2\ell)j^{0} \,\alpha _2 Q_2 J_2 \hspace{0.05cm}\shorteq\hspace{0.05cm} 0 \, \, \, J \hspace{0.05cm}\shorteq\hspace{0.05cm} J_{\alpha_1 1}},
\hspace{0.3cm}
\end{eqnarray} 
\begin{eqnarray}
\label{eq:one-m}
\ketm{\gamma_{\beta} J} = \ketm{\mathcal{C}_{\beta}}
\hspace{8.5cm}
      \nonumber  \\[1ex]
 \equiv
\ketm{\gamma_{\alpha_1} J_{\alpha_1} \,\, (n_1\ell)j^{1}\,\alpha' _1 Q'_1 J'_1 \hspace{0.05cm}\shorteq\hspace{0.05cm} j \, (J'_{\alpha_1 1}) \, (n_2\ell)j^{1} \,\alpha' _2 Q'_2 J'_2 \hspace{0.05cm}\shorteq\hspace{0.05cm} j  \, \, \, J \hspace{0.05cm}\shorteq\hspace{0.05cm} J_{\alpha_1 1}}
\end{eqnarray}
taking into account  (6.8.39) of Ref. \cite{IAN}
\begin{eqnarray}
\label{eq:one-a2}
\left(j^{1} \: \alpha^{\prime} Q^{\prime} J^{\prime} \shorteq\hspace{0.05cm} j \: |\} \:j^2 \: \alpha QJ \right) = \frac{1+(-1)^{J}}{2}
\end{eqnarray}
the reduced matrix element of the one-particle scalar operator (\ref{eq:one-b}) becomes
\begin{eqnarray}
\label{eq:one-n}
   \redmem{\mathcal{C}_{\alpha}}{\, \widehat{F} ( n_1 \ell  j, n_2 \ell j ) \,}{\mathcal{C}_{\beta}}
\hspace{6.5cm}
   \nonumber \\[1ex]
= (-1)^{1+J_{\alpha_1}+J} \,
\sqrt{2\left[ J_{1}, J'_{\alpha_1 1} \right]}
\, 
\left\{
   \begin{array}{lll}
     j            & j              & J_{1} \\
     J_{\alpha_1} & J & J'_{\alpha_1 1}
   \end{array}
\hspace{-0.1cm}\right\} 
 \frac{1+(-1)^{J_1}}{2}.
\end{eqnarray}

From (\ref{eq:one-n}), the reduced matrix element of the one-particle scalar operator between the following configuration state functions 
\begin{eqnarray}
\label{eq:one-w}
\bram{\gamma_{\alpha} J} = \bram{\mathcal{D}_{\alpha}}
\hspace{9.2cm}
      \nonumber  \\[1ex]
 \equiv
\bram{\gamma_{\alpha_1} J_{\alpha_1} \hspace{-0.05cm}\shorteq\hspace{0.05cm} 0 \, (n_1\ell)j^{2}\,\alpha _1 Q_1 J_1 \, (J_{\alpha_1 1} \hspace{0.05cm}\shorteq\hspace{0.05cm} J_1) \, (n_2\ell)j^{0} \,\alpha _2 Q_2 J_2 \hspace{0.05cm}\shorteq\hspace{0.05cm} 0 \, \, \, J \hspace{0.05cm}\shorteq\hspace{0.05cm} J_1},
\hspace{0.4cm}
\end{eqnarray} 
\begin{eqnarray}
\label{eq:one-o}
\ketm{\gamma_{\beta} J} = \ketm{\mathcal{D}_{\beta}}
\hspace{9.2cm}
      \nonumber  \\[1ex]
 \equiv
\ketm{\gamma_{\alpha_1} J_{\alpha_1} \hspace{-0.05cm}\shorteq\hspace{0.05cm} 0 \,\, (n_1\ell)j^{1}\,\alpha' _1 Q'_1 J'_1 \hspace{0.05cm}\shorteq\hspace{0.05cm} j \, (J'_{\alpha_1 1} \shorteq\hspace{0.05cm} j) \, (n_2\ell)j^{1} \,\alpha' _2 Q'_2 J'_2 \hspace{0.05cm}\shorteq\hspace{0.05cm} j \, \, \, J \hspace{0.05cm}\shorteq\hspace{0.05cm} J_1}
\hspace{0.4cm}
\end{eqnarray} 
has the form
\begin{eqnarray}
\label{eq:one-p}
   \redmem{\mathcal{D}_{\alpha}}{\, \widehat{F} ( n_1 \ell  j, n_2 \ell j ) \,}{\mathcal{D}_{\beta}}
= \sqrt{2} \, \, \, \frac{1+(-1)^{J_1}}{2}.
\end{eqnarray}

From (\ref{eq:one-n}) and (\ref{eq:one-p}), we obtain the following relationship
\begin{eqnarray}
\label{eq:rel-C_D}
   \redmem{\mathcal{C}_{\alpha}}{\, \widehat{F} ( n_1 \ell  j, n_2 \ell j ) \,}{\mathcal{C}_{\beta}}
\hspace{8.0cm}
   \nonumber \\[1ex]
= (-1)^{1+J_{\alpha_1}+J} \,
\sqrt{\left[ J_{1}, J'_{\alpha_1 1} \right]} 
\left\{
   \begin{array}{lll}
     j            & j              & J_{1} \\
     J_{\alpha_1} & J & J'_{\alpha_1 1} 
   \end{array}
\hspace{-0.1cm}\right\} 
 \redmem{\mathcal{D}_{\alpha}}{\, \widehat{F} ( n_1 \ell  j, n_2 \ell j ) \,}{\mathcal{D}_{\beta}}.
\hspace{0.5cm}
\end{eqnarray}

As in Eqs. (\ref{eq:one-c}) and (\ref{eq:one-d}), and Eqs. (\ref{eq:one-l}) and (\ref{eq:one-m}), the reduced matrix element of the one-particle scalar operator Eqs. (\ref{eq:one-b}) between the following configuration state functions
\begin{eqnarray}
\label{eq:one-r}
\bram{\gamma_{\alpha} J} = \bram{\mathcal{E}_{\alpha}}
\hspace{8.5cm}
      \nonumber  \\[1ex]
 \equiv
\bram{\gamma_{\alpha_1} J_{\alpha_1} \,\, (n_1\ell)j^{0}\,\alpha _1 Q_1 J_1 \shorteq\hspace{0.05cm} 0 \, (J_{\alpha_1 1} \shorteq\hspace{0.05cm} J_{\alpha_1}) \, (n_2\ell)j^{2} \,\alpha _2 Q_2 J_2 \, \, \, J}
\hspace{0.4cm}
\end{eqnarray} 
and
\begin{eqnarray}
\label{eq:one-s}
\ketm{\gamma_{\beta} J} = \ketm{\mathcal{E}_{\beta}}
\hspace{8.5cm}
      \nonumber  \\[1ex]
 \equiv
\ketm{\gamma_{\alpha_1} J_{\alpha_1} \,\, (n_1\ell)j^{1}\,\alpha' _1 Q'_1 J'_1  \shorteq\hspace{0.05cm} j \, (J'_{\alpha_1 1}) \, (n_2\ell)j^{1} \,\alpha' _2 Q'_2 J'_2 \hspace{0.05cm}\shorteq\hspace{0.05cm} j  \, \, \, J}
\hspace{0.8cm}
\end{eqnarray}
can be expressed as
\begin{eqnarray}
\label{eq:one-t}
   \redmem{\mathcal{E}_{\alpha}}{\, \widehat{F} ( n_2 \ell  j, n_1 \ell j ) \,}{\mathcal{E}_{\beta}}
\hspace{6.5cm}
   \nonumber \\[1ex]
= (-1)^{1+J_{\alpha_1}+J} \,
\sqrt{2\left[ J_{2}, J'_{\alpha_1 1} \right]}
\, 
\left\{
   \begin{array}{lll}
     j            & j & J_{2} \\
     J_{\alpha_1} & J & J'_{\alpha_1 1}
   \end{array}
\hspace{-0.1cm}\right\}
\frac{1+(-1)^{J_2}}{2}.
\end{eqnarray}
As we can see, using the algebraic expression of the reduced matrix element 
$\redmem{\mathcal{C}_{\alpha}}{\, \widehat{F} ( n_1 \ell j, n_2 \ell j ) \,}{\mathcal{C}_{\beta}}$, to write down the algebraic expression of the reduced matrix element $\redmem{\mathcal{E}_{\alpha}}{\, \widehat{F} ( n_2 \ell j, n_1 \ell j ) \,}{\mathcal{E}_{\beta}}$, one only has to replace $J_1$ by $J_2$ in the expression (\ref{eq:one-n}), and vice versa to get the expression (\ref{eq:one-n}) from (\ref{eq:one-t}). We can therefore write
\begin{eqnarray}
\label{eq:one-t2}
   \redmem{\mathcal{C}_{\alpha}}{\, \widehat{F} ( n_1 \ell  j, n_2 \ell j ) \,}{\mathcal{C}_{\beta}}
\begin{array}{c}
^{(J_1 \longleftrightarrow J_2)} \\[-8pt]
\vspace{0.05cm}
 =	\\
\end{array} 
   \redmem{\mathcal{E}_{\alpha}}{\, \widehat{F} ( n_2 \ell  j, n_1 \ell j ) \,}{\mathcal{E}_{\beta}}.
\end{eqnarray}

The reduced matrix element between configuration state functions
\begin{eqnarray}
\label{eq:one-v}
\bram{\gamma_{\alpha} J} = \bram{\mathcal{F}_{\alpha}}
\hspace{8.5cm}
      \nonumber  \\[1ex]
 \equiv
\bram{\gamma_{\alpha_1} J_{\alpha_1}\hspace{-0.05cm}\shorteq\hspace{0.05cm}0 \,\, (n_1\ell)j^{0}\,\alpha _1 Q_1 J_1 \shorteq\hspace{0.05cm} 0 \, (J_{\alpha_1 1} \shorteq\hspace{0.05cm} 0) \, (n_2\ell)j^{2} \,\alpha _2 Q_2 J_2 \, \, \, J \shorteq\hspace{0.05cm} J_2}
\end{eqnarray}
and
\begin{eqnarray}
\label{eq:one-x}
\ketm{\gamma_{\beta} J} = \ketm{\mathcal{F}_{\beta}}
\hspace{8.5cm}
      \nonumber  \\[1ex]
 \equiv
\ketm{\gamma_{\alpha_1} J_{\alpha_1}\hspace{-0.05cm}\shorteq\hspace{0.05cm}0
\,\, (n_1\ell)j^{1}\,\alpha' _1 Q'_1 J'_1 \shorteq\hspace{0.05cm} j \, (J'_{\alpha_1 1} \shorteq\hspace{0.05cm} j) \, (n_2\ell)j^{1} \,\alpha' _2 Q'_2 J'_2 \hspace{0.05cm}\shorteq\hspace{0.05cm} j  \, \, \, J}
\end{eqnarray}
is then
\begin{eqnarray}
\label{eq:one-z}
   \redmem{\mathcal{D}_{\alpha}}{\, \widehat{F} ( n_1 \ell  j, n_2 \ell j ) \,}{\mathcal{D}_{\beta}}
\begin{array}{c}
^{(J_1 \longleftrightarrow J_2)} \\[-8pt]
\vspace{0.05cm}
=	\\
\end{array} 
   \redmem{\mathcal{F}_{\alpha}}{\, \widehat{F} ( n_2 \ell  j, n_1 \ell j ) \,}{\mathcal{F}_{\beta}}.
\end{eqnarray}
This means that the reduced matrix element $\redmem{\mathcal{D}_{\alpha}}{\, \widehat{F} ( n_1 \ell  j, n_2 \ell j ) \,}{\mathcal{D}_{\beta}}$
expression is equal to the reduced  matrix element
$\redmem{\mathcal{F}_{\alpha}}{\, \widehat{F} ( n_2 \ell  j, n_1 \ell j ) \,}{\mathcal{F}_{\beta}}$ expression by replacing $J_1$ with $J_2$ in expression (\ref{eq:one-p}).
From (\ref{eq:one-t}) and (\ref{eq:one-z}), we obtain the following relationship
\begin{eqnarray}
\label{eq:rel-E_F}
   \redmem{\mathcal{E}_{\alpha}}{\, \widehat{F} ( n_2 \ell  j, n_1 \ell j ) \,}{\mathcal{E}_{\beta}}
\hspace{8.0cm}
   \nonumber \\[1ex]
= (-1)^{1+J_{\alpha_1}+J} \,
\sqrt{\left[ J_{2}, J'_{\alpha_1 1} \right]}
\, 
\left\{
   \begin{array}{lll}
     j            & j & J_{2} \\
     J_{\alpha_1} & J & J'_{\alpha_1 1}
   \end{array}
\hspace{-0.1cm}\right\}
\redmem{\mathcal{F}_{\alpha}}{\, \widehat{F} ( n_2 \ell  j, n_1 \ell j ) \,}{\mathcal{F}_{\beta}}.
\hspace{0.5cm}
\end{eqnarray}

From the above expressions, it is easy to obtain expressions for the reduced matrix elements between configuration state functions where both the bra and ket functions do not share a common member $\gamma_{\alpha_1} J_{\alpha_1}$. For example, configuration state functions $\bram{\gamma_{\alpha} J} = \bram{\mathcal{A}_{\alpha}}$ and $\ketm{\gamma_{\beta} J} = \ketm{\mathcal{A}_{\beta}}$ without this member are
\begin{eqnarray}
\label{eq:one-ca}
\bram{\gamma_{\alpha} J} = \bram{\mathcal{A'}_{\alpha}}
%
 \equiv
\bram{ (n_1\ell)j^{w_1}\,\alpha _1 Q_1 J_1 \, (J_{\alpha_1 1}) \, (n_2\ell)j^{w_2} \,\alpha _2 Q_2 J_2 \, \,\,  J}
\hspace{1.0cm}
\end{eqnarray} 
and 
\begin{eqnarray}
\label{eq:one-da}
\ketm{\gamma_{\beta} J} = \ketm{\mathcal{A'}_{\beta}}
%
 \equiv
\ketm{ (n_1\ell)j^{w_1-1}\,\alpha' _1 Q'_1 J'_1 \, (J'_{\alpha_1 1}) \, (n_2\ell)j^{w_2+1} \,\alpha' _2 Q'_2 J'_2 \, \,\, J}.
\end{eqnarray} 

The reduced matrix element between these configurations can be written using (\ref{eq:one-h}).
\begin{eqnarray}
\label{eq:one-h2}
   \redmem{\mathcal{A}_{\alpha}}{\, \widehat{F} ( n_1 \ell  j, n_2 \ell j ) \,}{\mathcal{A}_{\beta}}
   \begin{array}{c}
^{(J_{\alpha_1} \shorteq\hspace{0.04cm} \, 0)} \\[-8pt]
\vspace{0.05cm}
~~=	\\
\end{array}
   \redmem{\mathcal{A'}_{\alpha}}{\, \widehat{F} ( n_1 \ell  j, n_2 \ell j ) \,}{\mathcal{A'}_{\beta}}.
\end{eqnarray}
The same principle is used to obtain reduced matrix elements for the remaining configuration state functions $\bram{\mathcal{B'}_{\alpha}}$, $\bram{\mathcal{C'}_{\alpha}}$, and $\bram{\mathcal{E'}_{\alpha}}$.

The following reduced matrix elements are equal to the complex-conjugated reduced matrix elements.
\begin{eqnarray}
\label{eq:one-kt}
 \redmem{\mathcal{B_{\alpha}}}{\, \widehat{F} ( n_1 \ell  j, n_2 \ell j ) \,}{\mathcal{B}_{\beta}} 
=
 \redmem{\mathcal{B_{\beta}}}{\, \widehat{F} ( n_2 \ell  j, n_1 \ell j ) \,}{\mathcal{B}_{\alpha}},
\end{eqnarray}
\begin{eqnarray}
\label{eq:one-pt}
 \redmem{\mathcal{D}_{\alpha}}{\, \widehat{F} ( n_1 \ell  j, n_2 \ell j ) \,}{\mathcal{D}_{\beta}} 
= 
 \redmem{\mathcal{D}_{\beta}}{\, \widehat{F} ( n_2 \ell  j, n_1 \ell j ) \,}{\mathcal{D}_{\alpha}},
\end{eqnarray}
\begin{eqnarray}
\label{eq:one-zt}
  \redmem{\mathcal{F}_{\alpha}}{\, \widehat{F} ( n_2 \ell  j, n_1 \ell j ) \,}{\mathcal{F}_{\beta}}
=
  \redmem{\mathcal{F}_{\beta}}{\, \widehat{F} ( n_1 \ell  j, n_2 \ell j ) \,}{\mathcal{F}_{\alpha}}.
\end{eqnarray}
The following relationship also applies
\begin{eqnarray}
\label{eq:one-nt}
 \redmem{\mathcal{C}_{\alpha}}{\, \widehat{F} ( n_1 \ell  j, n_2 \ell j ) \,}{\mathcal{C}_{\beta}}
 \hspace{-0.6cm}
\begin{array}{c}
^{(J_{\alpha _1 1} \leftrightarrow J'_{\alpha _1 1}) (J'_{\alpha _1 1} \leftrightarrow J_{\alpha _1 1})} \\[-3pt]
^{(J_1 \leftrightarrow J'_1)} \\[-7pt]
\vspace{0.6cm}
 =	\\
\end{array}
\hspace{-0.7cm}
  \redmem{\mathcal{C}_{\beta}}{\, \widehat{F} ( n_2 \ell  j, n_1 \ell j ) \,}{\mathcal{C}_{\alpha}},
\end{eqnarray}
\begin{eqnarray}
\label{eq:one-tt}
 \redmem{\mathcal{E}_{\alpha}}{\, \widehat{F} ( n_2 \ell  j, n_1 \ell j ) \,}{\mathcal{E}_{\beta}}
   \begin{array}{c}
^{(J'_{\alpha _1 1} \longleftrightarrow J_{\alpha _1 1})} \\[-3pt]
^{(J_2 \leftrightarrow J'_2)} \\[-7pt]
\vspace{0.6cm}
 =	\\
\end{array}
\redmem{\mathcal{E}_{\beta}}{\, \widehat{F} ( n_1 \ell  j, n_2 \ell j ) \,}{\mathcal{E}_{\alpha}}.
\end{eqnarray}
To summarize, the main expressions for the reduced matrix element of the one-particle scalar operator required for the orbital biorthonormal transformation are given in Eqs. (\ref{eq:one-h}), (\ref{eq:one-k}), and (\ref{eq:one-n}). The remaining expressions can be derived from these basic expressions. Table~\ref{tab:CSF_transformation} list the expressions for the reduced matrix elements required for the orbital transformations associated with the individual generator types.

\begin{table}[!ht]
\caption{
\label{tab:CSF_transformation}
 Reduced matrix elements for different types of CSF that require orbital transformation.}
\vspace{-0.4cm}
\begin{center}
\begin{tabular}{cc|cc|cc} \hline 
\multicolumn{2}{c|}{Configuration state functions} & \multicolumn{2}{c|}{Types of generators} & \multirow{2}{1.30cm}{Expression}&\\ \cline{1-2} \cline{3-4}
\multicolumn{1}{c}{Bra} & \multicolumn{1}{c|}{Ket} & \multicolumn{1}{c}{Bra} & \multicolumn{1}{c|}{Ket} &  &\\
\hline
$\bram{\mathcal{B}_{\alpha}}$ & $\ketm{\mathcal{B}_{\beta}}$  & 1 & 1 & (\ref{eq:one-k}) &\\
& & 2& 2& &\\
& & 3& 3& &\\
& & 3& 4& &\\
$\bram{\mathcal{C}_{\alpha}}$ & $\ketm{\mathcal{C}_{\beta}}$  & 3 & 4 & (\ref{eq:one-n}) &\\
$\bram{\mathcal{D}_{\alpha}}$ & $\ketm{\mathcal{D}_{\beta}}$  & 3 & 4 & (\ref{eq:one-p}) &\\
$\bram{\mathcal{E}_{\alpha}}$ & $\ketm{\mathcal{E}_{\beta}}$  & 3 & 4 & (\ref{eq:one-t}) &\\
$\bram{\mathcal{F}_{\alpha}}$ & $\ketm{\mathcal{F}_{\beta}}$  & 3 & 4 & (\ref{eq:one-z}) &\\
\hline
\end{tabular}
\end{center}
\end{table}

\end{document}